\documentstyle[aps,epsf,eqsecnum,tighten,preprint]{revtex}
\begin{document}
\title{Spin dynamics and transport in
gapped one-dimensional
Heisenberg antiferromagnets at nonzero temperatures}
\author{Kedar Damle and Subir Sachdev}
\address{
Department of Physics, P.O. Box 208120,
Yale University, New Haven, CT 06520-8120}
\date{October 29, 1997}
\preprint{cond-mat/9711014}
\maketitle
\begin{abstract}
We present the theory of nonzero temperature ($T$) spin dynamics and
transport in one-dimensional Heisenberg antiferromagnets 
with an energy gap $\Delta$.
For $T \ll \Delta$, we develop a semiclassical picture of thermally excited
particles.  Multiple inelastic collisions between the particles 
are crucial,
and are described by a two-particle ${\cal S}$-matrix which has
a {\em super-universal}
form at low momenta. This is established by computations on the $O(3)$ $\sigma$-model,
and strong and weak coupling expansions (the latter using a Majorana fermion
representation)
for the two-leg $S=1/2$ Heisenberg antiferromagnetic ladder. 
As an aside, we note that the strong-coupling calculation reveals a $S=1$, two
particle bound state which leads to the presence of a second peak in the
$T=0$ inelastic neutron scattering (INS) 
cross-section for a range of values of momentum transfer. 
We obtain exact, or numerically exact, universal expressions 
for the thermal broadening of the quasi-particle
peak in the INS cross-section, for the magnetization transport,
and for the field dependence of the NMR relaxation rate $1/T_1$ of
the effective semiclassical model:
these are expected to be asymptotically exact 
for the quantum antiferromagnets.
 The results for $1/T_1$ are compared with
the experimental findings of Takigawa {\em et.al.} and the agreement
is quite good. In the regime
$\Delta < T < \mbox{(a typical microscopic exchange)}$ 
we argue that a complementary description in terms of semiclassical waves applies,
and give some exact results for the thermodynamics and dynamics.
\end{abstract}
\pacs{PACS numbers:}
\newpage
\section{Introduction}
\label{intro}
For more than a decade now, much effort has been devoted to understanding the
properties of a variety of insulating one-dimensional Heisenberg antiferromagnets. By now, the 
basic facts about these systems are very well
established: Heisenberg antiferromagnetic (HAF) chains with integer spins at
each site exhibit a gap in their excitation spectrum while those with half-
integer spins are gapless~\cite{brohgeilo,cowlgeilo}. Among the spin-$1/2$
ladder compounds, those with an even number of legs exhibit a gap just
like the integer spin chains~\cite{greg,brohprep,azuma} while ladders with an
odd number of legs are gapless analogous to the half-integer chains~\cite{azuma}.

Theoretically, the universal low-energy properties of the gapped systems are well described by 
the one dimensional quantum ${\it{O}}(3)$ non-linear $\sigma$-model (NL$\sigma $M) without any 
topological term~\cite{hald,sorenaffl,sierra}. A lot is known exactly about
this field theory~\cite{polya,zamal,smirn} and this is directly useful in
understanding the gapped systems. The spectrum of the $\sigma$-model consists
of a triplet of massive spin-$1$ particles as the lowest energy excitations
followed by multiparticle continua with no bound states.
Many zero temperature
($T$) properties of the gapped systems, including low frequency dynamic correlations, can be 
explained using the exact information available on
the $\sigma$-model~\cite{egwestaffl}. On the other hand, until very recently
exact results for $T>0$ were restricted to static, thermodynamic properties~\cite{tsvel} while 
many experimental observables (such as the
inelastic neutron scattering (INS) crosssection and NMR relaxation rates) directly
probe dynamical correlations at non-zero temperature.

(The universal low-energy properties of the gapless systems have been
treated via a mapping to a certain critical field theory~\cite{affleshouch}. In contrast to the 
NL$\sigma$M, powerful
techniques that exploit the conformal invariance of the theory can 
be used to determine exactly for $T>0$ some dynamical correlators that are
directly probed by NMR experiments~\cite{hjs,ssnmr,starykh}. Similar methods
have been used to obtain results for $T>0$ on static properties as well~\cite{afftakegg}. 
Transport properties have also been
studied recently~\cite{narozhny,zotos,mccoy}, with results that are quite
different from those we shall obtain here for gapful systems.)

This paper shall deal exclusively with the $T>0$ dynamical properties of
gapped Heisenberg spin chains.
A portion of our results have appeared earlier in a short report~\cite{sskd},
where we presented them in the context of the continuum NL$\sigma$M, but did not
fully discuss their range of applicability. Here we shall take a more
general point of view of working directly with lattice Heisenberg antiferromagnets.
The main, and essentially only, requirement on the spin chain being studied
is that it have an energy gap and that its low-lying 
excitations consist of a triplet
of spin-1 particles with the dispersion
\begin{equation}
\varepsilon(k) = \Delta + \frac{c^2 k^2}{2 \Delta} + {\cal O}(k^4) \ldots
\label{dispersion}
\end{equation}
Here $k$ is being measured from an antiferromagnetic wavevector $Q = \pi/a$
($k = q- Q$, and $a$ is the lattice spacing),
and we have introduced a velocity $c$ to parametrize the mass of the particles
as $\Delta /c^2$. This is in keeping with the `relativistic' spectrum of the $O(3)$
$\sigma$-model $\varepsilon(k) = ( \Delta^2 + c^2 k^2 )^{1/2}$, although most
of our results will {\em not} rely on this relativistic form.
Gapped spin chains with a spontaneously broken translational symmetry (spin-Peierls
order) can have spin-1/2 particle excitations: we shall not deal with this case explicitly,
although we believe most of our results can also be extended to these systems.

The energy gap $\Delta$ is an important energy scale which shall play a 
central role in our analysis. The majority of our results will be in the
regime $T \ll \Delta$ (we shall use units with $\hbar = k_B = 1$ throughout) which we now 
discuss.

\begin{figure}
\epsfxsize=3.5in
\centerline{\epsffile{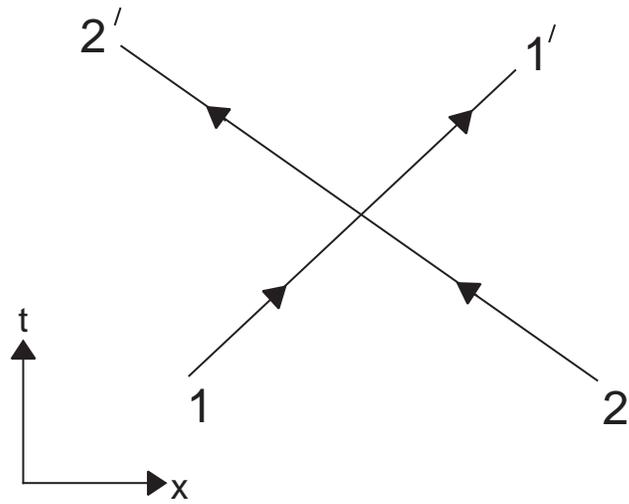}}
\vspace{0.2in}
\caption{Two particle collision described by the ${\cal S}$-matrix
(\protect\ref{smatrix}). The momenta before and after the collision are
the same, so the figure also represents the spacetime trajectories of
the particles. }
\label{fig1}
\end{figure}

In this regime, there is a dilute gas of excited particles present,
and their motion and collisions dominate the dynamical properties we study.
In particular their spacing $\sim c e^{\Delta /T}/(\Delta T)^{1/2}$ is much larger than
their thermal de-Broglie wavelength $\sim c/(\Delta T)^{1/2}$. As argued in
Ref~\cite{ssyoung}, these
particles can be treated classically except when two of them collide. Such
two-particle collisions need to be treated quantum-mechanically and are described by an ${\cal 
S}$-matrix, which is, in general,
a complicated function of the particle momenta and spin orientations.
Conservation of total momentum and energy implies that momenta before and after
collisions have to be the same, and $O(3)$ invariance and unitarity impose
further constraints, but a fairly complex structure is still permitted--we
will see some explicit examples in this paper. However, the r.m.s. 
thermal velocity of a particle $v_T = (T/\Delta)^{1/2} c \rightarrow 0$
as $T/\Delta \rightarrow 0$ and thus we need the ${\cal S}$-matrix only in
the limit of vanishing incoming (and outgoing) momenta. One of the central ingredients
in our computations will be our claim that in this limit, all of the complexity disappears,
and the ${\cal S}$-matrix has a {\em super-universal\/} form; for the scattering
event shown in Fig~\ref{fig1} we have (here $\mu _i = x,y,z$ are the three possible
values of the $O(3)$ spin label)
\begin{equation}
 {\cal S}^{\mu_1 \mu_2}_{\mu^{\prime}_1,
\mu^{\prime}_2} (k_1, k_2; k^{\prime}_1,k^{\prime}_2)
= (-1) \delta_{\mu_1 \mu^{\prime}_2}\delta_{\mu_2 \mu^{\prime}_1}
(2\pi)^2 \delta(k_1 - k^{\prime}_1) \delta (k_2 - k_2^{\prime})~;
\label{smatrix}
\end{equation}
Notice especially the opposite pairing of momentum and spin labels: crudely speaking,
the momenta go ``through'' the collision, while the spins ``bounce off''---this 
dichotomy will be crucial to our considerations. We dub this limiting value of
the ${\cal S}$-matrix
`super-universal' as it requires only that the lowest lying excitations above
the gap satisfy (\ref{dispersion}) at low momenta.
The value, however, does not depend on parameters such as $c$ and $\Delta$. Moreover, we expect 
this limiting result to hold even at the lattice
level for generic microscopic models of one-dimensional antiferromagnets with massive
spin one excitations; we shall see one explicit example that bears out this
expectation later in the paper.

With this simple form
of the ${\cal S}$-matrix in hand, we will use the semiclassical techniques of
Ref~\cite{ssyoung} to analyze dynamical properties of spin
fluctuations near $q=0$ and $q=Q$ in terms of the motion of the dilute gas
of quasiparticles.

We will begin by discussing the properties of the ${\cal S}$-matrix for two-particle
scattering in the limit of low momenta
in Section \ref{sec:begin}. In Section~\ref{sec:sigma} we consider the ${\cal S}$ matrix for the 
$O(3)$ non-linear
sigma model. This has been computed for all momenta by Zamolodchikov and Zamolodchikov,
and we shall show that the zero momentum limit does indeed satisfy (\ref{smatrix}). 

However, the $\sigma$-model is a continuum theory;
it would be much more satisfactory to be able to directly see that
(\ref{smatrix}) holds for some specific microscopic model,  
and explicitly verify that lattice effects do not
affect the simple structure of this limit. One such model is the $S=1/2$, two-leg
Heisenberg antiferromagnetic ladder with inter-chain exchange $J_{\parallel}$
and intra-chain exchange $J_{\perp}$. The properties of the ladder can be analysed
using a `strong-coupling' expansion~\cite{strngcpaps} in powers of $J_{\parallel}/J_{\perp}$ for 
the microscopic lattice Hamiltonian of the
system.
In Section~\ref{zerot} we shall explicitly verify (\ref{smatrix}) for
vanishing velocities in this lattice model within this strong-coupling expansion. 
Parenthetically, we note that our strong-coupling analysis also
allows us to make predictions about interesting features in the $T=0$ dynamic structure factor
$S(q,\omega)$ which are specific to the system considered. In particular, we find that, to second 
order in $J_{\parallel}/J_{\perp}$, 
a two-particle $S=1$ bound state gives rise to a {\em second} peak (in addition to the usual peak
coming from the stable single particle excitations of the system) in
$S(q,\omega )$ for a range of values of $q$ around $Q$. This should be of relevence to
inelastic neutron scattering (INS)
experiments on the ladder compounds and it is hoped that they experimentally
verify the existence of this effect.

In Section~\ref{fuku} we study the complementary `weak-coupling' expansion in 
powers of $J_{\perp}/J_{\parallel}$ for the two-leg ladder.
As was shown in Ref~\cite{shelton}, this expansion leads to a description of the low-energy, 
long-distance properties of the ladder
in terms of an effective field theory of a triplet of massive Majorana fermions.
The Hamiltonian for the Majorana fermions
also has a four-fermion coupling which has generally been ignored
in previous treatments. In the absence of this scattering, the Majorana fermions
are free, and the resulting ${\cal S}$-matrix does {\em not} obey (\ref{smatrix}). In this
paper, we consider the effect of the four-fermion coupling in perturbation theory.
We show that this expansion suffers from severe infra-red problems which have
to be resolved by an infinite-order resummation. The structure of the divergences
is very similar to those also present in the large-$N$ expansion of the $\sigma$-model  above,
and we find that the resulting resummed ${\cal S}$-matrix of the Majorana
fermions does indeed obey the analogue of (\ref{smatrix}). So neglecting the four-fermion
coupling~\cite{kishifuku}, (or even treating it in an unresummed
manner at finite order in perturbation theory) 
is a very bad approximation at low momenta,
and we expect that corresponding divergences in the perturbative evaluation of  the spin-spin 
correlation function invalidate the dynamical results of Ref~\cite{kishifuku} at low $T$.

In Section~\ref{sec:lowt} we shall turn to a discussion of the dynamical properties
in the regime $0<T \ll \Delta$. Our results apply universally to all
gapped one-dimensional antiferromagnetic systems with spin one quasiparticles; indeed they rely 
{\em only} on the dispersion (\ref{dispersion})
and the ${\cal S}$-matrix in (\ref{smatrix}). All our results will be 
expressed solely in terms of the parameters $c$ and $\Delta$, the temperature, $T$ and the 
external field $H$.

In Section~\ref{finitet} we study the dynamics of the staggered component
(with wavevector $q$ close to $Q$)
of the fluctuations in the spin density. More precisely, we study the
dynamical structure factor $S(q,\omega)$ for $q$ close to $Q$. Apart from
some overall factors, this directly gives the INS crosssection at the
corresponding values of momentum and energy transfer.
At $T=0$, the dynamical
structure factor has a sharply defined $\delta$-function peak at 
$\omega = \varepsilon(q-Q)$ for $q$ near $Q$. 
This peak can be thought of as arising from the ballistic propogation of
the stable quasiparticle of the system. At non-zero temperatures, the peak broadens as
the quasiparticle suffers collisions with other thermally excited
particles. The main objective of 
Section~\ref{finitet} is to describe the precise lineshape of the
quasiparticle peak in the dynamic structure factor for $T>0$.

In the $\sigma$-model approach, the staggered components of the spin density are represented by 
the
antiferromagnetic order parameter field $\vec n$.
We will use the semiclassical method of
Ref-\cite{ssyoung} to calculate the space and time dependent $2$-point correlation
function of the $\vec n$ field for $T>0$. 
This allows us to calculate
the thermal broadening
of the single particle peak in the dynamical structure factor $S(q,\omega )$
for wavevectors $q$ near $ Q$. In particular, we find that the dynamic structure
factor in the immediate vicinity of the quasiparticle peak at ($q=Q$, $\omega = \Delta $) may be
written in a reduced scaling form as
\begin{equation}
S(q,\omega )= \frac {{\cal A}cL_{t}}{\pi^2 \Delta}{\Phi}\left (\frac{\omega - 
\varepsilon(k)}{L_{t}^{-1}} \right )~,
\label{scalinspik}
\end{equation}
where $k=q-Q$, ${\cal A}$ is the (non-universal) quasiparticle amplitude of the
spin one excitations of the system, $L_t =\sqrt{\pi}e^{\Delta/T}/3T $
is the typical time spent by a thermally excited quasiparticle between collisions with other 
particles, and $\Phi$ is a {\em completely
universal} function that we determine numerically in this
paper. Notice that temperature enters this
scaling form only through $L_t$.
We claim, though this is not rigorously
established, that these results for the broadening are asymptotically exact for
$T \ll \Delta $: all corrections to the line-width are expected to be
suppressed by positive powers
of $T/\Delta$. Some evidence for the exactness of our results emerges from
consideration of simpler systems where exact results
for the line-broadening are available from the 
quantum inverse scattering method~\cite{korepin}; as we shall see in Section~\ref{finitet},
our semiclassical results are in perfect agreement~\cite{ssyoung} with these.
It is hoped that experimental studies of the temperature
dependence of the INS cross-section in this regime will confirm these results,
particularly the simple scaling form (\ref{scalinspik}).

In Section~\ref{hdep} we turn to the correlations of the conserved
magnetization density, or dynamic fluctuations near $q=0$, for $T \ll \Delta$. 
Unlike the staggered case, the
overall magnitude of the magnetization density fluctuations is universal
and given by $T \chi_u$, where $\chi_u$ is the uniform susceptibility of the system (the
non-universal overall scale of the staggered component is reflected
for instance by the presence of the overall constant ${\cal A}$ in (\ref{scalinspik})).
In this temperature regime, we have the
well-known result for $\chi_u$~\cite{tsvel}:
\begin{equation}
\chi_u = \frac{1}{c} \left( \frac{2 \Delta}{\pi T} \right)^{1/2}
e^{-\Delta /T}
\label{chilowt}
\end{equation}
We shall study 
the dynamics of the magnetization density in Section~\ref{hdep}~\cite{sskd}. 
We shall show that the long-time correlations of an effective semiclassical model
are characterized by {\em spin diffusion}, and obtain the following 
result for its low $T$ spin-diffusion constant $D_s$:
\begin{equation}
D_s = \frac{c^2}{3 \Delta } e^{\Delta /T}
\label{dslowt}
\end{equation}
Using the Einstein relation for the spin conductivity $\sigma_s = D_s \chi_u$, we obtain
from (\ref{chilowt}) and (\ref{dslowt})
\begin{equation}
\sigma_s = \frac{c}{3} \left( \frac{2}{\pi \Delta T} \right)^{1/2};
\end{equation}
Notice that the exponentially large factor $e^{\Delta/T}$ has dropped out,
and $\sigma_s$ diverges with an inverse square-root power in $T$ as $T \rightarrow 0$.
The semiclassical model is possesses an infinite
number of local conservation laws: in Appendix~\ref{integrable}, we discuss
how the existence of spin diffusion can be compatible with these local conservation
laws. However, these results do not rigorously allow us to conclude
that the ultimate long-time correlations of the underlying
gapped quantum spin chain are diffusive. This has to do with
a subtle question of order of limits: we computed the ${\cal S}$
matrix (\ref{smatrix}) in the limit $T/\Delta \rightarrow 0$, and then
used it to evaluate the long-time limit of correlations of the magnetization, whereas
in reality the limits should be taken in the opposite order. 
What we can claim is that our results will apply for all times upto
a time scale which is larger than the collision time $L_t$ by a factor
which diverges with a positive power of $\Delta/T$ as $T \rightarrow 0$;
there is a substantial time window in this regime where we have established
that the spin correlations are diffusive. 
For the generic gapped quantum spin chain, we can reasonably 
expect that the ultimate long-time
correlations are indeed diffusive, and the only consequences of the omitted
terms in the ${\cal S}$ matrix are subdominant corrections to the value of
$D_s$ in (\ref{dslowt}) which are suppressed by powers of $T/\Delta$.
For the continuum NL$\sigma$M with a relativistically invariant
regularization (this is unphysical for any experimental application), 
the issue is a little more subtle: this model does possess
additional non-local conserved quantities~\cite{luscher2}, but we consider it
unlikely that these will modify the long time limit~\cite{note1}.
On the experimental side, however, diffusive behaviour of the magnetization density
has already been convincingly demonstrated in the $S=1$ one-dimensional
antiferromagnet AgVP$_2$S$_6$ by the NMR experiments of Takigawa {\em et.al.}~\cite{taki}.

As has been argued earlier~\cite{sagiaffl}, the dynamic fluctuations near
$q=0$ provide the dominant contribution
to the NMR relaxation rate $1/T_1$ for $T \ll \Delta$. Thus, knowing the space and time
dependent two-point correlation function of the conserved magnetization
density, we are able to compute the field and temperature
dependent $1/T_1$ in this regime.
We shall see that the overall scale of $1/T_1$ is set
by the ratio $T \chi_u/\sqrt{D_S}$. As was pointed out to
us by M. Takigawa~\cite{sskd}, this immediately leads to an activation
gap for $1/T_1$ given as
\begin{equation}
\Delta_{1/T_1} = \frac{3}{2} \Delta~.
\label{nmrgap}
\end{equation}
This difference between the activation gaps for $\chi_u$ and $1/T_1$ appears to clear-up puzzling 
discrepencies in the experimental
literature~\cite{taki,shimizu,azuma} for the value of the energy gap in these systems obtained
from Knight-shift susceptibilty measurements on the one hand, and $1/T_1$ NMR relaxation rates 
on the other; a systematic tabulation of the activation gaps for a large number
of gapped spin chains~\cite{ssgeilo} 
does indeed show a trend consistent with (\ref{nmrgap}).
The crucial factor of $3/2$ clearly arises from the exponential divergence
in $D_s$.
This diffusive behaviour we find arises entirely from intrinsic inelastic scattering between
the quasi-particles. In real systems there will also be contributions
from elastic scattering off inhomogenities which will eventually saturate the
divergence of $D_s$ as $T \rightarrow 0$. However, because of the strong spin scattering
implied by (\ref{smatrix}) the effects of inelastic scattering is 
particularly strong in $d=1$, and can easily dominate inhomogenities in clean samples.

We will give a detailed account of the calculations leading up to our expression
for
$1/T_1$ (some details on the method used are relegated to Appendix~\ref{jep}) 
and then go on to compare the theoretical predictions for the field dependence
of $1/T_1$
with the extensive experimental data of Takigawa {\em et.al.}~\cite{taki}.
We will see that our results 
({\em without} any adjustable parameters, except for a field independent background rate) 
agree with the data extremely well for a range of
intermediate temperatures. At the lowest temperatures for which data is
available, the quality of the fit deteriorates significantly and the
$1/\sqrt{H}$ divergence predicted at small fields seems to get cutoff, 
presumably by some spin-dissipation mechanism present in the real system.
At the present time, we are unable to incorporate this dissipation in any
serious way in our approach. However, following Ref~\cite{taki}, we can phenomenologically 
introduce
some spin-dissipation in our results for the long-time limit of the
autocorrelation function and obtain a corresponding expression for the
field dependence of $1/T_1$. This allows us to fit the data at the
lowest temperatures with a phenomenological form that has one additional adjustible
parameter corresponding to the spin-dissipation rate. We also present
results for the temperature dependence of this effective rate. 

\begin{figure}[t]
\epsfxsize=4.5in
\centerline{\epsffile{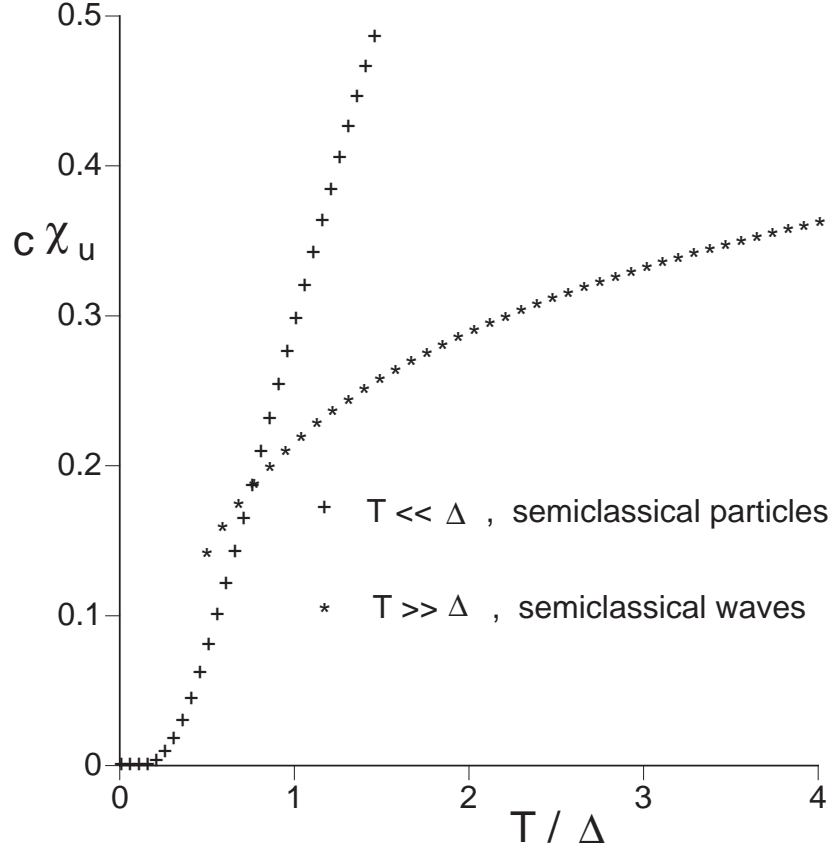}}
\vspace{0.2in}
\caption{Low and high temperature asymptotics for the uniform
susceptibility $\chi_u$ of the continuum $O(3)$ non-linear $\sigma$-model. 
At $T=0$, there is an energy gap $\Delta$ to all excitations, and $c$
is the velocity defined by (\protect\ref{dispersion}). 
The expression in Eqn~{\protect \ref{chilowt}} gives the
low temperature asymptotics while Eqn~{\protect \ref{chihight}} is used
for the high temperature asymptotics.
Any lattice antiferromagnet will have a very high temperature ($T>J$
where $J$ is a typical microscopic exchange constant)
Curie susceptibility $\sim 1/T$ which is not shown: the high temperature
limit of the continuum theory will apply for $\Delta < T < J$.}
\label{fig2}
\end{figure}

Finally, in Section~\ref{hight} we will turn to the regime $T \gg \Delta$.
We will do this in the context of the continuum $O(3)$ $\sigma$-model only.
Any continuum theory is applicable to real lattice experiments only below
some energy scale, and a natural choice for this energy scale is a typical
exchange constant $J$. So more specifically, we shall be studying the regime
$\Delta \ll T \ll J$. For $T \gg J$ we expect the spins to behave independently,
and the system exhibits a Curie susceptibility. It is an open question whether
the window of temperatures $\Delta \ll T \ll J$ with universal behavior
exists at all in any given system,
and the answer will surely depend upon details of the microscopics. 
It is unlikely to be present for $S=1$ spin chains, but appears
quite possible for $S=2$ spin chains~\cite{meisel}.
The static properties of this regime were first studied by Joliceour and Golinelli~\cite{joli} 
using the $N=\infty$ limit of the $O(N)$ $\sigma$ model 
of Ref~\cite{CSY}. We shall present here an 
exact treatment of static and dynamic properties for the case of general $O(N)$;
the numerical values of the $N=3$ static results are significantly different
from the earlier $N=\infty$ results.
We shall show that the antiferromagnetic correlations decay with a correlation
length $\xi$, which to leading logarithms in $\Delta/T$ is given at $N=3$ exactly by
\begin{equation}
\xi = \frac{c}{2 \pi T} \ln \left( \frac{32 \pi e^{-(1+\gamma)} T}{\Delta}
\right)~,
\label{xihight}
\end{equation}
where $\gamma$ is Euler's constant.
We also obtain the exact uniform susceptibility 
\begin{equation}
\chi_u = \frac{1}{3 \pi c} \ln \left( \frac{32 \pi e^{-(2+\gamma)} T}{\Delta}
\right)
\label{chihight}
\end{equation}
(notice the argument of the logarithm differs slightly from (\ref{xihight})).
It is interesting to compare the two asymptotic results (\ref{chilowt}) and (\ref{chihight}),
and we have done that in Fig~\ref{fig2}. It is reassuring to find that the two results
are quite compatible for $T \approx \Delta$. This suggests that one of 
either the $T \ll \Delta$ or $T \gg \Delta$ asymptotics are always appropriate.
We shall also consider the nature of spin transport in the $\Delta \ll T \ll J$ regime,
and show that it is related to transport in a certain classical statistical 
problem of deterministic non-linear
waves. We have not established whether spin diffusion exists or not in this classical 
problem; if the correlations were diffusive, however,
we are able to precisely predict the $T$ dependence of the spin diffusivity:
\begin{equation}
D_s = {\cal B} \frac{T^{1/2} [\xi (T)]^{3/2}}{[3\chi_{u} (T)/2]^{1/2}}.
\end{equation}
Here ${\cal B}$ is an undetermined universal number, and $\xi(T)$, $\chi_u (T)$ are
given in (\ref{xihight},\ref{chihight}). 

Notice the complementarity in the two $T$ regimes discussed above: 
the description for $T \ll \Delta$ was in terms of semiclassical
particles, while that for $\Delta \ll T \ll J$ is in terms of semiclassical non-linear
waves.

\section{Zero temperature properties}
\label{sec:begin}

The primary purpose of this section will be to establish the ${\cal S}$-matrix
by a variety of methods. We will begin in Section~\ref{sec:sigma}
by using the relativistic $O(3)$ $\sigma$ model. In Section~\ref{zerot} we will consider
the strong-coupling expansion of the two-leg ladder in powers of $J_{\parallel}/J_{\perp}$.
This section will also present supplementary results on some interesting
features in the $T=0$ INS cross-section of the strongly coupled ladder
arising from the presence of a $S=1$, two-particle bound state in its
spectrum. Finally, Section~\ref{fuku} will 
consider the complementary $J_{\perp}/J_{\parallel}$ expansion.

\subsection{$O(3)$ $\sigma$ model}
\label{sec:sigma}
Let us begin with a brief review of the $\sigma$-model as an effective
field theory for the low-energy properties of the gapped systems (for a 
more extensive discussion see \cite{affgeilo} and references therein).
The imaginary time ($\tau$) action of the $\sigma$-model is
\begin{equation}
{\cal L} = \frac{c}{2g} \int_0^{1/T} d\tau dx \left[ (\partial_x n_{\alpha}
)^2 + \frac{1}{c^2}(\partial_{\tau} n_{\alpha} - i
\epsilon_{\alpha \beta \gamma} H_{\beta} n_{\gamma}
)^2  \right]
\label{sigmaaction}
\end{equation}
where $x$ is the spatial co-ordinate, $\alpha,\beta,\gamma = 1,2,3$ are $O(3)$ vector
indices over which there is an implied summation, $\epsilon_{\alpha\beta\gamma}$ is the
totally antisymmetric tensor, $c$ is a velocity,
$H_{\alpha}$ is an external magnetic field (we have absorbed a factor of the
electronic magnetic moment, $g_e \mu_B$, into the definition of the field $H$.), and the 
partition function is obtained
by integrating over the unit vector field $n_{\alpha} (x,\tau)$, with
$n_{\alpha}^2 (x,\tau) = 1$. 
The dimensionless coupling constant $g$ is determined
by the underlying lattice antiferromagnet
at the momentum scale $\Lambda \sim \mbox{inverse lattice spacing}$ to
be $g \sim 1/S$ where $S$ is the spin at each site in the
original lattice system. The $\sigma$-model is used to make statements about
physics at length scales $\gg \Lambda ^{-1}$ and time scales $\gg (c \Lambda)^ {-1}$; this 
physics is {\em universally} characterized by the dimensionful
parameters $c$, $H$, $T$, and $\Delta$, the energy gap at $T=H=0$. Though the
magnitude of $\Delta$ is determined by non-universal lattice scale quantities
($\Delta \sim c \Lambda e ^{-2\pi/g}$ for small $g$), the long distance
physics of the $\sigma$-model depends on these lattice scale effects only through the value of 
$\Delta $, and has no direct dependence on $g$ or $\Lambda$. Also, the energy-momentum dispersion 
of the stable
particle-like excitation of this model is given by $\varepsilon(k) ={\sqrt {\Delta^2
+ k^2c^2}}$, and there is a triplet of them.
The conserved density of this model corresponding to its ${\it O}(3)$ symmetry
is the magnetization density $M_{\alpha}(x,\tau)=\delta L/\delta H_{\alpha}(x,\tau)$. In the 
Hamiltonian formalism, this is represented by the
operator $M_{\alpha}(x)$.

Finally, to make contact with the lattice antiferromagnet, we must have a prescription for 
representing the spin-density
operator $s_{\alpha}(x)$ of the lattice system in terms of the operators of the $\sigma$-model.
It is most convenient to do this in terms of Fourier components. We have:
\begin{displaymath}
s_{\alpha}(k+Q) \propto n_{\alpha}(k)
\end{displaymath}
(recall that $Q=\pi/a$) and
\begin{displaymath}
s_{\alpha}(q) = M_{\alpha}(q)
\end{displaymath}
for $| q |$, $|k|$ much smaller than some microscopically determined scale
$\sim \Lambda$. The missing proportionality constant in the first relation
is non-universal and related to the magnitude of the spin at each site in
the original lattice system.
Thus, the $\sigma$-model allows us to represent spin fluctuations near
$q=Q$ (these being the low-energy degrees of freedom) {\em and}
near $q=0$. This is of course because the $q=0$ component of the spin
density is the conserved charge corresponding to the ${\it O}(3)$ symmetry
of the system, and as such must be included in any description of the slow
modes.

The exact ${\cal S}$ matrix of the collision of two particles in the $\sigma$-model
was computed in a seminal paper of Zamolodchikov and Zamolodchikov. For the
scattering event shown in Fig~\ref{fig1} it is (recall $\mu_i = x,y,z$ are the three possible
values of the $O(3)$ spin label):
\begin{eqnarray}
{\cal S}^{\mu_1 \mu_2}_{\mu^{\prime}_1,
\mu^{\prime}_2} (k_1, k_2; k^{\prime}_1,k^{\prime}_2)
= && (2\pi)^2\delta(k_1 - k^{\prime}_1) \delta (k_2 - k_2^{\prime}) \left[
\sigma_1 ( \theta) \delta_{\mu_1 \mu_2} \delta_{\mu^{\prime}_1 \mu^{\prime}_2}\right.
\nonumber \\
&&~~~~~~~~~~~~~~~\left.
+ \sigma_2 ( \theta) \delta_{\mu_1 \mu^{\prime}_1} \delta_{\mu_2 \mu^{\prime}_2}
+ \sigma_3 ( \theta) \delta_{\mu_1 \mu^{\prime}_2} \delta_{\mu_2 \mu^{\prime}_1}\right]
\label{smatrix1}
\end{eqnarray}
where $\theta = \theta_1 - \theta_2$ is the `rapidity',
$k_i = (\Delta/c) \sinh \theta_i$ for $i=1,2$, and $O(3)$ invariance
guarantees a total lack of $H$ dependence in the result. The functions $\sigma$ in 
(\ref{smatrix1})
are
\begin{eqnarray}
\sigma_1 ( \theta ) &=& \frac{2 \pi i \theta}{(\theta + \pi i)(\theta - 2 \pi i)} \nonumber \\
\sigma_2 ( \theta ) &=& \frac{\theta(\theta - \pi i)}{(\theta + \pi i )(\theta - 2 \pi i)} 
\nonumber \\
\sigma_3 ( \theta ) &=& \frac{-2 \pi i (\theta -  \pi i)}{(\theta +  \pi i)(\theta - 2 \pi i)} 
\label{smatrix2}
\end{eqnarray}
Now notice the structure of the limit $\theta \rightarrow 0$ which is important
for our purposes in the region $T \ll \Delta$: we find that $\sigma_{1,2} (\theta
\rightarrow 0) = 0$, while $\sigma_3 (\theta \rightarrow 0 ) = -1$. 
This establishes the key result (\ref{smatrix}) for this continuum model.

\subsection{Strongly-coupled two-leg ladders}
\label{zerot}

In this section we concentrate on the properties of a particular model
system, the spin-$1/2$, $2$-leg Heisenberg antiferromagnetic ladder, to which
the low-energy phenomenology of the preceeding section is expected to apply. The Hamiltonian
of the system may be written as
\begin{equation}
{\cal H} = \sum_{i}{\vec S}_{I}(i)\cdot {\vec S}_{II}(i) + g\sum_{i}\left({\vec S}_{I}(i)\cdot 
{\vec S}_{I}(i+1)+{\vec S}_{II}(i)\cdot {\vec S}_{II}(i+1)\right) ~.
\label{2legham}
\end{equation}
Here, the ${\vec S}_{I}(i)$ and ${\vec S}_{II}(i)$ are spin-$1/2$ operators
at site i along the two chains that make up the ladder, $g$ is a dimensionless coupling constant 
equal to the ratio of the antiferromagnetic bond along the individual chains $J_{\parallel }$ to 
the bond along the rungs of the ladder $J_{\perp }$ and we have set the bond strength $J_{\perp 
}$ along the rungs to be unity; this defines our unit of energy.
We will analyze this model in the limit of small $g$; this `strong-coupling'
expansion~\cite{strngcpaps} is expected to be qualitatively correct for all
$g$. For $g=0$, we just have a system of isolated rungs with the two spins
on each rung coupled antiferromagnetically. The ground state is a product
state with each rung in a singlet state. The lowest lying excited states form
a degenerate manifold with precisely one rung promoted to the
triplet state. Perturbative corrections in $g$ would presumably make this triplet `particle' hop 
around producing a single-particle band of
triplet excitations as the lowest lying excited states. Thus we expect that
our perturbative analysis will be most conveniently performed in a
representation that directly describes the state of {\em individual rungs}
of the ladder. With this in mind, we switch to the `bond-operator' formalism
introduced in Ref~\cite{ssbhatt}. Following Ref~\cite{ssbhatt}, we write
the spin operators as:
\begin{eqnarray}
S_{I}^{\alpha}(i) & = & \frac{1}{2}\left(s^{\dagger}(i)t_{\alpha}(i) +
t_{\alpha }^{\dagger}(i)s(i)-i\epsilon_{\alpha \beta \gamma}t_{\beta}^{\dagger}(i)
t_{\gamma}(i)\right )~, \label{bonddefI} \\
S_{II}^{\alpha}(i) & = & \frac{1}{2}\left(-s^{\dagger}(i)t_{\alpha}(i) -
t_{\alpha }^{\dagger}(i)s(i)-i\epsilon_{\alpha \beta \gamma}t_{\beta}^{\dagger}(i)
t_{\gamma}(i)\right )~,
\label{bonddefII} 
\end{eqnarray}
where $\alpha$, $\beta$, and $\gamma$ are vector indices taking the values
$x$,$y$,$z$, repeated indices are summed over, and $\epsilon$ is the totally
antisymmetric tensor. $s^{\dagger}(i)$ and $t_{\alpha}^{\dagger}(i)$ are respectively creation 
operators for singlet and
triplet bosons at site $i$ (in the previous section we had used ${\vec s}(x)$ to 
denote the spin density of the lattice system; here we shall use
${\vec \sigma}(x)$ to denote the same and reserve $s$ for the singlet boson operator). The 
restriction that physical states on
a rung are either singlets or triplets leads to the following constraint on
the boson occupation numbers at each site:
\begin{displaymath}
s^{\dagger}(i)s(i)+t_{\alpha}^{\dagger}(i)t_{\alpha}(i)=1~.
\end{displaymath}
The spin density is given by
\begin{displaymath}
\sigma_{\alpha}(i) = -i\epsilon_{\alpha \beta \gamma}t_{\beta}^{\dagger}(i)
t_{\gamma}(i)~. 
\end{displaymath}
It is also convenient to define
\begin{displaymath}
\phi_{\alpha}(i)=s^{\dagger}(i)t_{\alpha}(i) +t_{\alpha }^{\dagger}(i)s(i)~.
\end{displaymath} 
The Hamiltonian in terms of these operators is given as
\begin{equation}
{\cal H} ={\cal H}_{0} + {\cal V}~;
\label{bondopham1}
\end{equation}
where
\begin{equation}
{\cal H}_{0}=\sum_{i}\left( -\frac{3}{4}s^{\dagger}(i)s(i) + \frac{1}{4}
t_{\alpha}^{\dagger}(i)t_{\alpha}(i)\right)~,
\label{bondopham2}
\end{equation}
and
\begin{equation} 
{\cal V}=\frac{g}{2}\sum_{i}\left(\phi_{\alpha}(i)\phi_{\alpha}(i+1)-\sigma_{\alpha}(i)
\sigma_{\alpha}(i+1)\right)~.
\label{bondopham3}
\end{equation}

In this representation, the ground state for $g=0$ is just the state with
every site occupied by a singlet boson. To zeroth order in $g$, the
lowest excited states form a degenerate manifold with a triplet boson
(of arbitrary polarization) replacing the singlet particle at precisely one site. Higher excited 
states also form degenerate manifolds labelled by the
number of singlet particles that are replaced by triplet bosons.In what follows, we
will describe states by the number (which can only be zero or one) and
polarization of the triplet particles at each site, the singlet occupation
numbers being determined by the constraint. Thus we will loosely refer to
the state with no triplet particles as the `vacuum'. At this order in $g$,
the physical particle-like excitation of the system is created at site $i$ by the action of 
$\phi_{\alpha}(i)$ on the vacuum, and thus coincides with the
bare triplet particle. In general at higher orders in $g$, we expect that
the physical single particle states of the system will contain an admixture of
states with more than one bare particle present. Similarly, the physical vacuum will also have a 
component with non-zero bare particle number.

In fact, it is quite convenient to
make a canonical transformation (determined order by order in $g$) to an auxillary problem in 
which the  physical
particle states do not contain admixtures of states with different bare particle number. The 
Hamiltonian of the auxillary problem is related to the
original one by a similarity transformation. The energy eigenvalues obtained in this manner of 
course give the
energy levels of the original Hamiltonian. However, to recover the corresponding
wavefunctions, one has to undo the effects of the canonical transformation. We will use this 
convenient formulation of perturbation theory below as we discuss the strong-coupling expansion.

The auxillary Hamiltonian in this approach is given by
\begin{equation}
{\tilde{\cal H}} = e^{iW}{\cal H}e^{-iW}~,
\label{auxham}
\end{equation}
where $W$ is the hermitian operator that generates our canonical transformation.
We choose $W$ to meet the following criteria:
\begin{itemize}

\item
The matrix elements of ${\tilde {\cal H}}$ between states with differing
numbers of bare particles should be zero to a given order in $g$. Note that
this implies that the elementary excitations of the auxillary Hamiltonian are just the bare 
particles. However their dynamics, and their mutual interactions (in multiparticle sectors) are 
determined by the restriction of ${\tilde {\cal H}}$ to the appropriate subspace of definite 
particle number.This restriction gives the corresponding energy levels of the original 
Hamiltonian ${\cal H}$ correct to that order in $g$. This then serves as our effective 
Hamiltonian for the corresponding sector of the original problem.

\item
This does not completely specify $W$. We therefore also require that $W$ have
zero matrix elements to a given order in $g$ between any two states populated by the same number 
of bare particles.

\end{itemize}

These criteria fix $W$ uniquely order by order in $g$ and in general we
have an expansion for $W$ that reads: $W=g\left(W_{1}+gW_2+\ldots~~\right)$~.
It is quite straightforward to use this procedure to generate
an expansion in $g$ for the effective Hamiltonians in the one and two
particle sector of the original problem (the `effective Hamiltonian' in
the physical vacuum sector is just a constant equal to the ground state
energy calculated to the relevant order in $g$). Solving for the eigenstates
and eigenvalues of these effective Hamiltonians is just a simple exercise
in elementary quantum mechanics. If the eigenstates of the spin-ladder
are of interest (as they will be when we calculate $S(q,\omega)$ perturbatively)
, we will have to obtain them from the eigenstates $|\psi \rangle$ of the effective Hamiltonian
using
\begin{equation}
|\psi \rangle_{physical}=e^{-iW}|\psi\rangle~.
\label{physestates}
\end{equation}

After this preamble, we turn to the actual calculations. As we have mentioned
earlier, the scattering matrix in the low-energy limit is a crucial input to the semiclassical 
calculations at non-zero temperature, and it is
therefore interesting to have results for it in our microscopic model. So, to 
begin with, let us look at the two-particle sector and work out the scattering properties of the 
physical particles.

First we need to find the effective Hamiltonian for the two-particle sector.
To first order in $g$, this is just given by the restriction of ${\cal H}_{0}+{\cal V}$ to the 
two particle subspace. Instead of introducing
a lot of cumbersome notation to write this down, we will just list the amplitudes of the various 
processes that are allowed in this two body problem:
\begin{itemize}
\item
Each particle can hop one site to the left or the right with amplitude $g/2$
except when the neighbouring site in question is occupied by the other particle.
\item
When the two particles are at neighbouring sites, there is a non-zero
amplitude for spin rotation. Consider the state $|i,\alpha_{1};(i+1),\alpha_{2}
\rangle$ which has one particle at $i$ with polarization $\alpha_1$ (which
can be any one of $x$,$y$,$z$) and another particle at $i+1$ with polarization
$\alpha_{2}$. The amplitude to make a transition from this to the state 
\mbox{$|i,\beta_{1};(i+1),\beta_{2}\rangle$} is \mbox{$\left(-g\epsilon_{\gamma 
\beta_{1}\alpha_{1}}\epsilon_{\gamma \beta_{2}\alpha_{2}}\right)/2$}~.
\end{itemize}
To solve for the scattering states of this two-body problem, it is more convenient to work in a
basis in which we label the spin part of the two particle states by the  total angular momentum 
$J$ and the
value of its $z$ component $J_z$. The spin rotation amplitude now becomes just a $J$ dependent 
nearest neighbour potential which takes the values $g/2$, $-g/2$,
and $-g$ for $J=2$,~$1$,~and $0$ respectively. Note that the potential energy
is independent of $J_z$ as one would expect from rotational invariance.
It is now quite simple to find the scattering eigenstates in
each channel. The spatial wavefunction in channel $J$ may be written as
\begin{equation}
\psi(x_1,x_2)= {\hat 
P}_{J}\{\left(e^{ik_{1}x_{1}+ik_{2}x_{2}}+r_{J}(k_{1},k_{2})e^{ik_{2}x_{1}+ik_{1}x_{2}}\right)
\theta(x_{2}-x_{1})\}~,
\label{scattwf}
\end{equation}
where ${\hat  P}_{J}$ is the symmetrizing operator for $J=2$,~$0$ and
the antisymmetrizing operator for $J=1$, and $r_{J}$ is the reflection coefficient that 
completely specifies the scattering properties of the
particles. For the $r_{J}$ we have:
\begin{eqnarray}
r_{2} & = & -\frac{e^{-ika}-2\cos \left(k_{cm}a/2\right)}{e^{ika}- 2\cos 
\left(k_{cm}a/2\right)}~, \label{r2} \\
r_{1} & = & -\frac{e^{-ika}+2\cos \left(k_{cm}a/2\right)}{e^{ika}+ 2\cos 
\left(k_{cm}a/2\right)}~, \label{r1} \\
r_{0} & = & -\frac{e^{-ika}+\cos \left(k_{cm}a/2\right)}{e^{ika}+ \cos \left(k_{cm}a/2\right)}~; 
\label{r0}
\end{eqnarray}
where $k=(k_{1}-k_{2})/2$~, $k_{cm}=k_{1}+k_{2}$~, and $a$ is the lattice
spacing along the length of either of the two chains that make up the
ladder system. Note that $k_{1}$ and $k_{2}$ both range over the interval
$(0,2\pi/a)$. The energy of the scattering state labelled by $\{k_{1},k_{2}\}$
(the energy of the ground state being set to zero) is given by
\begin{displaymath}
E(k_1,k_2)=2 + g\cos(k_{1}a) + g\cos(k_{2}a)~.
\end{displaymath}
This is consistent with the first order result~\cite{strngcpaps} for the single particle 
dispersion relation:~$E(k) =
1+g\cos(ka)$.

The next step is to use these results for the reflection coefficients to obtain the ${\cal S}$-
matrix for this two-body problem in the limit of low velocities. Low
velocities imply values of $k_1$ and $k_2$ in the vicinity of the band minimum
at $\pi/a$, i.e $k$ close to zero and $k_{cm}$ close to $2\pi/a$. Both $r_2$ and $r_1$ have the 
limiting value $-1$ as $k\rightarrow 0$,~$k_{cm}\rightarrow 2\pi/a$. However, $r_{0}$ is singular 
in the vicinity of $k=0$, $k_{cm}=2\pi/a$; its value depends on the order in which the two limits 
$k\rightarrow 0$ and
$k_{cm}\rightarrow 2\pi/a$ are taken. This is somewhat disconcerting as
we expect a well-defined low-velocity limit which agrees with the predictions
of the $O(3)$ NL$\sigma$M field theory.

To identify the source of our problem,
let us look more closely at the expression for $r_0$. We notice that
$r_0$, considered a function of the complex variable $k$, has a pole in the
upper half-plane for a range of values of $k_{cm}$. This indicates the
presence of a bound state in the $J=0$ channel for the corresponding
values of $k_{cm}$. This bound state hits threshold, i.e its binding 
energy goes to zero, as $k_{cm} \rightarrow 2\pi/a$. It is the presence of
a bound state at threshold that causes the singular behaviour of the reflection
coefficient in the limit $k\rightarrow 0$,~$k_{cm}\rightarrow 2\pi/a$. Clearly,
if there were a range of $k_{cm}$ around $2\pi/a$ for which there was no
singlet bound state, then we would not have this difficulty. It turns out (as we
shall briefly outline later) that extending our calculation to the next order in $g$ leads us to 
precisely this conclusion and gives a well-defined limiting value of $-1$ for $r_0$ as
$k\rightarrow 0$,~$k_{cm}\rightarrow 2\pi/a$.

This result can now be used to obtain the ${\cal S}$-matrix of our auxillary two-body
problem. We are interested, however, in the ${\cal S}$-matrix that describes the
scattering of the physical particle-like excitations of the spin-ladder. Thankfully, it is quite 
easy to see that though the wavefunctions of the
two problems are related by a canonical transformation, the purely `off-diagonal' form of $W$ 
implies that the two are the same at least to first order in $g$. Transforming to the basis used 
in (\ref{smatrix}), we see that the ${\cal S}$-matrix in the low velocity limit is indeed given 
by~(\ref{smatrix})
Thus, this super-universal form of the ${\cal S}$-matrix holds for our lattice model and lends 
support
to the idea that it is a generally valid consequence of just the slow motion
of the particles and is in no way dependent on the special properties
of the continuum $\sigma$-model.

To wind up this part of our discussion, let us now summarize the calculation
of the reflection coefficients to first order in $g$. We need to find the
effective Hamiltonian of our auxillary two-body problem to second order
in $g$. This involves first working out $W_1$ and then using this to obtain
the effective two-body Hamiltonian. To ${\cal O}(g^2)$, we generate in this
manner an additional next-nearest neighbour hopping term and some
additional $J$ dependent nearest-neighbour interactions. We skip the details as they are somewhat 
tedious and not particularly illuminating. The
$J=0$ reflection coefficient (correct to ${\cal O}(g)$) obtained in this manner
is given as: 
\begin{equation}
r_0= -\frac{e^{-ika}+\cos \left(k_{cm}a/2\right)-g\cos\left(k_{cm}a\right)\left(3e^{-
ika}+e^{ika}\right)/4}{e^{ika}+\cos \left(k_{cm}a/2\right)-
g\cos\left(k_{cm}a\right)\left(3e^{ika}+e^{-ika}\right)/4}~.
\label{r0g}
\end{equation}
From this, it is easy to see that there is no pole in the upper half $k$-plane
as long as $|k_{cm}a-2\pi| < \sqrt{8g}$. This means that there is no
singlet bound state possible in this range of $k_{cm}$. This is consistent
with our expectation that at the very lowest energies, the two-particle
spectrum should be free of bound states in order to match the
predictions of the $\sigma$-model. Moreover, (\ref{r0g}) has a well-defined
low-velocity limit of $-1$ as claimed earlier.

The foregoing analysis has shown that the two particle sector has a spin
$S=0$ bound state which leads to some interesting threshold singularities
for the scattering matrix. Examining our expressions for the
reflection coefficients, we notice that there is in fact a bound
state in the $S=1$ channel as well (actually, there is also a $S=2$
anti-bound state; we will not delve further into that aspect of the
spectrum here). Now, a $S=1$ excited state can have observable consequences
for the INS cross-section of a system and we might expect to see
some interesting features in the same as a result of this.

With this motivation, let us turn to the perturbative calculation of the dynamic structure factor
at $T=0$. We pick a coordinate system in which the two chains that make up our
ladder are parallel to
the $x$ axis and have $y$ coordinates of $+d/2$ and $-d/2$ respectively, where
$d$ is the distance between them (for simplicity, we are assuming here that the rungs of the
ladder are perpendicular to its legs). The spins along a chain are located at
$x$ values equal to integer multiples of $a$. We denote the position of
each spin in the $x$-$y$ plane by ${\vec R}$. We define ${\vec P} = (q_x,q_y)$~.
The $T=0$ dynamic structure factor may be written as
\begin{equation}
S({\vec P},\omega)=\frac{1}{2\pi}\int dt\frac{a}{2L}\sum_{{\vec R}~{\vec R}^{\prime}}
\langle \Phi _{0}|{\hat S}_{{\vec R}}^{z}(t) {\hat S}_{{\vec R}^{\prime}}^{z}(0)|\Phi_{0}\rangle 
e^{-i{\vec P}\cdot ({\vec R}-{\vec R}^{\prime })+i\omega t}~;
\label{Sforladd}
\end{equation}
where $|\Phi_{0}\rangle$ is the ground state of the system, $L$ is the
length of each chain, and ${\hat S}_{\vec R}$ denotes the spin operator at
${\vec R}$ in the Heisenberg representation. Our strategy is to write down
the usual spectral representation for (\ref {Sforladd}) and then evaluate
it perturbatively. Actually,
a complete calculation of the second order contribution would involve the
eigenstates with more than two-particles present;  below we will ignore this
complication and confine ourselves to calculating the contribution of the
one and two-particle sectors, correct to the appropriate order in
$g$.

The spin operator at any site is a sum of two terms: a single particle
piece coming from the $\phi _{\alpha}$ , and a two particle part coming
from the spin density operator $\sigma _{\alpha}$. From the structure of the
strong-coupling expansion, it is clear that the single particle part does not
have matrix elements between the ground state and any state in the two
particle sector; similarly the two-particle piece does not have matrix
elements between the ground state and any state in the single-particle sector.
Thus, keeping only the contributions from the one and two particle sectors, we can write to 
second order in $g$:
\begin{eqnarray}
S({\vec P},\omega) & = & \frac{1}{2}\sum_{1-particle~states}\delta(\omega-E_{1})|\langle 
\Phi_{1}|\phi_{z}(-q_x)|\Phi_{0}\rangle |^2\sin^2\left(\frac{q_yd}{2}\right)  \nonumber \\
                   &   & + ~\frac{1}{2}\sum_{2-particle~states}\delta(\omega-E_{2})|\langle 
\Phi_{2}|\sigma_{z}(-q_x)|\Phi_{0}\rangle |^2\cos^2\left(\frac{q_yd}{2}\right)~;
\label{spectralform}
\end{eqnarray}
where $|\Phi_1\rangle$ and $|\Phi_2\rangle$ denote one and two particle
states respectively, and $\phi_z(q_x)$ and $\sigma_z(q_x)$ denote the
discrete Fourier transforms of $\phi_z(x)$ and $\sigma_z(x)$. Let us digress for
a moment and think in terms of the inelastic neutron scattering cross-section
for a process with momentum transfer ${\vec P}$ and energy tranfer $\omega$; this coincides with 
the dynamic structure factor apart from some geometrical
factors. This scattering can of course produce a single spin-one particle in
the spin system. But there is also a non-zero amplitude for producing a pair
of these particles close to each other (as is clear from the actual
calculations described later). This is the origin of the second term in
(\ref{spectralform}).

Now, these two pieces contribute to the structure factor over very distinct intervals
along the frequency axis. While it is in principle possible to calculate both
terms correct to ${\cal O}(g^2)$, we will confine ourselves below to
calculating the leading perturbative correction for each value of $\omega$. Thus, we will 
calculate the single particle piece only to first order in
$g$, while doing a full second order calculation for the two-particle piece.
Below, we give a brief outline of the calculation and then
discuss our final results.

To calculate the single particle piece, we first need to determine the
ground state and the physical one-particle state wavefunctions correct
to ${\cal O}(g)$. This involves using
$W$ correct to first order to obtain the physical wavefunctions from
the wavefunctions of the corresponding auxillary problem (for the one-particle
sector, these are just plane waves to all orders in $g$; this follows from
translational invariance). A simple calculation then gives the one-particle
piece as
\begin{equation}
S_1({\vec P},\omega)=\frac{1}{2}\left(1-
g\cos\left(q_xa\right)\right)\sin^2\left(\frac{q_yd}{2}\right)\delta\left
(\omega-E\left(q_x\right)\right)~,
\label{S1}
\end{equation}
where $E\left(q_x\right)=1+g\cos\left(q_xa\right)$.

Turning to the two-particle
piece, we see that one can actually ignore the distinction between the physical $2$-particle
wavefunction and the wavefunction of the auxillary two-body problem. Moreover,
it suffices to consider the auxillary problem to first order in $g$. Also,
since the ground state has spin zero and we are looking at the matrix
elements of a vector operator, we need to consider only the triplet ($J=1$)
channel of the auxillary problem. The only subtlety lies in the fact that
we need to consider the bound state contribution as well as the usual
contribution of the scattering states. From (\ref{r1}), we see that this
bound state exists for $\pi < k_{cm}a < \pi + \pi/3$ and for $3\pi-\pi/3 < 
k_{cm}a < 3\pi$ (remember $k_{cm}$ ranges from $0$ to $4\pi/a$). Thinking in
terms of an inelastic neutron scattering event with momentum transfer $q_x$
in the fundamental domain ($0$, $2\pi/a$), we see that this bound state can
be excited for $\pi < q_x a < \pi+\pi/3$ {\em and} for $\pi-\pi/3 < q_xa < \pi$.
In the latter case, momentum is conserved modulo a reciprocal lattice
vector of $2\pi/a$. Of course, in addition to the bound state contribution
there is a background term coming from the scattering states in this
channel. Again, the two particles can be created in the scattering state
either with total momentum $k_{cm}$ equal to the momentum transfer $q_x$, or
with the two differing by a reciprocal lattice vector of $2\pi/a$.

\begin{figure}[t]
\epsfxsize=4.0in
\centerline{\epsffile{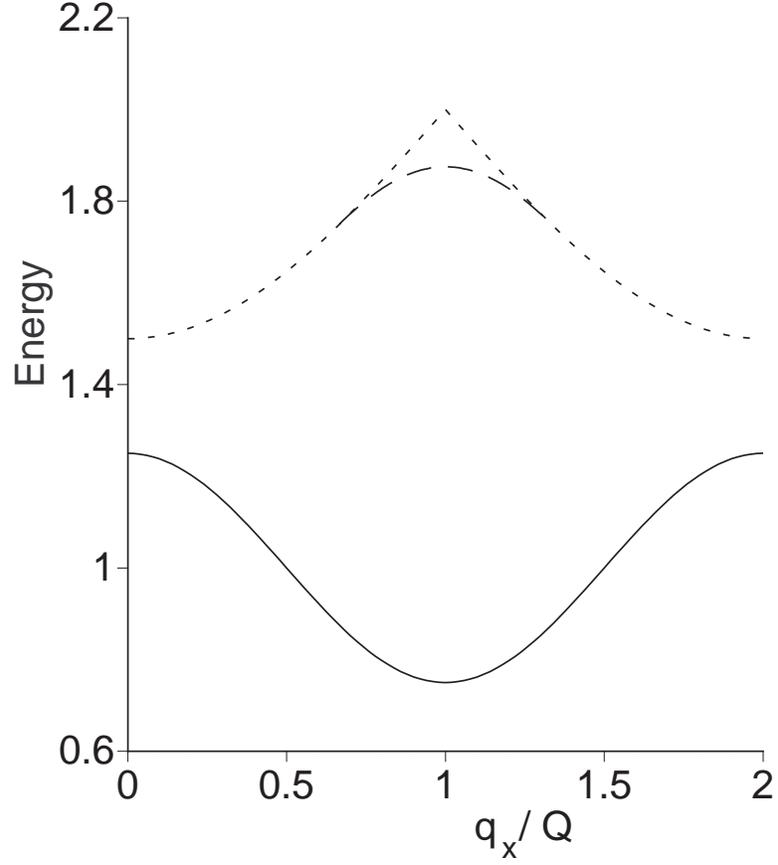}}
\vspace{0.2in}
\caption{Positions in $\omega$ of the single particle peak (solid line), 
bound state peak (long-dash line), and the bottom of the two particle continuum
(short-dash line) in $S({\vec Q},\omega)$ plotted as a function of $q_x$ for the
strongly-coupled ladder (a typical value of $g=0.25$ is used).}
\label{fig3}
\end{figure}

\begin{figure}
\epsfxsize=4.0in
\centerline{\epsffile{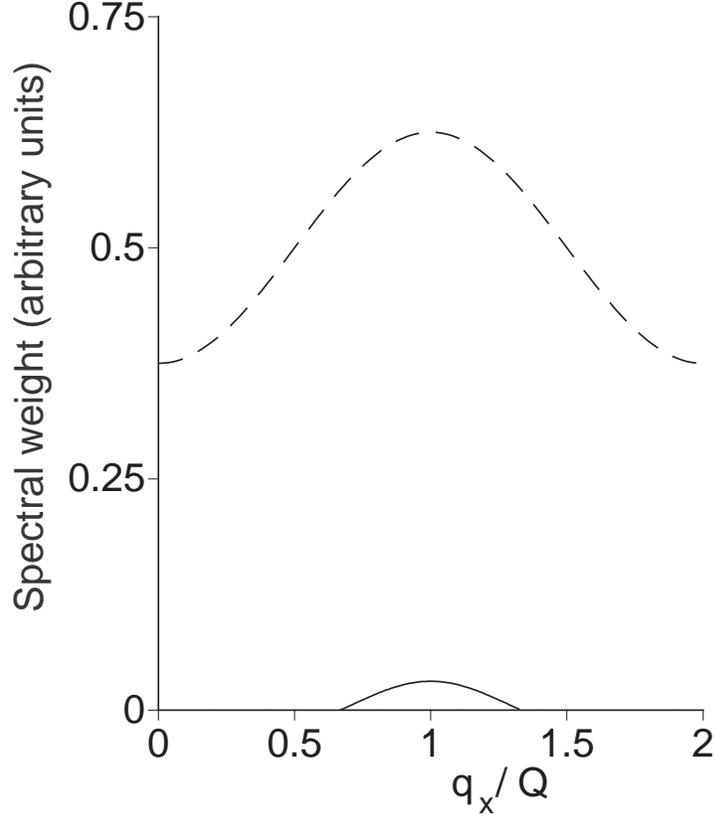}}
\vspace{0.2in}
\caption{Spectral weight in the single particle peak (dashed line) and
the bound state peak (solid line) in $S({\vec Q},\omega)$ for a
strongly coupled ladder (a typical value of $g=0.25$ is used). Note that
the two curves actually correspond to {\em different} values of
the transverve momentum $q_y$ chosen to maximize the respective spectral
weights: the single particle part is shown for $q_y = \pi/d$ while the
bound state part is shown for $q_y=0$ ($d$ is the spacing along the
rung of the ladder).}
\label{fig4}
\end{figure}

The actual
calculations are quite elementary and we proceed directly to the results for
the two-particle contributions. The bound state contribution for $2\pi/3 < q_xa < 4\pi/3$ may be 
written as:
\begin{equation}
S_{B}({\vec P},\omega)=\frac{g^2}{2}\cos^2\left(\frac{q_yd}{2}\right)
\sin^2\left(\frac{q_xa}{2}\right)\left(1-
4\cos^2\left(\frac{q_xa}{2}\right)\right)\delta\left(\omega-E_B(q_x)\right)~,
\label{Sb}
\end{equation}
where $E_B(q_x) = 2-g(1+4\cos^2(q_xa/2))/2$~.
On the other hand, the scattering states give rise to the following background
contribution for $|\omega -2|< +2g|\cos(q_xa/2)|$~:
\begin{equation}
S_{sc}({\vec 
P},\omega)=\frac{g}{\pi}\cos^2\left(\frac{q_yd}{2}\right)\sin^2\left(\frac{q_xa}{2}\right)\frac{
\sqrt{4g^2\cos^2(q_xa/2)-(\omega-2)^2}}{g+2(\omega-2)+4g\cos^2(q_xa/2)}~.
\label{Ssc}
\end{equation}
Note that for $q_xa =2\pi/3$~or~$4\pi/3$, there is a square-root divergence
at the lower threshold to the continuum in $\omega$; these are precisely the
values of $q_x$ for which the binding energy of the triplet bound-state
goes to zero. This enhanced scattering can thus be thought
of as arising from the presence of the triplet bound state at threshold. The
salient features of these results are summarized in Fig~\ref{fig3} and Fig~\ref{fig4}. 
Fig~\ref{fig3} is a plot of the positions along the $\omega$ axis of the single particle peak,
the bound state peak, and the bottom of the two-particle continuum
as a function of $q_x$. In Fig~\ref{fig4}, we show the spectral weight in the single particle and 
bound
state peaks as a function of $q_x$.

Thus, we see that that the existence of a triplet bound state of two elementary
spin-one excitations leads to some interesting features in the
dynamic structure factor. Actually, qualitatively similar features, again
arising from a triplet bound-state, had been predicted earlier~\cite{uhrig} in the alternating 
one-dimensional Heisenberg antiferromangnetic
chain. Recent INS experiments~\cite{garret,tntprp} on
$($VO)$_2$P$_2$O$_7$ do indeed see a second sharply defined peak in the
dynamical structure factor for a range of values of $q_x$. While this
compound had been previously thought to be a good example of a spin-ladder~\cite{whogoofed}, more 
recent work~\cite{tntprl2} has favoured the
alternating chain model~\cite{beltran} and the INS results have been interpreted~\cite{tntprp} in 
terms of the additional bound state
contribution predicted in Ref~\cite{uhrig}. Thus, our results may
not be of direct relevance to this particular experimental system. However,
our work does predict that a second peak in the INS cross-section should
be seen in strongly coupled ladder systems and it is quite possible
that the feature persists to all orders in the
perturbation expansion we have employed. It would be interesting to 
confirm this effect by looking at other systems that are more convincingly
modelled by a simple ladder Hamiltonian and it is hoped that future
experiments do indeed see the effects coming from the bound state.

\subsection{Weakly-coupled two-leg ladders}
\label{fuku}

In this section, we analyze the ladder system (\ref{2legham}) in the
complementary weak-coupling limit: $J_{\perp} \ll J_{\parallel}$. An elegant
mapping developed by Shelton {\em et.al.}~\cite{shelton} allows one to express the
low-energy, long-distance properties of the model in terms of a continuum theory
of weakly-interacting massive Majorana (real) fermions. We will analyze the low-energy
scattering properties of the spin one excitations of the weakly-coupled
ladder by working in this Majorana fermion representation.

We begin with a brief review of the Majorana fermion representation. We will not
attempt here to describe in any detail the procedure used~\cite{shelton} to
arrive at this field-theoretic representation. Instead, we will be content
with a rather telegraphic summary of the principal steps involved. To begin
with, one writes down the usual, free, massless bosonic theory~\cite{affleshouch} for the low-
energy
properties of each of the two $S=1/2$ Heisenberg antiferromagnetic chains that
make up the ladder. The interchain exchange $J_{\perp}$ is then turned on,
introducing a local, isotropic (in spin space) coupling between the
spin-density operators of each chain in the bosonic representation. This has
two pieces to it: one coupling the staggered parts of the spin densities with
each other and the other doing the same for the uniform component. Now, one
works with symmetric and antisymmetric combinations of the two boson fields
(one for each chain) and transcribes everything to a fermionic representation,
introducing one Dirac fermion for the symmetric combination and another for
the antisymmetric combination in the usual manner (for a readable account of the relevant 
machinery of Abelian bosonization, see for instance the review~\cite{shankar} by Shankar). The 
last step is to write each Dirac
fermion as two Majorana fermions. If one leaves out the uniform part of the
coupling to begin with, the theory in terms of the Majorana fermions is,
remarkably enough, a free-field theory. The staggered part of the coupling
just provides a mass $\Delta$ to each of the two Majorana fermions obtained
from the symmetric combination of the bosons, while the two Majorana fermions
obtained from the antisymmetric combination acquire masses $\Delta$ and$-3\Delta$ respectively 
(the actual energy gap is given by the absolute
value of the mass). The three Majorana fermions with mass $\Delta$ form
the spin one triplet we expect on general grounds, and the fourth Majorana
fermion represents a high-energy singlet mode that will not be very important
for our purposes. The mass parameter $\Delta$ of the theory is proportional
to $J_{\perp}$ with the proportionality constant being non-universal. Finally,
turning on the coupling between the uniform part of the spin densities gives us
a four-fermion interaction term between these massive Majorana fermions which will play a crucial 
role in our
analysis of the ${\cal S}$-matrix.

The procedure outlined above gives us the following Hamiltonian for the
effective field theory written in terms of Majorana fermions:
\begin{equation}
{\cal H} = \sum_{a=x,y,z}{\cal H}_{\Delta}({\xi}^a) + {\cal H}_{-3\Delta}(\rho)+{\cal H}_{I}~;
\label{mjnham1}
\end{equation}
here the ${\xi}^a$ and $\rho$ are Majorana fermion fields with anticommutation
relations given as
\begin{eqnarray}
{\{}{\rho}_{R}(x),{\rho}_{R}(y){\}} & = & \delta(x-y)~, \nonumber \\
{\{}{\rho}_{L}(x),{\rho}_{L}(y){\}} & = & \delta(x-y)~, \nonumber \\
{\{}{\xi}_{R}^{a}(x),{\xi}_{R}^{b}(y){\}} & = & \delta_{ab}\delta(x-y)~, \nonumber \\
{\{}{\xi}_{L}^{a}(x),{\xi}_{L}^{b}(y){\}}  & = & \delta_{ab}\delta(x-y)~, \label{anticoms}
\end{eqnarray}
with all other anticommutators being equal to zero, ${\cal H}_{m}(\phi)$
is defined in general as
\begin{equation}
{\cal H}_{m}(\phi) = \frac {iv_{F}}{2}\int dx ({\phi}_L{\partial}_x{\phi}_L-
{\phi}_R{\partial}_x{\phi}_R -m{\phi}_R{\phi}_L)~,
\label{massmajham}
\end{equation}
with $v_F \sim J_{\parallel}a$ and the interaction term ${\cal H}_I$ may be written as
\begin{equation}
{\cal H}_{I}=g\int dx{\{}{\xi}_{R}^{x}{\xi}_{L}^{x}{\xi}_{R}^{y}{\xi}_{L}^{y}+
{\xi}_{R}^{y}{\xi}_{L}^{y}{\xi}_{R}^{z}{\xi}_{L}^{z}+{\xi}_{R}^{z}{\xi}_{L}^{z}{\xi}_{R}^{x}{\xi}
_{L}^{x}-({\xi}_{R}^{x}{\xi}_{L}^{x} + {\xi}_{R}^{y}{\xi}_{L}^{y} + 
{\xi}_{R}^{z}{\xi}_{L}^{z})\rho_R\rho_L{\}}~,
\label{Hintformaj}
\end{equation}
with $g \sim J_{\perp}a$. Note that each Majorana fermion is a two component object, the two 
components being labelled with the subscripts $R$ and $L$ to denote the `right-moving' and `left-
moving' parts. To make contact
with the original spin-ladder, we  also need a prescription for expressing the
spin-operators of the ladder in terms of the Majorana fermions. In sharp
contrast to the $\sigma$-model, {\em only} the uniform part of the spin-density
operator has a local representation in terms of the fermions; the components
of the spin-density near $q=Q$ can be expressed only in terms of {\em highly
non-local} functions of the fermi-fields~\cite{shelton}. We have the following
expressions~\cite{shelton} for the uniform parts, ${\vec J}_1$ and ${\vec J}_2$, of the spin 
density on each chain:
\begin{eqnarray}
J_{1}^{a}(x) & = & \frac{i}{2}(\frac{1}{2}{\epsilon}^{abc}{\xi}_{\nu}^{b}(x){\xi}_{\nu}^{c}(x) + 
{\xi}_{\nu}^{a}(x){\rho}_{\nu}(x))~, \nonumber \\
J_{2}^{a}(x) & = & \frac{i}{2}(\frac{1}{2}{\epsilon}^{abc}{\xi}_{\nu}^{b}(x){\xi}_{\nu}^{c}(x) - 
{\xi}_{\nu}^{a}(x){\rho}_{\nu}(x))~,
\label{repunispin}
\end{eqnarray}
where the index $\nu$ takes on values $R$ or $L$ and repeated indices are
summed over. Note that the field $\rho$ corresponding to the non-universal
high-energy singlet mode drops out of the expression for the uniform
part of the total spin-density of the ladder which can then be expressed
entirely in terms of the spin one triplet fields.

We shall find it convenient, when it comes to actually doing any calculations, to rewrite all of 
the foregoing in terms of fermionic creation and annihilation operators. These are defined as 
follows: Let ${\hat \xi}_{\nu}^{a}(p)$
and ${\hat \rho}_{\nu}(p)$ denote the Fourier transforms of ${\xi}_{\nu}^{a}(x)$
and ${\rho}_{\nu}(x)$ respectively. We write
\begin{eqnarray}
{\hat \xi}_{\nu}^{a}(p) & = & f_{\nu}(p)t_{a}(p) + {\bar f}_{\nu}(-p)t_{a}^{\dag }(-p)~, 
\nonumber \\
{\hat \rho}_{\nu}(p) & = & g_{\nu}(p)s(p) + {\bar g}_{\nu}(-p)s^{\dag }(-p)~,
\label{defsandt}
\end{eqnarray}
where $t_{a}(p)$ and $s(p)$ are the fermionic annihilation operators corresponding
to the triplet and singlet modes respectively and $f_{\nu}(p)$ and $g_{\nu}(p)$
are complex-valued functions of $p$ which we specify below. These creation
and annihilation operators obey the usual anticommutation relations:
\begin{eqnarray}
{\{}t_{a}(p),t_{b}^{\dag}(q){\}} & = & 2\pi \delta_{ab}\delta (p-q)~, \nonumber \\
{\{}s(p),s^{\dag}(q){\}} & = & 2\pi \delta (p-q)~,
\label{stdanticoms}
\end{eqnarray}
with all other anticommutators equal to zero. In terms of these operators, the
non-interacting part of the Hamiltonian reads
\begin{equation}
{\cal H}_{0} = \int_{-\infty}^{\infty}\frac{dp}{2\pi} \varepsilon(p)t_{a}^{\dag}(p)
t_{a}(p) + \int_{-\infty}^{\infty}\frac{dp}{2\pi} \varepsilon_{s}(p)s^{\dag}(p)
s(p)~,
\label{nonintmajham}
\end{equation}
where $\varepsilon(p) = (p^2v_{F}^2 + \Delta^2)^{1/2}$, $\varepsilon_s(p) = (p^2v_{F}^2 + 
9\Delta^2)^{1/2}$, and the repeated index $a$ is summed over. The functions $f_{\nu}$ and 
$g_{\nu}$ are actually chosen to ensure that the
non-interacting Hamiltonian has this simple diagonal form
in terms of the creation and annihilation operators; this choice guarantees that
the operators $s^{\dag}$ and $t_{a}^{\dag}$, as defined in (\ref{defsandt}), create the true 
quasiparticles of the non-interacting system. The expressions
for $f_{\nu}$ and $g_{\nu}$ are best written as follows:
\begin{eqnarray}
f_{R}(p) & = & u_{\Delta}(p)~~~p>0~, \nonumber \\
f_{R}(p) & = & iv_{\Delta}(p)~~~p<0~, \nonumber \\
f_{L}(p) & = & {\bar f}_{R}(-p)~~~\forall p~, \nonumber \\
g_{R}(p) & = & u_{(-3\Delta)}(p)~~~p>0~, \nonumber \\
g_{R}(p) & = & iv_{(-3\Delta)}(p)~~~p<0~, \nonumber \\
g_{L}(p) & = & {\bar g}_{R}(-p)~~~\forall p~; 
\label{fandgdef}
\end{eqnarray}
here the functions $u_m(p)$ and $v_m(p)$ are defined in general as
\begin{eqnarray}
u_m(p) & =& \cos(\theta_m(p)/2)~, \nonumber \\
v_m(p) & = & \sin(\theta_m(p)/2)~,
\label{defuv}
\end{eqnarray}
with the angle $\theta_m(p)$ being specified by $\cos(\theta_m(p)) =
v_F|p|/(m^2+v_{F}^{2}p^2)^{1/2}$, $\sin(\theta_m(p)) = m{\mbox{sgn}}(p)/(m^2+v_{F}^2p^2)^{1/2}$. 
Now, we can rewrite the
interaction term in normal ordered form with respect to these
singlet and triplet creation and annihilation operators. The quadratic
terms so generated give the first order correction to the masses of
the singlet and triplet modes (this correction has already been calculated in
Ref~\cite{shelton} by other means). The quartic term left over, has, in
addition to the usual, normal-ordered, particle-number conserving piece, other
pieces that involve pair creation and destruction. The full expressions
are somewhat messy and we refrain from displaying them here. However, and this
is key, we will need only a very simple part (corresponding to the
low momentum limit of the particle-number conserving piece) of this quartic term for the 
calculation
of the ${\cal S}$-matrix in the low-momentum limit; our method of writing everything in
terms of the creation and annihilation operators has the
advantage of identifying and isolating this piece at the very outset.
Finally, as an aside, we note
that the total spin operator of the system may be written in terms of the
triplet operators as
\begin{equation}
S_{tot}^{a}=i\epsilon^{abc}\int_{-\infty}^{\infty}\frac{dp}{2\pi}t_{b}^{\dag}(p)t_{c}(p)~;
\label{totspin}
\end{equation}
this confirms that the triplet creation operator $t_{a}^{\dag}$ does
indeed create a single spin one quasiparticle (with polarization $a$) of the non-interacting 
system.

With all of this in mind, let us turn to the analysis of the scattering
properties of this model. As we are hoping to calculate the ${\cal S}$-matrix
perturbatively in the coupling $g$, it is convenient to write
${\cal S} = {\bf 1} + i{\cal T}$. The `transition-matrix'~~${\cal T}$ can then
be calculated perturbatively using the standard field-theoretic prescription
that relates it to the corresponding amputated, connected Green's
functions of the theory. Let us make this precise for the case we
are interested in: namely, a scattering process in which the initial
state consists of two particles, one with momentum $k_1$ and spin polarization
$\mu_1$, and the other with momentum $k_2$ and spin polarization $\mu_2$, and
the final state has two particles labelled by $(k_{1}^{\prime },{\mu}_{1}^{\prime})$ and 
$(k_{2}^{\prime },{\mu}_{2}^{\prime})$. Note that
we are now {\em not} talking about the bare particles of the non-interacting
theory, but the actual physical quasiparticle states of the system, correct
to the relevant order in the perturbative expansion in $g$. The corresponding
matrix element, ${\cal S}^{\mu_1 \mu_2}_{\mu^{\prime}_1,
\mu^{\prime}_2} (k_1, k_2; k^{\prime}_1,k^{\prime}_2) \equiv \langle 
k^{\prime}_1~\mu^{\prime}_1,k^{\prime}_2~\mu^{\prime}_2|{\cal S}|k_1~\mu_1,
k_2~\mu_2 \rangle $, may then be written as
\begin{eqnarray}
{\cal S}^{\mu_1 \mu_2}_{\mu^{\prime}_1,
\mu^{\prime}_2} (k_1, k_2; k^{\prime}_1,k^{\prime}_2) = &&(2\pi )^2\delta_{\mu_1 \mu^{\prime}_1} 
\delta_{\mu_2 \mu^{\prime}_2}
\delta(k_1 - k^{\prime}_1) \delta (k_2 - k_2^{\prime}) ~~~+ \nonumber \\ 
&&~~~~~(2\pi)^2\delta(E_f-E_i)\delta(k_f-k_i)i{\cal M}^{\mu_1 
\mu_2}_{\mu^{\prime}_1,\mu^{\prime}_2} (k_1, k_2; k^{\prime}_1,k^{\prime}_2)~,
\label{defM}
\end{eqnarray}
where $E_f=\varepsilon(k_{1}^{\prime})+\varepsilon(k_{2}^{\prime})$ and 
$E_i=\varepsilon(k_1)+\varepsilon(k_2)$ are the final and initial energies respectively, $k_f$ 
and $k_i$ are the total momenta in the final and
initial states respectively, and ${\cal M}$ is the `reduced' matrix
element (with energy and momentum conserving $\delta$ functions removed) for
the process under consideration.

We now specialize to the case $k_1=k$, $k_2 = -k$ ($k>0$); this
special case allows us to make our basic point (regarding the infrared
divergences present in a perturbative calculation of the scattering
properties) while keeping the calculations simple. In this case, we may
decompose the scattering matrix as follows:
\begin{eqnarray}
{\cal S}^{\mu_1 \mu_2}_{\mu^{\prime}_1,
\mu^{\prime}_2} (k, -k; k^{\prime}_1,k^{\prime}_2)
= && \delta( k^{\prime}_1-k) \delta (k_2^{\prime}+k) \left[
S_1 ( k) \delta_{\mu_1 \mu_2} \delta_{\mu^{\prime}_1 \mu^{\prime}_2}\right.
\nonumber \\
&&~~~~~~~~~~~~~~~\left.
+ S_2 ( k) \delta_{\mu_1 \mu^{\prime}_1} \delta_{\mu_2 \mu^{\prime}_2}
+ S_3 ( k) \delta_{\mu_1 \mu^{\prime}_2} \delta_{\mu_2 \mu^{\prime}_1}\right]~.
\label{decompS}
\end{eqnarray}
Now, energy and momentum conservation
in one dimension provide enough constraints on the two-body problem to ensure that the allowed
final states have the same set of momentum labels as the initial state. This
allows us to convert the overall energy and momentum conserving $\delta$ functions in the second 
term of (\ref{defM}) to $\delta$ functions that
identify $k_1^{\prime}$ with $k$ and $k_2^{\prime}$ with $-k$. In the
process, we of course introduce additional kinematic factors coming from
the Jacobian (we are basically using $\delta(f(x)) = \delta(x)/|f^{\prime}(x)|$~).
Using this, we can write
\begin{eqnarray}
S_1(k) & = & \left(\frac{\varepsilon(k)}{2kv_{F}^{2}}\right)iM_1(k)~, \nonumber \\
S_2(k) & = & 1 + \left(\frac{\varepsilon(k)}{2kv_{F}^{2}}\right)iM_2(k)~, \nonumber \\
S_3(k) & = & \left(\frac{\varepsilon(k)}{2kv_{F}^{2}}\right)iM_3(k)~, 
\label{relSM}
\end{eqnarray}
where $M_1$, $M_2$, and $M_3$ are defined in terms of the following decomposition for ${\cal M}$:
\begin{eqnarray}
{\cal M}^{\mu_1 \mu_2}_{\mu^{\prime}_1,
\mu^{\prime}_2} (k, -k; k,-k)
& = & \left[
M_1 ( k) \delta_{\mu_1 \mu_2} \delta_{\mu^{\prime}_1 \mu^{\prime}_2}+ M_2 ( k) \delta_{\mu_1 
\mu^{\prime}_1} \delta_{\mu_2 \mu^{\prime}_2}
+ M_3 ( k) \delta_{\mu_1 \mu^{\prime}_2} \delta_{\mu_2 \mu^{\prime}_1}\right]~.
\label{decompM}
\end{eqnarray}

The relations (\ref{relSM}) are useful because there is a simple diagrammatic
prescription for the perturbative evaluation of ${\cal M}$. According to this
standard field theoretic prescription~\cite{peskin}, $i{\cal M}^{\mu_1 \mu_2}_{\mu^{\prime}_1,
\mu^{\prime}_2} (k_1, k_2; k_1^{\prime},k_2^{\prime})$ is proportional to the sum of all 
`amputated' (factors corresponding to external legs omitted), fully connected, one particle 
irreducible diagrams contributing to the time ordered four-point function with two incoming 
external lines and two outgoing external
lines. The incoming lines must carry momenta $k_1$ and $k_2$, frequencies
$\omega_1$ and $\omega_2$ set to their respective `on-shell' values of
$\varepsilon(k_1)$ and $\varepsilon(k_2)$, and spin labels $\mu_1$ and $\mu_2$
respectively. The outgoing lines must carry momenta $k_1^{\prime}$ and
$k_2^{\prime}$, frequencies again set to their on-shell values of
$\varepsilon(k_1^{\prime})$ and $\varepsilon(k_2^{\prime})$, and spin labels $\mu_{1}^{\prime}$ 
and $\mu_{2}^{\prime}$ respectively. Denoting the
sum of all such diagrams schematically by $\Gamma_4$, we can
write
\begin{equation}
i{\cal M}^{\mu_1 \mu_2}_{\mu^{\prime}_1,
\mu^{\prime}_2} (k_1, k_2; k_1^{\prime},k_2^{\prime}) = (\sqrt{Z})^4 
\Gamma_4(k_1~\mu_1,k_2~\mu_2;k^{\prime}_1~\mu^{\prime}_1,k^{\prime}_2~\mu^
{\prime}_2)~,
\label{schempres}
\end{equation}
where the field-strength renormalization factor $Z$ comes into
play because the singlet and triplet creation operators $s^{\dag}$
and $t_a^{\dag}$ create the bare particles, while we are asking questions
about the scattering properties of the physical quasiparticle excitations.
We will not be very careful here about the precise definition of $Z$; it
will soon become apparent that this does not play any role in the
calculation we do.

\begin{figure}
\epsfxsize=4.0in
\centerline{\epsffile{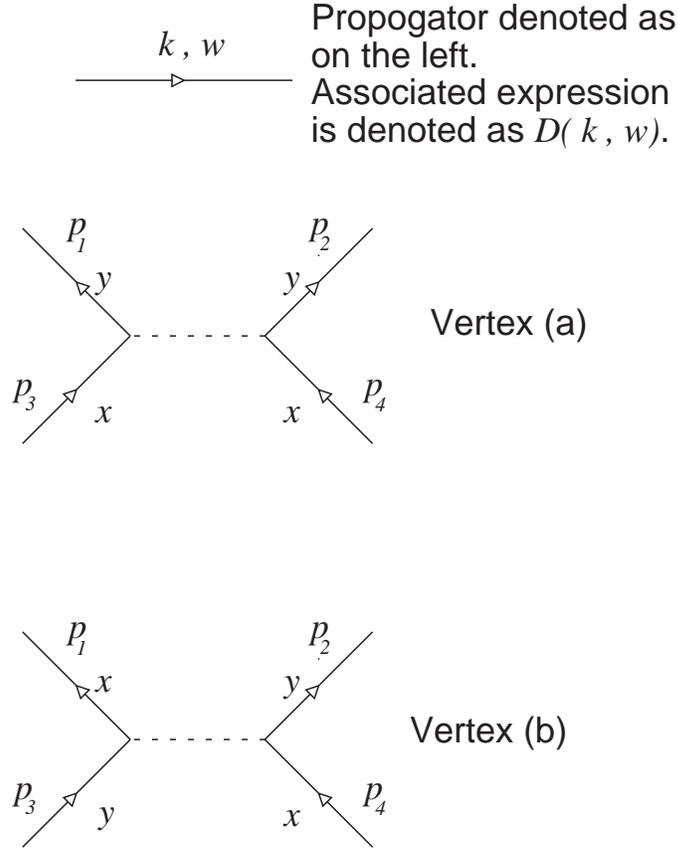}}
\vspace{0.2in}
\caption{The Feynman rules we need for the calculation described in
Section~{\protect {\ref{fuku}}}. The propogator $D(k,\omega)$ is
given as $D(k,\omega) = i/(\omega -\varepsilon(k) +i\eta)$.  The factor corresponding to the 
vertex
(a) is $ig(f_R(p_3){\bar f}_R(p_1)f_L(p_4){\bar f}_L(p_2)+R \longleftrightarrow L)/2$~. The 
factor corresponding to (b) is $ig(f_L(p_4){\bar f}_R(p_1)-f_R(p_4){\bar f}_L(p_1))(f_L(p_3){\bar 
f}_R(p_2)-f_R(p_3){\bar f}_L(p_2))$~.}
\label{fig5}
\end{figure}

Before we set about calculating $\Gamma_4$, we need to specify our 
conventions regarding the diagrammatic representation of perturbation theory.
As shown in Fig~\ref{fig5}, we denote the propogator of the triplet particle by a solid line with 
an arrow carrying momentum $k$, frequency $\omega$
and spin label $\mu$; this has a factor of~~$i/(\omega -\varepsilon(k) + i\eta )$
associated with it. It turns out that we do not need to consider any diagrams
that have lines corresponding to singlet particles and we will not bother
to introduce a diagrammatic representation for their propogator. We also
display our diagram convention for the four point vertices of the theory in the
same figure; again, only the particle number conserving vertices in which
all four lines involved correspond to triplet particles have been
assigned a diagram as the others will not play a role in what follows. One
type of vertex, labelled (a) in Fig~\ref{fig5},  depicts a process in which two particles of 
momentum $p_3$ and $p_4$, both with
spin label $\mu =x$ scatter into a final state populated by two particles
with momenta $p_1$ and $p_2$, and spin label $\mu = y$. The full momentum
dependent factor associated with this diagram is also shown below it. We will
need only a very simple low-momentum limit of this expression in most of what follows.
The other kind of vertex, labelled (b) in Fig~\ref{fig5}, shows
incoming particles with labels $(p_3~y)$ and $(p_4~x)$ scattering into a
final state populated by particles with labels $(p_1~x)$ and $(p_2~y)$ respectively. Again, the 
full momentum dependent factor is displayed alongside
for completeness. We will mostly need only the value of this factor when all four
momenta equal zero; this is given simply by $-ig$. Of course, all other
vertices of the same type, but having different spin labels that can be obtained
from these using the $O(3)$ symmetry of the problem, have the same factors
associated with them.

\begin{figure}
\epsfxsize=4.0in
\centerline{\epsffile{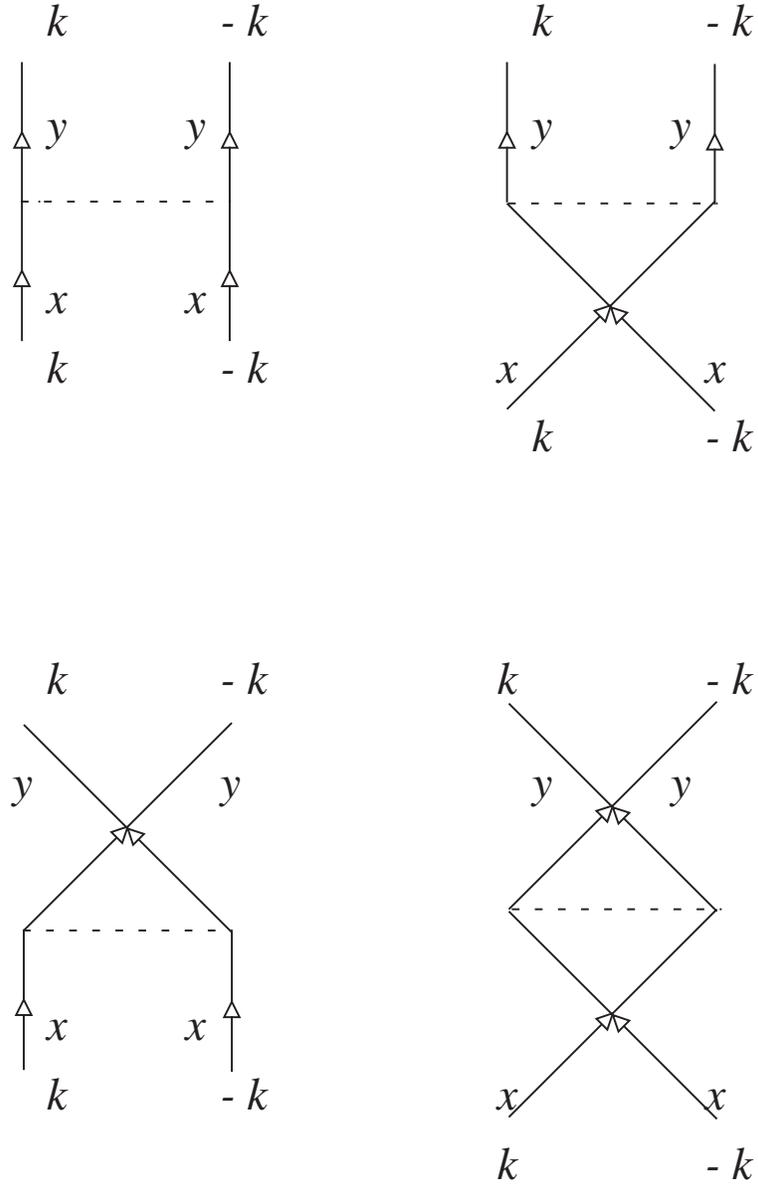}}
\vspace{0.2in}
\caption{Feynman diagrams contributing to $\Gamma_4(k~x,-k~x;k~y,-k~y)$ to
first order in $g$. All external lines carry on-shell frequencies
corresponding to the momentum labels shown.}
\label{fig6}
\end{figure}

\begin{figure}
\epsfxsize=2in
\centerline{\epsffile{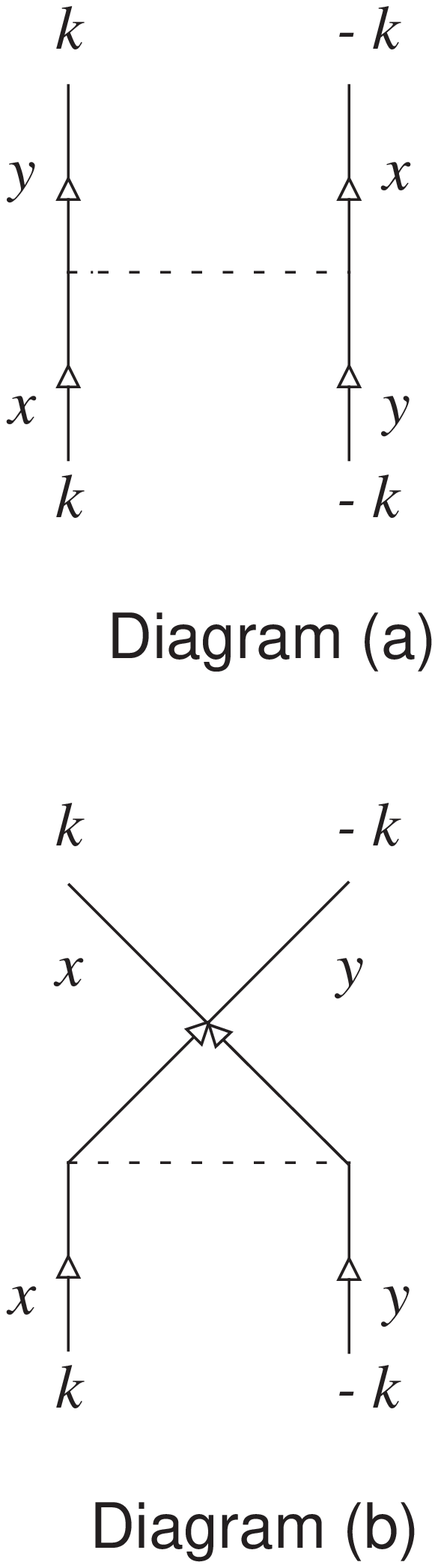}}
\vspace{0.2in}
\caption{Feynman diagrams contributing to $\Gamma_4(k~x,-k~y;k~y,-k~x)$ (diagram (a)) and 
$\Gamma_4(k~x,-k~y;k~x,-k~y)$ (diagram (b)) to
first order in $g$. All external lines carry on-shell frequencies
corresponding to the momentum labels shown.}
\label{fig7}
\end{figure}

We are now in a postition to do some calculations. We begin by noting
that, apart from the overall factor of $(\sqrt{Z})^4$ which we are ignoring
for now, $iM_1(k)$, $iM_2(k)$ and $iM_3(k)$ are equal to $\Gamma_4(k~x,-k~x;k~y,-k~y)$~,
$\Gamma_4(k~x,-k~y;k~x,-k~y)$~, and $\Gamma_4(k~x,-k~y;k~y,-k~x)$~ respectively.
It is quite simple to calculate these three quantities to leading order in
$g$. The diagrams contributing to $iM_1$is shown in Fig~\ref{fig6}, while those contributing to 
$iM_2$ and $iM_3$ are shown in Fig~\ref{fig7}. Evaluating these `tree-level' amplitudes, we 
obtain
\begin{eqnarray}
\Gamma_4(k~x,-k~x;k~y,-k~y) & = & ig\frac{k^2v_{F}^{2}}{\varepsilon^2(k)}~, \nonumber \\
\Gamma_4(k~x,-k~y;k~x,-k~y) & = & ig\frac{\Delta^2}{\varepsilon^2(k)}~,\nonumber \\
\Gamma_4(k~x,-k~y;k~y,-k~x) & = & -ig~.
\label{Oggamma}
\end{eqnarray}
As long as we are interested in only the first order result for ${\cal S}$,
we can set $Z=1$ and directly use these expressions to get the following
results for the leading low $k$ behaviour of $S_1$, $S_2$, and $S_3$ correct
to first order in $g$:
\begin{eqnarray}
S_1(k) & = & \frac{ig}{2v_F}\left(\frac{kv_F}{\Delta}\right)~, \nonumber \\
S_2(k) & = & 1+\frac{ig}{2v_F}\left(\frac{\Delta}{kv_F}\right)~, \nonumber \\
S_3(k) & = & -\frac{ig}{2v_F}\left(\frac{\Delta}{kv_F}\right)~.
\label{OgS}
\end{eqnarray}
We immediately see that the perturbative expansion cannot be trusted
in the low-momentum limit because of the infrared divergences
present in the expressions for $S_2$ and $S_3$. The structure of this
first order result is seen to be qualitatively similar to the
${\cal O}(1/N)$ result for the two-particle ${\cal S}$-matrix of the
$O(N)$ $\sigma$-model~\cite{zamal}. In the latter case, we know that
the exact value of the ${\cal S}$-matrix is perfectly well-behaved in
the $k \rightarrow 0$ limit and is in fact given by the super-universal
expression (\ref{smatrix}). To obtain the correct result in this limit
for our problem, we need to identify the {\em leading} infrared divergences
at {\em each order in} $g$ and perform a resummation. Now, we do not
expect any infrared divergences in the perturbation expansion of $Z$ and
as a result the prefactor of $(\sqrt{Z})^4$ in the expression for ${\cal M}$ does not contribute 
at
all to the terms that need to be resummed; we will forget about this
factor from now on.

\begin{figure}
\epsfxsize=4.5in
\centerline{\epsffile{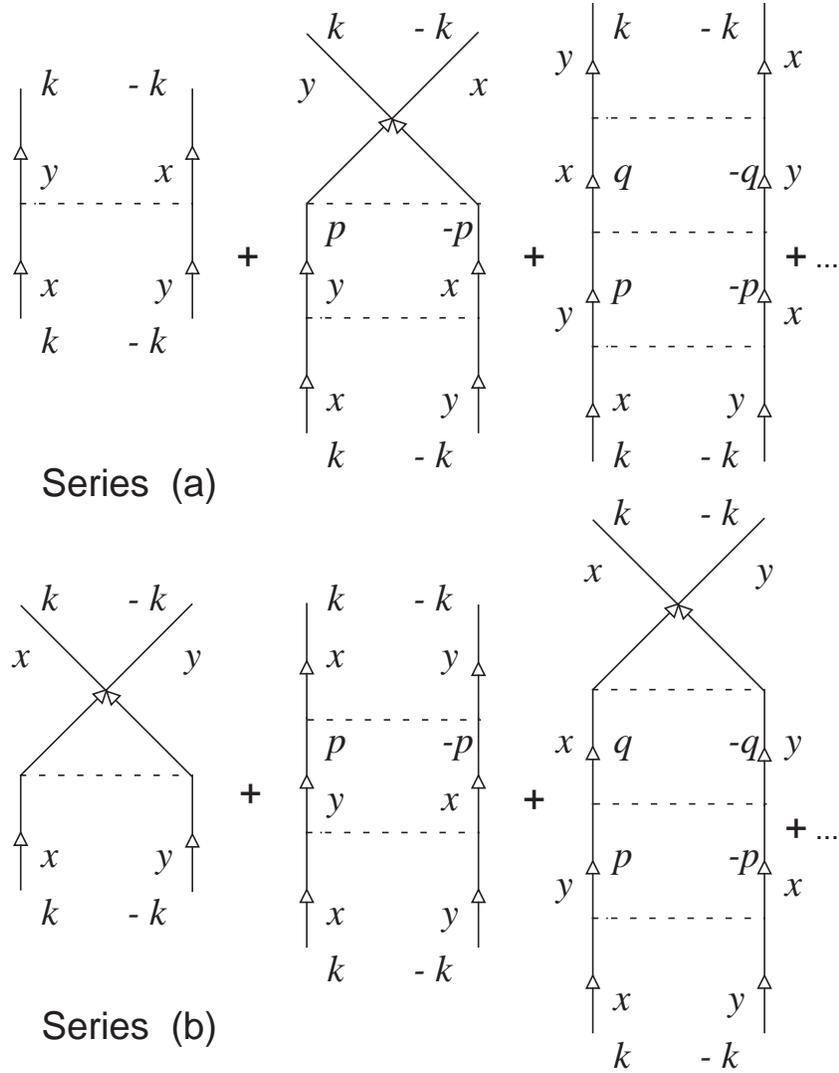}}
\vspace{0.2in}
\caption{Ladder series giving the leading infrared divergent terms
in the expansion for $iM_3$ (diagram (a)) and $iM_2$ (diagram (b)). All
external lines carry on-shell frequencies corresponding to the momentum
labels shown. The internal lines also carry frequency labels that are
not explicitly shown.}
\label{fig8}
\end{figure}

Let us now try and identify the leading infrared divergent diagrams at
each order in perturbation theory. First of all, it is clear, purely
from frequency and momentum conservation at each vertex, that no
diagrams involving pair creation or annihilation can provide the
leading divergence at any order. Moreover, only internal loops in
which both propogators involved point in the same direction give
a nonzero result on doing the integral over the frequency running through
the loop. A little thought should convince the reader that these two
constraints allow us to conclude that the ladder series shown in Fig~\ref{fig8} give the leading 
infrared divergent terms in
$S_2$ and $S_3$ to all orders in $g$. Turning our attention to $S_1$, we
see immediately that Fermi statistics guarantees that each vertex in the
analogous ladder series for $S_1$ has enough factors of momentum associated
with it to rule out any infrared divergence appearing in $S_1$. Our
task is thus reduced to evaluating the two series shown in Fig~\ref{fig8}.
To do this, we note that as far as the coefficient of the divergent piece is
concerned, we can ignore the momentum dependence of each vertex and simply
replace it with a factor of $-ig$. Each crossing of the fermion propogators
gives a factor of $-1$ and each loop integral gives $\Delta/2kv_F^2$. Putting
all this together and summing the resultant geometric series, we obtain
the following {\em non-perturbative} results for the low momentum
behaviour of $S_2$ and $S_3$:
\begin{eqnarray}
S_2(k) & = & \frac {2ikv_F^2}{g \Delta + 2ikv_F^2}~, \nonumber \\
S_3(k) & = & \frac{g \Delta}{g \Delta + 2ikv_F^2}~. 
\label{resum12}
\end{eqnarray}
An interesting feature of these results is the pole in the upper-half $k$
plane at $k=i\Delta g/2v_F^2$ which seems to suggest the presence of
a bound state. However, this region of $k$ space is definitely beyond the
domain of validity of (\ref{resum12}) and it is not clear what significance, if
any, to ascribe to this curious fact.

Turning to firmer ground, we see that
the foregoing implies that the low-momentum limit of the two particle
${\cal S}$-matrix is perfectly well-defined and is in fact given by
\begin{equation}
 {\cal S}^{\mu_1 \mu_2}_{\mu^{\prime}_1,
\mu^{\prime}_2} (k_1, k_2; k^{\prime}_1,k^{\prime}_2)
= \delta_{\mu_1 \mu^{\prime}_2}\delta_{\mu_2 \mu^{\prime}_1}
2\pi\delta(k_1 - k^{\prime}_1) 2\pi\delta (k_2 - k_2^{\prime})~.
\label{fermismatix}
\end{equation}
Note that apart from an overall factor of minus one, this is exactly
the super-universal form (\ref{smatrix}). The relative sign is simply
a consequence of fermi statistics and our choice of phase for the
final state of the scattering process. In any case, we will see that
when we use the super-universal form of the ${\cal S}$-matrix for
discussing spin transport, the overall phase is immaterial. On the other hand,
the overall factor of $-1$ in the superuniversal form (\ref{smatrix})
will be crucial when we work out the correlators of the staggered
component of the magnetization density. This may seem worrisome at first
sight. However, as we do not
have any local representation of the staggered component of the spin density in terms
of the Majorana fermion operators, there is no contradiction at all. In fact,
the semiclassical techniques used in Section~\ref{finitet} may also be applied
to the problem of calculating the finite temperature Green function of
the fermions; this would correspond to calculating the finite temperature
correlators of some {\em highly non-local} string operators of the
original spin system. However, as it is difficult to see how these
may be accessible at all to any experimental probes, we do not pursue
this line of thought any further.

Thus, we see that the low-momentum behaviour of the ${\cal S}$-matrix
in this fermionic representation of the weakly-coupled ladder is
consistent with the super-universal form (\ref{smatrix}), although this
behaviour is definitely not accessible to perturbation theory. This leads
us to believe that similar infrared divergences would invalidate
any perturbative calculation of dynamical properties at finite temperature
(when there will be a dilute gas of thermally excited particles present)
that uses this representation. In particular, this appears to indicate
that the results of Ref~\cite{kishifuku} for the NMR relaxation rate $1/T_1$
are incorrect at low $T>0$.

\section{Dynamics and transport for $0 < T\ll \Delta$}
\label{sec:lowt}

The results of this section are expected to apply to all gapped one-dimensional
antiferromagnets with massive spin-one quasiparticles. We will develop, what we believe 
is an exact semiclassical theory of dynamics
and transport for $T \ll \Delta$. We will consider fluctuations near $q=Q$
in Section~\ref{finitet}, and near $q=0$ in Section~\ref{hdep}.

\subsection{Thermal broadening of the single-particle peak in $S(q,\omega)$}
\label{finitet}

In this section, we present calculations leading up to
our results for the thermal broadening of the single particle peak in the
dynamics structure factor.

The inelastic neutron scattering cross-section provides a direct measure~\cite{brohgeilo} of the
dynamical structure factor $S(q,\omega )$ which is defined as
\begin{equation}
S(q,\omega)={1 \over {2 \pi}} \int dt e^{i \omega t}\langle \hat{ s}_{\alpha}(q,t)
\hat{s}_{\alpha}(-q,0)\rangle ;
\label{strucfacdefn}
\end{equation}
where $\hat{s}_{\alpha}(q,t)$ is the Heisenberg representation operator
corresponding to the component of the spin density at wavevector $q$, the
expectation values are with respect to the usual equilibrium density matrix and
summation over the repeated index $\alpha$ is implied (note that we are
assuming rotational invariance in spin space and working at $H=0$). We are
interested in the structure factor for $q$ close to $\pi/a$. In this case
we have
\begin{equation}
S(q,\omega)\propto {1 \over {2 \pi}} \int dt e^{i \omega t}\langle \hat{ n}_{\alpha}(k,t)
\hat{n}_{\alpha}(-k,0)\rangle ;
\label{strucfac1}
\end{equation}
where $k=q-\pi/a$. To get a feel for what (\ref{strucfac1}) looks like
at $T=0$, let us consider a particular lattice regularization of the
$\sigma$-model, defined by the quantum rotor Hamiltonian
\begin{displaymath}
{\cal H}={g \over  2}\sum_{i}{\bf{\hat{L}}}_{i}^2 -{1 \over g}\sum_{i}{\bf{\hat{n}}}
_{i}{\bf{.}} {\bf{\hat{n}}}_{i+1};
\end{displaymath}
where ${\bf{\hat{L}}}_{i}$ is the angular momentum operator of the rotor at site
$i$, ${\bf{\hat{n}}}_{i}$ is the unit vector that denotes the position
of the rotor at site $i$ and we have temporarily set $c=a=1$. It is not hard
to analyze the properties of this model in a large $g$, strong coupling
expansion; moreover this is expected to be qualitatively correct for all $g$
in one dimension~\cite{polyakov}.
To lowest order, we can easily see that
the ground state would just be a product state with each site being in
an eigenstate of ${\bf{\hat{L}}}$ with zero eigenvalue. The lowest
excited states would be a degenerate manifold corresponding to promoting
any one site to the $L=1$ state and thereby creating a `particle' at that
site. To first order in $1/g$, a hopping term would be generated in the effective
Hamiltonian for the single-particle sector, resulting in a band of one
particle excitations. To this order,  ${\bf{\hat{n}}}$ is just a sum
of creation and annihilation operators for the stable particle-like excitation
of the system. At higher orders in $1/g$, ${\bf{\hat{n}}}$ acting on the vacuum will also produce
multiparticle states, but there will always be some single particle
component. 
Reverting back to our continuum theory, we see that (\ref{strucfac1}) evaluated at $T=0$ would 
have a 
contribution $\sim \delta(\omega-\varepsilon(k))$ associated with the
stable particle. The next
contribution is actually a continuum above the $3$-particle threshold~\cite{unknown2}. Following 
\cite{ssyoung}, we shall now
focus exclusively on how this one-particle peak broadens as $T$ becomes
non-zero. Let us define
\begin{equation}
C(x,t)= \frac{1}{3}\langle \hat{ n}_{\alpha}(x,t)
\hat{n}_{\alpha}(0,0)\rangle~,
\label{defC}
\end{equation}
where the repeated index $\alpha$ is summed over.
Let $K(x,t)$ denote $C(x,t)$ evaluated at $T=0$ keeping only the
single particle contributions. We have
\begin{equation}
K(x,t)=\int {dp \over {2 \pi}}D(p)e^{ipx-i \varepsilon (p) t}~.
\label{Kexpression}
\end{equation}
Here $D(p)$ is a `form factor'. For our Lorentz invariant continuum model, 
\begin{equation}
D(p) = \frac{{\cal A} c}{2\varepsilon (p)}
\label{defcala}
\end{equation}
where ${\cal A}$ is a non-universal quasiparticle residue.
This gives $K(x,t)={\cal A}K_{0}(\Delta(x^2-c^2t^2)^{1/2}/c)/(2\pi)$,
with $K_{0}$ the modified Bessel function.

Now let us evaluate $C(x,t)$ for non-zero temperatures using the semiclassical
method of \cite{ssyoung}. First, it is convenient to switch to operators
${ n}_{+1}(x)$, ${ n}_{-1}(x)$ and ${ n}_{0}(x)$, defined as
\begin{displaymath}
{n}_{+1} = {n}_{-1}^{\dagger} = { n}_{x}-i{ n}_{y}~,
\end{displaymath}
and
\begin{displaymath}
{n}_{0}={n}_{z}~.
\end{displaymath}
${ n}_{+1}$ is a sum of a creation operator for particles with $z$-component
of spin $m$ equal to $+1$ and an annihilation operator for particles with
$m$ equal to $-1$. ${ n}_{0}$ is a sum of creation and
annihilation operators for particles with $m$ equal to
$0$. In the absence of an external field, we may write
\begin{equation}
C(x,t) = \langle \hat{ n}_{0}(x,t)
\hat{n}_{0}(0,0)\rangle~.
\label{Czero}
\end{equation}
We represent (\ref{Czero}) as a `double time' path integral, with the
$e^{-i{\hat {\cal H}}t}$ factor coming from the Heisenberg operator generating
paths that move forward in time, and the $e^{+i{\hat {\cal H}}t}$ producing
paths that move backward in time. We begin with an initial state which is
populated by thermally excited particles, the density of particles being
$\sim$ $e^{-\Delta/T}$ and their mean spacing being much larger than the
thermal de-Broglie wavelength $\sim$ $c/(\Delta T)^{-1/2}$. As argued in
\cite {sskd,ssyoung}, this means that the particles can be treated semiclassically. In
this semiclassical limit the dominant contribution to the Feynman sum comes
about when the paths going backward in time are exactly the time-reversed
counterparts of those going forward and all particles follow their classical trajectories between 
collisions~\cite{sskd,ssyoung}. Whenever two particles collide, energy and
momentum conservation is sufficient to determine the final momenta. However,
one cannot entirely ignore quantum effects of the collisions. The spins
of the particles after the collision as well as the phase picked up by
the wavefunction of the system as a result of the collision is determined
by the quantum mechanical scattering matrix (${\cal S}$). For $T \ll \Delta$,
the particles all move very slowly and we need only the super-universal
low-momentum limit (\ref{smatrix}).

All this leads to the
following description of $C(x,t)$ in this asymptotic limit~\cite{sskd}: At time
$t=0$ we begin with an initial state populated equally (for $H=0$) with three species 
(corresponding to the three values of spin projection $m$) of particles each uniformly 
distributed in space with density $\rho/3$, where the total
density $\rho$ is given as
\begin{displaymath}
\rho =  3\int \frac{dp}{2\pi} e^{-(\Delta  + c^2 p^2/2 \Delta)/T}
=3\sqrt{ \frac{T \Delta}{2\pi c^2} }  e^{-\Delta /T}.
\end{displaymath}
The velocities are distributed according to the classical Maxwell-Boltzmann
distribution function
\begin{displaymath}
{\cal P}(v) = \sqrt{  \frac {\Delta}{2\pi c^2 T}} e^{-\Delta v^2/2c^2T}~.
\end{displaymath}
Each particle in the initial state is assigned one of the three values of
$m$ with equal probability (assuming $H=0$). The operator $n_{0}(0)$ acting on this initial state 
creates at time $t=0$ one extra particle at $x=0$ with spin value equal to $0$ (the annihilation 
part of $n_{0}$ gives a contribution which is exponentially suppressed and is ignored here). 
These particles follow their classical
trajectories forward in time. At every collision, we pick up a factor
of $-1$ from the ${\cal S}$-matrix. At time $t$, a particle with spin projection of zero is 
annihilated at
$x$ by $n_{0}(x)$. The resulting state is then propogated backward in time
to $t=0$ and its overlap with the initial state calculated. $C(x,t)$ is then
given by the average of this overlap over the ensemble specified earlier.

A typical example of a space-time configuration of trajectories that
leads to a non-zero value for this overlap is shown in Fig~\ref{fig9}.
All trajectories in the figure except the dotted line denote space-time
paths that are traversed both forward and backward in time. The dotted line
is traversed only forward in time as the particle travelling on it is
destroyed at time $t$ by $n_{0}(x)$. A little thought convinces one that
this overlap is non-zero only when all particles colliding with a
particle travelling on the dotted trajectory have the same spin $m$
(equal to zero) as it does. Moreover, when this condition is satisfied, the
value of the overlap is just $(-1)^{n_{ l}}K(x,t)$ where $n_{l}$ is
the number of collisions that the dotted trajectory suffers. The factor of
$(-1)^{n_{l}}$ comes from the scattering matrix at each collision
between a particle on the dotted trajectory and other particles. All other
collisions occur in pairs (the second member of the pair coming from the
evolution backward in time) and thus do not contribute any phase factor.
The factor of $K(x,t)$ is just the relativistic amplitude for the propogation
of a single particle from $x=0$ at $t=0$ to position $x$ at time $t$.

All this implies that we can write
\begin{equation}
C(x,t)=R(x,t)K(x,t)
\label{Rdefine}
\end{equation}
which defines the `relaxation function' $R$. 
For the case where the particle has only {\em one} allowed value of its spin label, $m$,
it is possible to compute $R(x,t)$ analytically~\cite{ssyoung}, and the resulting
expression (\ref{Rdefine}) then agrees precisely with a computation using very sophisticated
quantum inverse scattering method~\cite{korepin}: this agreement gives us 
confidence that the physical approach developed here is asymptotically
exact at low temperatures.

Let us now turn to the calculation of $R$ for
the case of interest here.
We begin by writing a formal expression
for $R$ based on the foregoing semiclassical description. Let ${\bf \{} x_{k}(0){\bf  \}}$ be the 
positions of the thermally excited particles at time $t=0$. Let ${\bf \{} v_{k}{\bf  \}}$ be 
their initial velocities. Here $k$ is an index running
from $1$ to $N$, the total number of particles present in the initial state in
a system of size $L$. We label the initial positions with the convention that
$x_{k}(0) < x_{l}(0)$ for $k < l$. Let $X_{k}(t) \equiv x_{k}(0)+v_{k}t$ denote
the $k^{th}$ space-time trajectory (note that this is quite distinct from the
position of the $k^{th}$ particle at time $t$). Let $m_{k}(t)$ denote the
spin projection value of the particle travelling along the $k^{th}$ trajectory
at time $t$. The spin projections are randomly assigned to each particle at
time $t=0$ as described earlier and $m_{k}(t)$ at later times depends on which
particle is travelling on the $k^{th}$ trajectory at any given time. We have the
following expression for $R$:
\begin{equation}
R(x,t)= \langle \prod _{k} {\cal F}_{k} \rangle~,
\label{Rexpress}
\end{equation}
with
\begin{displaymath}
{\cal F}_{k} =  1-(1+\delta_{m_{k}(\tau_{k}),1}){\tilde \Theta }_{k}~;
\end{displaymath}
where
\begin{displaymath}
{\tilde \Theta }_{k} = \theta (x-X_{k}(t))\theta (x_{k}(0)) + \theta (X_{k}(t)-x)\theta (-
x_{k}(0))~ ,
\end{displaymath}
and
\begin{displaymath}
\tau_{k} = x_{k}(0)t/(x-v_{k}t)~.
\end{displaymath}
The angular brackets in (\ref {Rexpress}) denote averaging over the ensemble of
initial conditions specified earlier. 

Unfortunately, it does not seem possible to do the ensemble average
analytically. Using methods of Refs~\cite{jepson,lebowitz}, it
is possible to develop a `cumulant' expansion for the logarithm of $R$~\cite{ourattempt}. This
expansion, however, is essentially a short time expansion which is not uniformly
convergent, and thus not very useful for our purposes as we eventually need to
Fourier transform $C(x,t)$. It is also possible to develop a `mean-field'
approximation to this classical model that ignores the complicated correlations
between the $m_k(t)$ at different times~(see Appendix~\ref{mf}). This proves to be reasonably 
accurate
at least for the $R(0,t)$, though the high-accuracy numerics we describe
next show clear deviations from the mean-field results. So, although we have an asymptotically 
exact formulation for the non-zero temperature $C(x,t)$ at distances much larger than the thermal 
de-Broglie
wavelength and times much larger than $T^{-1}$, we need to numerically
determine the relaxation function $R$ to actually calculate anything accurately.
This is what we turn to next.

An important property of $R(x,t)$, which follows directly from (\ref{Rexpress}) is that it can be 
written in a
scaling form as $R(x,t)={\tilde R}({\tilde x}, {\tilde t})$ with
${\tilde x} =x/L_{x}$ and ${\tilde t} = t/L_{t}$ where 
\begin{displaymath}
L_{x} = \frac{1}{\rho} ~,
\end{displaymath}
and
\begin{displaymath}
L_{t} = {\frac {1}{\rho}}{\left (\frac{\Delta}{2c^2T}\right )^{1/2}}~.
\end{displaymath}
Thus it is most convenient for the numerics to measure length in units of
$L_x$ and time in units of $L_t$ and directly calculate ${\tilde R}$. We
start with a system size of $L = 400$ (in units of $L_x$) and impose periodic boundary 
conditions. The density in
these units is unity and so the initial state is populated by $400$ particles
with their initial positions drawn from a uniform ensemble. This system size
is large enough that finite-size effects are negligible for our purposes. 
Each particle is assigned a velocity from the classical thermal ensemble. In these new units this 
implies that we choose velocities from the distribution
\begin{displaymath}
{\tilde {\cal P}}({\tilde v}) = \frac {1}{\sqrt \pi } e^{-{\tilde v}^2}~.
\end{displaymath}
An important advantage of our method is that we do the average over
the spin values analytically. To do this, we note that it is possible to
reformulate the calculation of ${\tilde R}$ by writing
\begin{displaymath}
{\tilde R} = \langle T({\cal C})\rangle ~,
\end{displaymath}
where ${\cal C}$ denotes a given space time configuration of trajectories,
the angular brackets denote averages {\em only} over the initial positions and
velocities that define this configuration, and $T({\cal C})$ is defined as
\begin{displaymath}
T({\cal C}) = (-1)^{n_h} \left ({\frac {1}{3}}\right )^{n_p}~.
\end{displaymath}
Here, $n_h$ is the total number of collisions involving a particle travelling on 
the dotted trajectory of Fig~\ref{fig9} and $n_p$ is the number of {\em different} thermally 
excited
particles that have had collisions with a particle
travelling on the dotted trajectory. Now, $T({\cal C})=0$ for all configurations ${\cal C}$ in 
which
the presence of the extra particle (that starts out on the dotted trajectory)
affects the evolution of the various spin values $m_{k}(t)$. So we might as
well forget about the particle travelling on the dotted trajectory and consider
an auxillary spacetime diagram that now involves only the
thermally excited particles. We now agree to ignore the spin label on the
dotted line of Fig~\ref{fig9}; the dotted line now does $\em not$ denote
the trajectory of any particle. In terms of this picture we can define $n_h$ as the number of
times any solid line crosses the dotted line, and $n_p$ as the number of {\em different} 
thermally excited
particles that cross the dotted line.

With this new formulation in hand, calculating $T({\cal C})$ reduces to some simple book-keeping 
that keeps track of these two integers for a given configuration ${\cal C}$. We implement the 
ensemble average by averaging over $4 \times 10^6$
configurations drawn from the appropriate distribution. The combined absolute
error (statistical and finite-size) in ${\tilde R}({\tilde x},{\tilde t})$ for values of ${\tilde 
x}$,~${\tilde t}$ of interest to us is estimated to be
less than about $5 \times 10^{-4}$.

With ${\tilde R}$ available, it is a simple matter to numerically Fourier
transform the resulting $C(x,t)$ and obtain the dynamic structure factor
$S(q,\omega)$. Details of the numerical procedure used are relegated to 
Appendix~\ref{A}. Here we only comment on some conceptual issues involved and
discuss our results. 

There is an important subtlety associated with doing the Fourier transform
that needs to be first addressed. As discussed in Ref~\cite{ssyoung}, the
semiclassical result for $C(x,t)$ is valid as long as both $x$ and $t$ are
not very small; the results break down when $x \sim \lambda_T$ and $t \sim 
1/T$ ($\lambda_T$ being the thermal de-Broglie wavelength). However, the Fourier transform of $C$ 
(at wavevector $k=q-\pi/a$) is  an asymptotically valid
approximation to $S(q,\omega)$ {\em only} for $\omega$ close to $\varepsilon(k)$.
The reason for this can be understood by noting that the long-time
asymptotics of our form for $C(x,t)$ have an oscillatory character with
oscillations on the scale of $\Delta ^{-1}$. Put another way, it is
the spectral weight in the one-particle peak that plays a dominant role in
determining the long-time, large-distance asymptotics of $C(x,t)$ and 
so we can learn only about this feature in the spectral weight by Fourier
transforming our form for $C$.

With this caveat in mind, we have 
\begin{equation}
S(q,\omega)={1 \over 2\pi}\int\int  dtdx e^{i\omega t-ikx}K(x,t)R(x,t)~.
\label{broadS}
\end{equation}
where $k=q-\pi/a$.
We have not attempted to exhaustively map out $S(q,\omega)$, although
it would be quite straightforward to get more extensive numerical results should
they be of interest in some experimental context. Below we
confine ourselves to discussing our results for $S(q,\omega)$ for a 
couple of sample values of $q$.
Fig~\ref{fig10}
shows  scans in frequency across the quasiparticle peak in $S(q,\omega)$ for
$q=Q$ at four different values of temperature. It is interesting to note that when rescaled
by $L_t$ and plotted against a rescaled frequency variable $\delta {\tilde \omega} =(\omega - 
\Delta )L_t$, the three curves for
$\Delta/T = 3$, $4$, and $5$ seem to collapse on top of one another within
our numerical errors (which are conservatively estimated to be a few percent at the most). In
Fig~\ref{fig11}, we show a scan in wavevector across the same peak for $\omega =
\Delta$, again at the same four values of temperature. The curves at the lower
temperatures
again show scaling collapse; when rescaled by $L_t$ and plotted against the
rescaled variable $\hat {k}=kc \sqrt{L_{t}/\Delta}$, they seem
to all fall on top of one another. Moreover, the scaling curve in
Fig~\ref{fig11}, when plotted as a function of the independent variable $-{\hat {k}}^2/2$ 
coincides within our numerical error with the scaling function
of Fig~\ref{fig10} for $\delta {\tilde \omega} < 0$; this is displayed
in Fig~\ref{fig12}. While we do not have any reason to expect
that this scaling is generally true, all three observations may be put together in terms of
a scaling form that is valid {\em locally} in the neighbourhood
of the quasiparticle peak for $q=Q$; more formally we write
\begin{equation}
S(q,\omega )= \frac {{\cal A}cL_{t}}{\pi^2 \Delta}{\Phi}\left (\frac{\omega - 
\varepsilon(k)}{L_{t}^{-1}} \right )~.
\label{scale0}
\end{equation}

We also investigated $S(q,\omega)$ in the vicinity of the quasiparticle peak
corresponding to $q=Q+\Delta/c$ ; for this to be meaningful, we of course
need $\Delta/c$ to be  much less than the microscopic scale $\sim ~ a^{-1}$
beyond which our continuum theory does not work. We again tried to check if
the analogous scaling form,  
\begin{equation}
S(q,\omega )= \frac {{\cal A}cL_{t}}{\pi^2 \Delta}{ {\Phi}_{\Delta/c}}\left (\frac{\omega - 
\varepsilon(k)}{L_{t}^{-1}} \right )~,
\label{scale1}
\end{equation}
is approximately valid. Fig~\ref{fig13} shows scans in frequency across the
peak with $k$ held fixed at $\Delta/c$, for $\Delta/T = 2$, $3$, $4$, and $5$.
We see that the curves do not really appear to scale. In Fig~\ref{fig14}, we show scans in
wavevector, with $\omega$ held fixed equal to $\sqrt{2} \Delta$ for the same
values of the ratio $\Delta/T$. We plot the data (rescaled by $L_t$) against
the rescaled variable $\delta \hat {k}=cL_t(k-\Delta/c)$ (note the difference
in the choice of rescaling of the independent variable from the earlier case).
Again, in sharp contrast to the $q=\pi/a$ peak, we see that the curves do
not show any signs of scaling; our local scaling form is not a very good
way of organizing the data in this case.

These scaling properties are best understood as follows: Imagine developing $R(x,t)$ in
an expansion about $x=0$ for constant $t$ and then
calculating the Fourier integral in (\ref{broadS}). The zeroth order term
clearly gives us a result for $S(q,\omega)$ which is compatible with
the scaling form we have postulated for asymptotically low temperatures.
However, before we can trust this result, we need to check that the
corrections to the leading behaviour go to zero in the limit $T\rightarrow 0$. This
is where the difference between the two peaks we looked at becomes
apparent. It is easy to see that this is true
only for values of $q$ such that $c^2|k|L_t/\Delta L_x \rightarrow 0$
as $T \rightarrow 0$ and this explains why the scaling form (\ref{scale1})
does not work. Now consider the peak at $q=Q$: The zeroth order scaling result
has most of its weight in the region $|k| \le \sqrt{\Delta}/c\sqrt{L_t}$. For $|k| \sim 
\sqrt{\Delta}/c\sqrt{L_t}$, the corrections to this leading result do indeed go to zero and
this establishes the scaling form (\ref{scale0}). An interesting feature of
this result is that the scaling function $\Phi$ is {\em completely
determined} by the $x=0$ part $R(0,t)$ of the relaxation function. More precisely, we have
\begin{equation}
\Phi(z) = \frac{\pi}{4}\int _{-\infty}^{\infty}ds e^{izs}{\tilde R}(0,s)~.
\label{dirscalfn}
\end{equation}
A useful check on all of our calculations is thus to
compare the scaling function obtained in Fig~\ref{fig10} and Fig~\ref{fig12} with 
(\ref{dirscalfn})
evaluated numerically (it is possible to do this to a high accuracy; details
may be found in Appendix~\ref{A}). The results of such a comparison are shown in Fig~\ref{fig15} 
and the agreement is seen to be quite good. While the
numerical results for $R(0,t)$ show a clear deviation from the simple
exponential decay predicted by the `mean-field' theory referred to earlier, we
do find that the corresponding simple Lorentzian form for the Fourier transform: $\Phi(z) = \pi 
\alpha /2(\alpha^2 + z^2)$ (with $\alpha \approx 0.71$) provides an excellent approximation to 
the line-shape (the `mean-field' theory,
however, gives a value of $4/3\sqrt{\pi} \approx 0.7523$ for
$\alpha$---see Appendix~\ref{mf}).

We thus have results for the thermally broadened quasiparticle peak in 
$S(q,\omega)$; the accuracy of these in the asymptotic regime ($T \ll \Delta$)
is limited only by the computer time spent in numerically evaluating
the relaxation function and doing the Fourier transform. These results,
especially the scaling properties in the vicinity of the
peak corresponding to $q=Q$, should be of relevance to neutron
scattering experiments on gapped one-dimensional Heisenberg antiferromagnets performed at 
temperatures $T \ll \Delta$ and it is hoped that this study
provides a useful paradigm for organizing the experimental results. 

\subsection{Low temperature spin diffusion probed by $1/T_1$}
\label{hdep}
In this section, we shall present a detailed comparison of our results~\cite{sskd} for the
field ($H$) and temperature ($T$) dependence of the NMR relaxation
rate $1/T_1$ (in the regime $T$,~$H \ll \Delta$) with the experimental
data of Ref~\cite{taki} on the NMR relaxation rate in the compound
AgVP$_2$S$_6$ which is thought to be a $S=1$ one-dimensional antiferromagnet
with a large gap $\Delta \approx 300$ K and single-ion anisotropy energy of
about $4.5$ K~\cite{taki}. We will ignore this anisotropy for the
most part in our theoretical analysis (although we are forced to
phenomenologically introduce spin-dissipation into our theorerical
results in order to fit the data of Ref~\cite{taki} at low temperatures, we do not
have any theory that gives the detailed temperature dependence of this spin dissipation
rate starting from the anisotropic coupling term in the Hamiltonian).

For completeness, let us begin with a detailed review of the
calculations leading up to our expression for $1/T_1$.
The NMR relaxation rate is given in general by an expression of the
form
\begin{equation}
\frac{1}{T_1} =\sum_{\alpha=x,y}\sum_{\beta,\gamma=x,y,z} \int \frac{dq}{2\pi} A_{\alpha 
\beta}(q)A_{\alpha \gamma}(-q)S_{\beta\gamma}(q,\omega_N)~;
\label{gennmrrate}
\end{equation}
where $S_{\beta\gamma}(q,\omega)$ is the Fourier transform of the spin-spin
correlation function (the subscripts refer to the {\em O}(3) indices of the
spin operators), $\omega_N=\gamma_N H$ is the nuclear Larmor frequency ($\gamma_N$ is the nuclear 
gyromagnetic ratio), the field $H$ points in
the $z$ direction and $A_{\alpha\beta}$ are the hyperfine coupling constants.
The $q$ integral in (\ref{gennmrrate}) is dominated by values of $q$ near
$0$~\cite{sagiaffl} and we can thus work out the field and temperature
dependence of $1/T_1$ knowing the $T>0$ correlators of the conserved
magnetization density of the {\em O}(3) NL$\sigma$M field theory. This is
what we turn to next.

We define the correlation functions
\begin{eqnarray}
C_{u,zz}(x,t) & = & \langle {\hat{ L}}_{z}(x,t){\hat {L}}_{z}(0,0)\rangle - 
\langle \hat{L} \rangle^2 \nonumber \\
C_{u,-+}(x,t) & = & \langle {\hat {L}}_{-}(x,t){\hat {L}}_{+}(0,0)\rangle~;
\label{defczzpm}
\end{eqnarray}
here the angular brackets denote averaging over the usual equilibrium
density matrix, ${\hat {L}}_{z}(x,t)$ is the Heisenberg representation operator
corresponding to the $z$ component of the magnetization density, and ${\hat {L}}_{\pm}$ are
operators corresponding to the circularly polarized components
of the magnetization density defined as ${\hat{ L}}_{\pm} \equiv {\hat {L}}_x \pm
i{\hat {L}}_y$. 
As argued in Refs~\cite{sskd,ssyoung}, these correlation functions in the
asymptotic regime may be evaluated by writing down a double-time path
integral representation for them and evaluating it semiclassically.

This
leads to the following prescription~\cite{sskd} for $C_{u,zz}(x,t)$
: At time
$t=0$ we begin with an initial state populated with three species (corresponding to the three 
values of spin projection $m$) of particles each uniformly distributed in space with densities 
given respectively by
\begin{displaymath}
\rho_m = \int \frac{dp}{2\pi} e^{-(\Delta -mH + c^2 p^2/2 \Delta)/T}
= \sqrt{ \frac{T \Delta}{2\pi c^2} }  e^{-(\Delta-mH) /T},
\end{displaymath}
and with velocities distributed according to the classical Maxwell-Boltzmann
distribution function
\begin{displaymath}
{\cal P}(v) = \sqrt{  \frac {\Delta}{2\pi c^2 T}} e^{-\Delta v^2/2c^2T}~.
\end{displaymath}
Each particle in the initial state is assigned one of the three values of
$m$ with probability $f_m = e^{mH/T}/(1+2\cosh (H/T))$~. The operator ${\hat L}_z(0)$ merely 
keeps track of the local value of
the $z$ component of the spin. Acting on the initial state, it measures the 
$z$ component of the magnetization density in the initial state at position $x=0$.
These particles then follow their classical
trajectories forward in time. At every collision, the particles retain their
spin labels. In addition, the state picks up a factor
of $-1$ from the ${\cal S}$-matrix at each collision. At time $t$, the operator ${\hat L}_z(x)$ 
measures the
value of the $z$ component of magnetization density at position $x$. The state is then propogated 
backward in time
to $t=0$ and its overlap with the initial state calculated. $C_{u,zz}(x,t)$ is then
given by the average of this overlap over the ensemble specified earlier. As
all collisions have a time-reversed counterpart, the phase of the scattering
matrix does not matter here and  the overlap we are interested in equals the
two-point correlation function of the classical observable
\begin{equation}
\varrho_z(x,t)=\sum_{k}m_k\delta(x-x_k(t))~;
\label{varrho}
\end{equation}
where we are labelling particles consecutively from left to right with an
index $k$, $x_k(t)$ denotes the position of the $k^{th}$ particle at time $t$, and
$m_k$ is the $z$ component of the spin of the $k^{th}$ particle.
This correlation function is calculated using the ensemble of initial conditions
outlined above. The dynamics governing the time evolution of the $x_k$ is 
just that of particles moving ballistically except for elastic collisions in
which they retain their spin values.

Thus we can write
\begin{equation}
C_{u,zz}(x,t)=\sum_{k,l} \langle m_k\delta(x-x_k(t))m_l\delta(x_l(0))\rangle
- \langle \varrho_z \rangle^2 ~,
\label{zzcorr}
\end{equation}
here the angular brackets refer to averaging over the ensemble of spin labels $m_k$, initial
velocities $v_k(0)$, and initial positions $x_k(0)$ specified earlier. Now as
the spin-projections $m_k$ are not correlated with the initial positions
or velocities, the averages factorize. The correlators of the $m_k$ are
easily evaluated as:
\begin{equation}
\langle m_km_l\rangle = A_1 + A_2\delta_{kl}~,
\label{mmcorr}
\end{equation}
where $A_1=(f_1-f_{-1})^2$ and $A_2=f_1+f_{-1}-(f_1-f_{-1})^2$ are simple,
dimensionless, known functions of $H/T$ only. Using (\ref{mmcorr}) we have
\begin{eqnarray}
&& C_{u,zz} (x , t ) = A_1 \left( \left\langle \rho(x,t) \rho(0,0)
\right\rangle -
\rho^2\right)
\nonumber \\ &&~~~~~~~~+ A_2 \sum_k \left\langle \delta(x - x_k (t))
\delta(x_k(0))
\right\rangle
\label{ab}
\end{eqnarray}
where $\rho(x,t) = \sum_k \delta( x- x_k (t))$ is the spacetime dependent
total density, all averages are now with respect to initial positions and
velocities,
and $\rho \equiv \langle \rho(x,t) \rangle = \sum_m\rho_m$.
The two-point correlators of $\rho (x,t)$ are also easy to evaluate: if
the spin labels are neglected, the collisions have no effect and correlators
of the total denstiy can be obtained by considering an ideal gas
of point particles. The second correlator in (\ref{ab}), multiplying $A_2$,
is more difficult: it involves the self two-point correlation of a given
particle $k$, which follows a complicated trajectory.
Fortunately, precisely this correlator was considered
three decades ago by Jepsen~\cite{jepson} and a little later by
others~\cite{lebowitz}; they showed that, at sufficiently long times, this
correlator has a Brownian motion form. In Appendix~\ref{jep}, we give a self-
contained
summary of Jepsen's calculation. Here we
just write down the final results~\cite{sskd} for the correlation function:
\begin{equation}
C_{u,zz} (x,t) = \rho^2 \left[
 A_1 F_1 \left( \frac{|x|}{L_x}
, \frac{|t|}{L_t} \right) + A_2 F_2 \left( \frac{|x|}{L_x} , \frac{|t|}{L_t} \right)
 \right]
\label{czz1}
\end{equation}
where $\rho^2 F_1$ is the connected density correlator of a classical ideal
gas in $d=1$,
\begin{equation}
F_1 ( \tilde{x}, \tilde{t} ) = e^{-\tilde{x}^2/\tilde{t}^2}/\tilde{t} \sqrt{\pi}~,
\label{czz2}
\end{equation}
and $\rho^2 F_2$ is the correlator of a given labeled
particle~\cite{jepson,lebowitz},
\begin{eqnarray}
&& F_2 ( \tilde{x}, \tilde {t} ) =
\Biggl[ \left(2 G_1 (u) G_1 (-u) + F_1( \tilde{x},
\tilde{t}) \right)  \nonumber \\
&&~~~~~~~~~~~~~~~~\times I_0 \left( 2 \tilde{t} \sqrt{G_2 (u) G_2 (-u)} \right)
\nonumber \\
&& ~~~~~~~~~~~~~~~~~+ \frac{G_1^2 (u) G_2 (-u) + G_1^2(-u) G_2 (u)}{\sqrt{G_2(u) G_2 (-u)}}
\nonumber \\
&&~~~~~~~~~~~~\times I_1 \left( 2 \tilde{t} \sqrt{G_2 (u) G_2 (-u)} \right) \Biggr]
e^{-(G_2(u) + G_2 (-u)) \tilde{t}}
\label{czz3}
\end{eqnarray}
with $u \equiv \tilde{x}/\tilde{t}$,
$G_1 (u) = \mbox{erfc} (u)/2$, and
$G_2 (u) = e^{-u^2}/(2 \sqrt{\pi}) - u G_1 (u)$. For $|\bar{t}| \ll |\bar{x}| \ll 1$, the 
function $F_2$ has the ballistic form
$F_2 (\bar{x}, \bar{t} ) \approx F_1 (\bar{x} , \bar{t})$, while for
$|\bar{t}| \gg 1, |x|$ it crosses over to the {\em diffusive \/} form
\begin{equation}
F_2 (\bar{x} , \bar{t}) \approx \frac{e^{-\sqrt{\pi}\bar{x}^2/2 \bar{t}}}{(4 \pi
\bar{t}^2)^{1/4}}~~~~\mbox{for large $\bar{t}$~.}
\label{larget}
\end{equation}
In the original dimensionful units, (\ref{larget})
implies a spin diffusion constant, $D_s$, given exactly by
\begin{equation}
D_s = \frac{c^2 e^{\Delta/T}}{ \Delta (1 + 2 \cosh(H/T))} ~.
\label{diffres}
\end{equation}
This result has been obtained by the solution of a classical
model which possesses an infinite
number of local conservation laws: in Appendix~\ref{integrable}, 
we explicitly show how the existence of these local conservation laws
is not incompatible with diffusive spin dynamics.
It must be noted that the result (\ref{diffres}) does not imply that we have 
rigorously established
that the ultimate long-time corelations of the quantum model
are also diffusive: the reasons for this and related comments were made earlier
in Section~\ref{intro} below (\ref{dslowt}).

Let us now summarize the calculation of the correlator of the transverse
components of the magnetization density.
The semiclassical prescription for evaluating $C_{u,-+}(x,t)$ is again
quite straightforward: We begin with an initial state chosen from the
same ensemble as before. ${\hat L}_{+}(0)$ acting on the initial state
gives zero unless there is a particle at $x=0$ with spin label $m =0$,~$-1$,
in which case it raises the $m$ value of that particle by $1$ and multiplies
the state by a factor of $\sqrt {2}$ (coming from the usual properties of
raising operators for the spin-one representation of the angular momentum
algebra). The resulting state is then
propogated forward in time with all the particles moving along their classical
trajectories as before. At time $t$, the operator ${\hat L}_-(x)$
acting on this state gives zero unless there is a particle at $x$ with spin
label $m =0$,~$1$, in which case it lowers the spin value of that particle by $1$ and again 
multiplies the state by a factor of $\sqrt{2}$. This state
is then propogated backward in time and its overlap with the initial state
calculated. $C_{u,-+}(x,t)$ is given by this overlap averaged over the
ensemble of initial conditions. Here, as before, the phase factor of $-1$ coming from
each collision does not matter as each collision has a time reversed
counterpart. Also, it is easy to see that in this case the overlap with
the initial state is zero {\em unless} ${\hat L}_{-}(x)$ {\em lowers the spin of
precisely the particle whose spin was raised by} ${\hat L}_{+}(0)$. Lastly, we
see that there is an overall factor of $e^{+iHt}$ coming from the
unitary time evolution as the total spin of the state during its
evolution forward in time is greater than the total spin during its
evolution backward in time by precisely one. Similar considerations
apply to $C_{u,+-}$. 
Putting all
of this together we see that
\begin{equation}
C_{u,\mp\pm}(x,t) = 2 \rho^2 e^{\pm i H t} A_{\mp}
F_2 \left( \frac{|x|}{L_x}
, \frac{|t|}{L_t} \right)
\label{ctrans}
\end{equation}
where $A_{\mp} \equiv f_0 + f_{\mp 1}$.

Now, we may express the NMR relaxation rate in terms of the correlation
functions of the conserved magnetization density as
\begin{equation}
\frac{1}{T_1}= \sum_{\alpha=x,y}\sum_{\beta,\gamma=x,y,z} A_{\alpha \beta}A_{\alpha 
\gamma}~S_{\beta\gamma}(\omega_N),
\label{nmrrateuniv}
\end{equation}
where the local dynamic structure factor $S_{\beta\gamma}(\omega_N)$ is defined
as
\begin{equation}
S_{\beta\gamma}(\omega_N) = \int dt e^{i\omega_N t}C_{u,\beta\gamma}(0,t)~;
\label{deflocalsw}
\end{equation}
note that we have neglected the $q$ dependence of the hyperfine couplings and
ignored the contribution of the antiferromagnetic spin fluctuations to the
integral over $q$ in (\ref{gennmrrate}). At this point we have to address
an important subtlety that arises in calculating the local dynamic structure
factor from the autocorrelation function. We are treating the spin dynamics semiclassically to
arrive at our expressions for the correlation functions. This gives rise to a characteristic 
$1/t$ divergence at
short times in the corresponding autocorrelation functions. This is
basically a signature of {\em classical} ballistic spin transport; at these
short time scales collisions play no role. As a result, the
integral as written is logarithmically divergent at short times. Our
semiclassical expressions for the correlation functions do not make
sense for very short times. This is natural as our whole approach has
been geared towards calculating these correlations at time scales much
larger than $1/T$ and length scales much larger than the thermal de broglie
wavelength; our method fails when both these conditions are simultaneously
violated~\cite{ssyoung}. The semiclassical expressions for $C_u (0,t)$ are
thus only valid for $t > \epsilon_t$ where $\epsilon_t$ is a short time
cutoff $\sim 1/T$. Introducing this short time cutoff will give a well-
defined result for $S_{\alpha\beta}(\omega_N)$ at the price of introducing an arbitrary
scale $\epsilon_t \sim 1/T$; this does not seem very promising as our
results for $S_{zz}(\omega_N)$ ($S_{\pm \mp}(\omega_N)$) will depend sensitively upon 
$\epsilon_t$ except for very small fields such that we are in the collision
dominated diffusive regime:
$\gamma_N H\ll  1/L_t$~($H \ll 1/L_t$). Note that the range of $H$ for which the
results are insensitive to the cutoff differs for the transverse components of
the local dynamic structure factor because of the overall oscillatory factor
of $e^{\pm iHt}$ in the corresponding autocorrelation functions (this factor
always dominates as
$\gamma_N \ll 1$). 
However, we can still use our approach to
compute the $S_{\alpha\beta}(\omega_N)$. The point is that, at very short times, the collisions
between the thermally excited particles do not matter, and the spin
dynamics is ballistic. This means that $S_{\alpha\beta}(\omega_N)$, for high frequencies 
$\omega_N$ (such that $\omega_N$ is much larger than the mean collision rate $\sim 1/L_t$), may 
be calculated exactly by doing a full quantum calculation for
a gas of non-interacting spin-one particles~\cite{sagiaffl}. Now, we can
expand our semiclassical result (obtained by using a cutoff $\epsilon_t$)
for $\omega_N  \gg 1/L_t$ and {\em match the leading term in this large $H$
expansion} with the {\em small $H$ limit} (for $S_{zz}$~($S_{\pm \mp}$) this would
be the regime $\gamma_NH \ll T$~($H \ll T$)) of the quantum calculation
of Ref~\cite{sagiaffl}. This, then, will uniquely fix $\epsilon_t$ and
give us results for the $S_{\alpha\beta}(\omega_N)$ that will work reasonably well even for 
$H\sim T $ (though
strictly speaking they are valid only in the range $\gamma_N H \ll T$~($H \ll T$) for
$S_{zz}$~($S_{\pm \mp}$)). 

To see explicitly how
this procedure works, consider $S_{zz}(\omega_N)$. It is
quite easy to see that the $\omega_N \gg 1/L_t$ limit
of the semiclassical $S_{zz}(\omega_N)$ is:
\begin{displaymath}
\frac{\Delta e^{-\Delta/T}}{\pi c^2}(e^{H/T}+e^{-H/T})\ln\left(\frac{e^{-
\gamma}}{\epsilon_t\omega_N}\right)~,
\end{displaymath}
where $\gamma\approx 0.577216$ is Euler's constant. 
The $\omega_N \ll T$ limit of the full quantum calculation reads~\cite{sagiaffl}:
\begin{displaymath}
\frac{\Delta e^{-\Delta/T}}{\pi c^2}(e^{H/T}+e^{-H/T})\ln\left(\frac{4Te^{-
\gamma}}{\omega_N}\right)~.
\end{displaymath}
Thus we can set $\epsilon_t = 1/4T$ to match the two logarithms. It is easy to check that the 
same choice works for the transverse correlators. It is now quite straightforward to do the $t$ 
integrals and obtain the following results~\cite{sskd} for
the local dynamic structure factor:
\begin{eqnarray}
S_{zz} ( \omega_N ) &&=
\frac{\rho}{c} \sqrt{\frac{2  \Delta}{\pi T}} \left[
A_1 \left\{ \ln (T L_t) + \Phi_1 ( \sqrt{\pi} |\omega_N| L_t) \right\} \right. \nonumber \\
&&~~~~~~~~~\left.+ A_2 \left\{ \ln (T L_t) + \Phi_2 ( \sqrt{\pi} |\omega_N| L_t ) \right\} 
\right]~,
\nonumber
\\ S_{\mp \pm} ( \omega_N ) &&=
\frac{2\rho A_{\mp}}{c} \sqrt{\frac{2  \Delta}{\pi T}} \left\{ \ln (T L_t) + \Phi_2 (
\sqrt{\pi} |\omega_N\pm  H| L_t) \right\}~.
\label{summaryofS}
\end{eqnarray}
The $\ln (T L_t)$ terms logarithmically violate the purely classical,
reduced scaling forms~\cite{CSY},
and were fixed using the matching procedure just outlined.
The scaling functions $\Phi_{1,2} (\Omega )$ were determined in Ref~\cite{sskd} to be
\begin{eqnarray}
\Phi_1 ( \Omega ) &&= \ln \left( \frac{4 \sqrt{\pi} e^{-\gamma}}{\Omega} \right)~,
\nonumber \\
\Phi_2 (\Omega ) &&= \Phi_1 ( \Omega ) + \frac{\pi [ ( \sqrt{4 + \Omega^2} + 2)^{1/2} -
\sqrt{\Omega} ]^2}{ 4 \sqrt{\Omega} ( \sqrt{4 + \Omega^2} + 2)^{1/2} } \nonumber \\
&&~~~~ - \ln \frac{( 1 + \Omega^2
/\Psi^2(\Omega))^{1/2}  (1 + \Psi(\Omega)) }{2\Omega}~;
\label{psis}
\end{eqnarray}
where $\gamma = 0.57721\ldots$ is Euler's constant, and $\Psi ( \Omega ) =
( \Omega \sqrt{1 + \Omega^2 / 4} - \Omega^2 / 2 )^{1/2}$. Note that the above expression for 
$\Phi_2(\Omega)$ clearly shows
the expected crossover from the large frequency ballistic behavior $\Phi_2 ( \Omega \rightarrow
\infty )  = \ln ( 1/\Omega)$, to the small frequency diffusive form
$\Phi_2 ( \Omega \rightarrow 0 ) = \pi /(2 \sqrt{ \Omega})$.

Let us now use all of this to make contact with the experimental results of
Ref~\cite{taki}. For this particular experimental setup, the expression for
$1/T_1$  simplifies and to a very good approximation we can write~\cite{taki}
\begin{equation}
\frac{1}{T_1} = \Gamma_1\times  S_{xx}(\omega_N)~;
\label{ratexx}
\end{equation}
here the relevant hyperfine coupling constant is known~\cite{taki} to
have the value $\Gamma_1 \approx (7.5 \times 10^5)~^{\circ} K~sec^{-1}$ (note that we have
used units such that $\hbar = k_B = 1$ in our computation of the correlation
functions and thus time is being measured in inverse Kelvins). To begin with,
we straightforwardly attempt to fit the field dependent $1/T_1$ with our
results. We use the values $\Delta = 320$ K and $c=3.32\Delta $ (we are working
in units where the lattice constant $a$ is set to one) extracted from
the susceptibility data~\cite{takib}. In actual fact, we introduce an additional, {\em field-
independent} background rate  $R_b$ that we add on to our theoretical result for $1/T_1$. This
serves as our fitting parameter; we choose it at each temperature to achieve
the best agreement with the results of Ref~\cite{taki}. We show the resulting
fits for $T=320$, $220$, and $160$ K in Fig~\ref{fig16}. We
see that the theoretical curves account for the field dependence of $1/T_1$
{\em extremely} well in this temperature range (of course the agreement for
$T=320$ K should not be taken too seriously as our theory is valid only for
temperatures smaller than the gap). In particular, the data
seems to clearly exhibit the theoretically predicted $1/\sqrt{H}$ divergence at low fields which 
is
a characteristic of diffusive spin dynamics. In Fig~\ref{fig17} and
Fig~\ref{fig18}, we compare
the theoretical predictions with the experimental data at $T =120$, $100$, $90$, $80$, $70$, and 
$60$ K. At these lower temperatures this
divergence seems to get cut off below some threshold field and the quality
of the fit deteriorates rapidly. This indicates the presence of some
spin-dissipation mechanism which becomes significant at these lower temperatures
and rounds off the diffusive $1/\sqrt{\omega}$ divergence in the local dynamic
structure factor. Both inter-chain coupling and single-ion anisotropy of
the intra-chain coupling are expected to contribute to the spin dissipation
rate.
However, we do not have any real theory that can work out the effects of
these terms in the Hamiltonian on the field and temperature dependence of $1/T_1$. 

We can only attempt to phenomenologically introduce some spin
dissipation in our theoretical results for the spin correlators. Following
\cite{taki}, we do this by simply introducing an exponential cutoff to the
long-time tail of the autocorrelation function; thus we write $ C^{\prime}_{u,xx}
(0,t)=e^{-\gamma  t}C_{u,xx}(0,t)$. It is straightforward, though somewhat tedious
to work out the corresponding local dynamic structure factor by doing the
Fourier transform and we will spare the reader the details. This now gives us a phenomenological 
result for $1/T_1$
with an additional tunable parameter $\gamma $. We choose this spin-dissipation
rate at each temperature to achieve the best fit with the data. The resulting
curves are shown in Fig~\ref{fig19} and Fig~\ref{fig20} for a few representative temperature 
values.
We see that it is possible to fit the data moderately well;
discrepencies are however clearly visible and it is not clear how much significance to attach to 
the sharp increase in $\gamma $ as the temperature is
lowered. The quality of our fit seems at first sight to be much worse than
the corresponding fit to a purely classical diffusive form employed in Ref~\cite{taki}.
However, it is important to note that the phenomenological model of Ref~\cite{taki} used the 
diffusion
constant as an additional fitting parameter; we do not have any such
freedom. Moreover, both the diffusion constant and the constant background rate
extracted from the fit in Ref~\cite{taki} take on unphysical values below about $100$ 
K~\cite{taki}. This is because, at these lower temperatures, we are
in the {\em ballistic} regime of spin-transport for a significant portion of
the $H$ axis and the contribution from the `free-boson logarithms' cannot
be neglected. As the crossover to the ballistic regime is already incorporated
in our form, the present results for the background rate do not suffer from
any such obvious problems (the diffusion constant of course is just given
by (\ref{diffres}) in our approach). In Fig~\ref{fig21} and Fig~\ref{fig22}, we plot the 
corresponding values of the spin-dissipation rate $\gamma$ and the background rate $R_b$
as a function of temperature. The spin dissipation rate is seen to
increase rapidly as the temperature is decreased. On the other hand, we see that
the temperature dependence of $R_b$ may be fit approximately by an activated
form with activation gap close to $3\Delta/2$.

\section{High $T$ region ($T\gg \Delta$) of the continuum $\sigma$ model}
\label{hight} 
We consider here the possibility that it may be possible to find
gapped spin chains which satisfy $\Delta \ll J$, where $J$ is a typical 
exchange constant. In this case, it becomes possible to access a 
higher temperature regime where a continuum field theory description
is possible in the regime $\Delta \ll T \ll  J$. In particular,
we expect that the continuum $\sigma$-model to apply in such a regime~\cite{joli}.
It is our hope that such a universal high $T$ regime can be experimentally
accessed in $S=2$ spin chains~\cite{meisel}. Moreover, the study of such
a high $T$ regime is of importance as matching its results with the
$T \ll \Delta $ theory can, in principle, help us estimate the values
of $T$ to which the low $T$ results can be applied.

An important property of this regime~\cite{joli} is that equal-time
two-point correlator
of $n_{\alpha}$, $C(x,0)$ (Eqn (\ref{defC})) decays at large $x$ with a
correlation length
\begin{equation}
\xi \sim \frac{c}{T} \ln (T/\Delta).
\label{hight1}
\end{equation}
We will shortly determine the exact values of the prefactor and
the argument of the logarithm in (\ref{hight1}).
At distances of order or shorter
than this correlation length we may crudely expect that a weak-coupling, 
spin-wave picture will hold, and 
excitations will have energy of order or smaller than $c \xi^{-1}$, which is
logarithmically smaller than the thermal energy $T$; in other words
\begin{equation}
\frac{c \xi^{-1}}{T} \sim \frac{1}{\ln(T/\Delta)} < 1
\label{1drot25}
\end{equation}
So the occupation
number of these spin-wave modes will then be
\begin{equation}
\frac{1}{e^{c\xi^{-1}/T} - 1} \approx \frac{T}{c \xi^{-1}} > 1
\label{1drot26}
\end{equation}
The last occupation number is precisely that appearing in a classical description
of thermally excited spin waves, which suggests that a classical {\em wave\/} description
should yield an appropriate picture of the dynamics of this high $T$ region.
However, notice that classical thermal effects are only logarithmically
preferred, and any predictions of a classical dynamical theory will only
be correct to leading logarithms.

We begin our analysis by  first focussing on the {\em equal-time} correlations in this
region. We shall use a method originally introduced by Luscher~\cite{luscher}.
The main idea of Luscher
is to develop an effective action for only the {\em zero Matsubara frequency} ($\omega_n = 0$)
components of $n_{\alpha}$ after integrating out all the $\omega_n \neq 0$
modes (the $\omega_n =0$ modes are related to the equal-time correlations
via the fluctuations-dissipation theorem and the Kramers-Kronig relations~\cite{hh}).
This is expected to yield the following partition function for
a $\tau$-independent field $n_{\alpha} (x)$:
\begin{equation}
Z =  \int {\cal D} n_{\alpha} (x) \delta( n_{\alpha}^2 - 1)  \exp\left(
- \frac{(N-1)\xi}{4} \int d x 
\left( \frac{d n_{\alpha} (x) }{d x} \right)^2 
\right)
\label{1drot28}
\end{equation}
We have now generalized to a field $n_{\alpha}$ with $N$ components, and
will quote many of our results for general $N$; the physical case if of course
$N=3$. The coupling constant in (\ref{1drot28}) is written in a form such
that $\xi$ is the exact correlation length: this follows from the 
easily computable exact correlations of $Z$ by interpreting it as the
quantum mechanics of a single quantum rotor.
The value of $\xi$ can be computed in a perturbation theory in $g$
on the quantum model (\ref{sigmaaction}): the $\omega_n \neq 0$ modes
can be integrated out using a now standard approach~\cite{polyakov}
\begin{equation}
\frac{(N-1)\xi T}{2} = \frac{c}{g} - c^2 (N-2) \int \frac{d k }{2 \pi}
T \sum_{\omega_n \neq 0} \frac{1}{c^2 k^2 + \omega_n^2} 
\label{1drot29}
\end{equation}
The integral on the right-hand-side is not ultraviolet convergent.
We evaluate it using the renormalization procedure discussed
by Brezin and Zinn-Justin~\cite{bzj}.
We introduce a momentum scale $\mu$ at which coupling constants
are defined, and generalize (\ref{sigmaaction}) to a model in
$d$ spatial dimensions. We now
define the renormalized dimensionless coupling
\begin{equation}
g_R ( \mu) = \mu^{\epsilon} Z_1 g,
\label{hight2}
\end{equation}
where $\epsilon \equiv 1 - d$, and the renormalization constant $Z_1$ is
determined in dimensional regularization to be~\cite{bzj}:
\begin{equation}
Z_1 = 1 - \frac{(N-2)}{2 \pi} \frac{g_R (\mu)}{\epsilon} + \ldots
\label{1drot32}
\end{equation}
We now need to 
express (\ref{1drot29}) in terms of $g_R$, and evaluate the 
integral on the right hand side in $d=1-\epsilon$ dimensions.
Let us display a few steps of the latter evaluation:
\begin{eqnarray}
T \sum_{\omega_n \neq 0} \int \frac{d^{1-\epsilon} k}{(2 \pi)^{1-\epsilon}}
\frac{1}{c^2 k^2 + \omega_n^2}
=&& \int \frac{d^{1-\epsilon} k}{(2 \pi)^{1-\epsilon}}
\left[T \sum_{\omega_n \neq 0} \frac{1}{c^2 k^2 + \omega_n^2}
- \int \frac{d \omega}{2 \pi} \frac{1}{c^2 k^2 + \omega^2 + T^2} \right] \nonumber\\
&&~~~~~~~~~~~~~~~~+ 
c^{1-\epsilon} \int \frac{d^{2-\epsilon} p}{(2 \pi)^{1-\epsilon}}
\frac{1}{c^2 p^2 + T^2} \nonumber \\
= && \frac{1}{c}
\left(\frac{T}{c} \right)^{-\epsilon} \left\{
\int \frac{d^{1-\epsilon} k}{(2\pi)^{1-\epsilon}} \left[
\frac{1}{2k} \coth \frac{k}{2} - \frac{1}{k^2} - \frac{1}{2\sqrt{k^2 + 1}} \right] 
\right. \nonumber \\
&&~~~~~~~~~~~~~~~~~~~~~\left. +  \frac{\Gamma (\epsilon/2)}{(4 \pi)^{1-\epsilon/2}}
\right\}
\label{hight3}
\end{eqnarray}
We are only interested in the poles in $\epsilon$ and the accompanying
constants, and to this accuracy
the first integral on the right hand side can be evaluated directly at $\epsilon=0$,
while the $\Gamma$ function yields a pole.
Now combining (\ref{hight2}), (\ref{1drot32}) and (\ref{hight3}) into
(\ref{1drot29}) we find that the poles in $\epsilon$ cancel (as they must), and
\begin{equation}
\frac{(N-1)\xi T}{2c} = \frac{1}{g_R (\mu)} - \frac{(N-2)}{2 \pi} 
\ln(c \mu/T\sqrt{{\cal G}}) 
\label{1drot34}
\end{equation}
where the constant ${\cal G}$ is
\begin{equation}
{\cal G} = 4 \pi e^{-\gamma} = 7.055507955\ldots.
\label{rotor20b}
\end{equation}
with $\gamma$ Euler's constant.
Now we use the conventional relationship between $\mu$ and the renormalization
group invariant $\Lambda_{\overline{MS}}$~\cite{bzj,luscher}
\begin{equation}
\Lambda_{\overline{MS}} = \mu \sqrt{{\cal G}}
\left( \frac{(N-2)}{2 \pi } g_R (\mu) \right)^{-1/(N-2)}
\exp\left( - \frac{2\pi }{ (N-2) g_R (\mu) } \right)
\label{1drot35}
\end{equation}
to eliminate the scale $\mu$ from (\ref{1drot34}). As expected, the coupling
$g_R (\mu)$ drops out of the resulting expression, and we get
\begin{eqnarray}
&&\xi = 
\frac{c (N-2)}{T \pi (N-1)} \left\{
\ln \left[
\frac{{\cal G} T}{c \Lambda_{\overline{MS}}} \right] \right. \nonumber \\
&&~~~~~~~~~~~~~~~~~~\left.
+ \frac{1}{(N-2)} \ln \ln \frac{T}{c \Lambda_{\overline{MS}}} 
+ {\cal O} \left( \frac{ \ln \ln (T/c \Lambda_{\overline{MS}})}{
\ln (T/c \Lambda_{\overline{MS}})} \right) \right\}.
\label{1drot37}
\end{eqnarray}
Finally, we can express this in terms of the $T=0$ gap $\Delta$ by
using the relationship between $\Lambda_{\overline{MS}}$ and $\Delta$
obtained using the Bethe ansatz solution of the $\sigma$-model~\cite{hasen}
\begin{equation}
\frac{\Delta}{c \Lambda_{\overline{MS}}} = 
\frac{(8/e)^{1/(N-2)}}{\Gamma(1 + 1/(N-2))}.
\label{1drot36}
\end{equation}
The results (\ref{1drot37},\ref{1drot36}) lead to the $N=3$ result
for $\xi$ quoted earlier in (\ref{xihight}).

Having obtained the classical model (\ref{1drot28}) for the equal-time correlations,
and the precise value of the coupling $\xi$ in (\ref{1drot37}), we now turn
to an examination of unequal time correlations in the high $T$ region $T \gg \Delta$.
We employ an approach related to that 
used in the study of the quantum $\sigma$-model in $d=2$ in Ref~\cite{thc}
in a {\em low\/} $T$ region; unlike (\ref{1drot28}), the equal time correlations
in $d=2$ were described by a theory that was not ultraviolet finite, and this
will lead to significant differences in the analysis and physical properties here.
 To obtain classical equations
of motions we clearly need to 
extend the classical Hamiltonian in (\ref{1drot28}) by including a kinetic energy
term, expressed in terms of a canonical conjugate momentum to $n_{\alpha}$. 
The obvious approach is to take the quantum equations of motion, and to simply
treat the variables as $c$-number classical degrees of freedom.
In particular, we treat the rotor-angular momentum $L$ as a classical variable,
and augment the classical Hamiltonian by the kinetic energy of rotational motion. 
The moment of inertia of the rotor
is related to the response of the system to a magnetic field $H$, and we therefore
need to study the behavior of $\chi_u$ in the $T \gg \Delta$ regime.

We will determine $\chi_u$ by strategy similar to that employed above in the
computation of $\xi$: first integrate out the non-zero frequency modes,
and then perform the average over the zero frequency fluctuations.
We choose an $H$ which rotates $n_{\alpha}$ in the 1--2 plane, and define
\begin{equation}
n_{\alpha} (x, \tau) = \sqrt{1 - \vec{\pi}^2 (x,\tau)} n_{\alpha} (x)
+ \sum_{a=1}^{N-1} \pi_a (x, \tau) e_{a \alpha} (x)
\label{hight4}
\end{equation}
where $n_{\alpha} (x)$, $e_{a \alpha} (x)$ are a set
of $N$ mutually orthogonal vectors in spacetime, and $\pi_a (x, \tau)$
represent the finite frequency degrees of freedom which must be
integrated out. We expand the partition
function to quadratic order in $H$, drop all terms proportional
to the spatial gradients of $n_{\alpha} (x)$ or $e_{a \alpha} (x)$
(these can be shown to be logarithmically subdominant to the terms
kept),
and find that the $H$ dependent
terms in the free energy density are
\begin{eqnarray}
&& - \frac{H^2}{2 c g} \left[
(n_1^2  + n_2^2  ) \left( 1 - \sum_a \langle \pi_a^2 \rangle \right)
+ \sum_{ab} \left( e_{a1} e_{b1}  + e_{a2}  e_{b2}  \right)
\langle \pi_a \pi_b \rangle \right. \nonumber \\
&& \left. - \frac{1}{cg}\sum_{abcd} (e_{a1} e_{b2}  - e_{a2}  e_{b1} )
(e_{c1} e_{d2}  - e_{c2}  e_{d1} )
\int dx d\tau \left\langle \pi_a \partial_{\tau} \pi_b (0,0);
\pi_c \partial_{\tau} \pi_d (x,\tau) \right\rangle \right]
\label{1drot38}
\end{eqnarray}
Evaluating the expectation values of the $\pi$ fields,
and using orthonormality of the vectors
$n_{\alpha}$, $e_{a \alpha}$, the expression (\ref{1drot38}) simplifies to
\begin{eqnarray}
&& - \frac{H^2}{2 c g} \left[
(n_1^2 + n_2^2 ) \left( 1 - c (N-2) g T \sum_{\omega_n \neq 0} \int \frac{dk}{2 \pi}
\frac{1}{c^2 k^2 + \omega_n^2} \right) \right. \nonumber \\
&&~~~\left. + 2 c g 
( 1 - n_1^2 - n_2^2) T \sum_{\omega_n \neq 0} \int \frac{dk}{2 \pi}
\frac{c^2 k^2 - \omega_n^2 }{(c^2 k^2 + \omega_n^2)^2} \right]
\label{1drot39}
\end{eqnarray}
Finally to obtain the susceptibility $\chi_u$, we have to evaluate
the expectation value of the zero frequency field $n_{\alpha}$ under
the partition function
(\ref{1drot28}). This simply yields $\langle n_1^2 \rangle = \langle n_2^2 \rangle 
= 1/N$. The first frequency summation is precisely the same as that
evaluated earlier for $\xi$ in (\ref{1drot29}), while the second is
explicitly finite in $d=1$ and can directly evaluated; in this manner we
obtain our final result for $\chi_u$:
\begin{eqnarray}
\chi_u (T) &=& \frac{2}{N} \left[ \frac{(N-1)T \xi }{2 c^2} - \frac{(N-2)}{2 \pi c} \right] 
\nonumber \\
&=& \frac{(N-2)}{N \pi c} \ln \left( \frac{{\cal G} T}{\Lambda_{\overline{MS}} e} \right)
\label{1drot40}
\end{eqnarray}
We have omitted the form of the subleading logarithms, which are the same as
those in (\ref{1drot37}). This result was quoted earlier in (\ref{chihight}).

We have now assembled all the information necessary to describe the effective
classical dynamics in the region $T \gg \Delta$. 
The classical partition function is given by the following phase-space functional
integral, which generalizes (\ref{1drot28}) (and we will now specialize the remainder
of the discussion to the special case $N=3$):
\begin{eqnarray}
&& Z =  \int {\cal D} n_{\alpha} (x) {\cal D} L_{\alpha} (x)
\delta( n_{\alpha}^2 - 1) \delta( L_{\alpha} n_{\alpha}) \exp\left(-\frac{{\cal H}_c}{T} 
\right) \nonumber \\
&& {\cal H}_c = \frac{1}{2} \int d x \left[ T \xi 
\left( \frac{d n_{\alpha} }{d x} \right)^2  
+ \frac{1}{\chi_{u \perp}} L_{\alpha}^2\right]
\label{1drot44}
\end{eqnarray}
where $L_{\alpha}$ is the classical angular momentum density, and $L_{\alpha}$, $n_{\alpha}$
are classical commuting variables. The second term in ${\cal H}_c$ was absent in
(\ref{1drot28}), and represents the kinetic energy of the classical rotors: integrating
out $L_{\alpha}$ we obtain (\ref{1drot28}).
By evaluating linear response to a field under which
\begin{equation}
{\cal H}_c \rightarrow {\cal H}_c -
\int dx H_{\alpha} L_{\alpha}.
\label{1drot44a}
\end{equation}
we find
\begin{equation}
\chi_u = \frac{2}{N}  \chi_{u \perp}
\label{1drot41}
\end{equation}
with $N=3$ (we have given, without proof, the expression for general $N$);
the factor of $2/3$ comes from the contraint $L_{\alpha} \cdot n_{\alpha} = 0$.
Using (\ref{1drot40}), we then have the value of $\chi_{u \perp}$. 

We can finally specify the manner in which time-dependent correlations
have to be computed in this effective classical model.
The classical equations of motion are the Hamilton-Jacobi equations of the
Hamiltonian ${\cal H}_c$, with Poisson brackets which are the continuum
classical limit of the quantum commutation relations :
\begin{eqnarray}
\left\{ L_{\alpha} (x) , L_{\beta} (x') \right\}_{PB} &=& 
\epsilon_{\alpha \beta \gamma} L_{\gamma} (x) \delta(x-x') \nonumber \\
\left\{ L_{\alpha} (x) , n_{\beta} (x') \right\}_{PB} &=& 
\epsilon_{\alpha \beta \gamma} n_{\gamma} (x) \delta(x-x') \nonumber \\
\left\{ n_{\alpha} (x) , n_{\beta} (x') \right\}_{PB} &=& 0.
\label{1drot45}
\end{eqnarray}
The equations of motion are
\begin{eqnarray}
\frac{\partial n_{\alpha}}{\partial t} &=& \left(\frac{1}{\chi_{u \perp}} \right)
\epsilon_{\alpha\beta\gamma} L_{\beta} n_{\gamma} 
\nonumber \\
\frac{\partial L_{\alpha}}{\partial t} &=& (T \xi) \epsilon_{\alpha\beta\gamma}
n_{\beta} \frac{\partial^2 n_{\gamma}}{
\partial x^2}
\label{1drot45a}
\end{eqnarray}
The classical correlation functions are obtained by averaging
these deterministic equations over an ensemble of initial conditions
specified by
(\ref{1drot44}). 
Note also that simple dimensional analysis of the differential equations
(\ref{1drot45a}) shows that disturbances travel with a characteristic velocity
$c(T)$ given by
\begin{equation}
c(T) = ( T \xi (T) / \chi_{u \perp} (T) )^{1/2},
\label{1drot42}
\end{equation}
Notice from (\ref{1drot37}) and (\ref{1drot40}) that to
leading logarithms $c(T) \approx c$, but the second term in the first equation
of (\ref{1drot40}) already shows that exact equality does not hold.

We complete the relationship of the quantum to the classical model,
by noting that there is also an additional wave-function renormalization
of the $n_{\alpha}$ field~\cite{polyakov,bzj} which appears
when the non-zero frequency modes are integrated out.
Our final result for the correlator $C$ in (\ref{defC}) then takes the form
\begin{equation}
C_{\alpha\beta} (x, t) = {\cal A} \widetilde{{\cal G}}
\left[ \ln \left(\frac{T}{\Delta} \right) \right]^{\frac{(N-1)}{(N-2)}}
\left\langle n_{\alpha} (x,t) n_{\beta} (x, t) \right\rangle_c
\label{1drot48}
\end{equation}
The subscript $c$ represents the classical average
specified by (\ref{1drot44}) and (\ref{1drot45a}).
The constant ${\cal A}$ is the $T=0$ quasi-particle residue which 
appeared in (\ref{defcala}).
The constant $\widetilde{{\cal G}}$ is an unknown universal number which cannot
be obtained by the present methods. It could, in principle, be obtained
from the Bethe-ansatz solution. There is no similar
renormalization of the correlator of the magnetization density , $C_u$ in (\ref{defczzpm}),
which is precisely equal to the two-point 
correlator of $L_{\alpha}$ under (\ref{1drot44}) and (\ref{1drot45a}).

It is now possible to perform a simple rescaling
and to show that the classical dynamics problem above
is free of any dimensionless
couplings, and is a unique, parameter-free theory. This will allow us
to completely specify the $T$ dependence of observables upto unknown numerical 
constants. Let us perform the following rescalings on 
(\ref{1drot44}) and (\ref{1drot45a})
\begin{eqnarray}
 x &=&  \overline{x} {\xi} \nonumber \\
 t &=&  \overline{t} \sqrt{\frac{\xi \chi_{u\perp}}{T}} \nonumber \\
 L_{\alpha} &=&  \overline{L}_{\alpha} 
\sqrt{\frac{T\chi_{u \perp}}{\xi}}
\label{1drot49}
\end{eqnarray}
Then the partition function (\ref{1drot44}) is transformed to
\begin{eqnarray}
&& \bar{Z} =  \int {\cal D} n_{\alpha} (\overline{x}) {\cal D} \overline{L}_{\alpha} 
(\overline{x})
\delta( n_{\alpha}^2 - 1) \delta( \overline{L}_{\alpha} \cdot n_{\alpha}) \exp\left(-{\cal 
H}_{\bar{c}} 
\right) \nonumber \\
&& {\cal H}_{\bar{c}} = \frac{1}{2} \int d \overline{x} \left[ 
\left( \frac{d n_{\alpha} }{d \overline{x}} \right)^2  
+  \overline{L}_{\alpha}^2\right]
\label{1drot50}
\end{eqnarray}
while the equations of motion become
\begin{eqnarray}
\frac{\partial n_{\alpha}}{\partial \overline{t}} &=& \epsilon_{\alpha\beta\gamma} 
\overline{L}_{\beta} n_{\gamma} 
\nonumber \\
\frac{\partial \overline{L}_{\alpha}}{\partial \overline{t}} &=&  \epsilon_{\alpha\beta\gamma} 
n_{\beta} \frac{\partial^2 n_{\gamma}}{
\partial \overline{x}^2}.
\label{1drot51}
\end{eqnarray}
Notice that coupling constants and parameters
have been scaled away, and (\ref{1drot50},\ref{1drot51})
constitute a unique problem that must be solved exactly.
The $T$ and $\Delta$ dependencies of all quantities arise
only through the rescalings defined in (\ref{1drot49}) and the results
(\ref{1drot37}) and (\ref{1drot40},\ref{1drot41}) for $\xi$ and
$\chi_{u \perp}$ given earlier. 
Complete description of the correlators now requires exact solution of
(\ref{1drot50},\ref{1drot51}).
The equal-time correlations are of-course known from (\ref{1drot50}):
\begin{eqnarray}
\left\langle
\overline{L}_{\alpha} (\overline{x},0) \overline{L}_{\beta} (0,0) \right\rangle_{\overline{c}} 
&=&
\frac{2}{3} \delta_{\alpha\beta} \delta(\overline{x})\nonumber \\
\left\langle
n_{\alpha} (\overline{x},0) n_{\beta} (0,0) \right\rangle_{\overline{c}}
&=& \frac{1}{3} \delta_{\alpha\beta} e^{-|\overline{x}|}
\label{1drot53}
\end{eqnarray}
Even though the equations of motion constitute an integrable
system with an infinite number of non-local conservation laws~\cite{fadeev,pohl}, 
it is not known how to analytically compute
correlations averaged over the initial conditions of a thermal ensemble,
or whether the correlator
$\langle\overline {L}_{\alpha} (\overline{x},\overline{t}) \overline{L}_{\beta} (0,0) 
\rangle_{\overline{c}}$
has a diffusive form at long times and distances. If diffusion did exist in the 
continuum equations (\ref{1drot53}),
the present analysis does
allows us to completely specify the $T$ dependence of the diffusion
constant; by a simple dimensional analysis of (\ref{1drot49}), we get
\begin{equation}
D_s ={\cal B} \frac{T^{1/2} [\xi (T)]^{3/2}}{[\chi_{u \perp} (T)]^{1/2}} 
\label{1drot54}
\end{equation}
where ${\cal B}$ is an unknown universal number,
and the $T$ dependencies of $\xi$ and $\chi_{u \perp}$ 
are in (\ref{1drot37}) and (\ref{1drot40},\ref{1drot41}).

In this context, it is interesting to note that recent measurements~\cite{kikuchietal} of the
field dependence of $1/T_1$ in the compound (VO)$_2$P$_2$O$_7$ at
temperatures $T \gg \Delta $ seem to
provide clear evidence for spin diffusion. However, the bulk of the
data is at temperatures comparable to the microscopic exchange constants
of the system and it is not clear if the foregoing description based on the
universal high temperature properties of the continuum field theory
is applicable in the temperature regime studied experimentally. 
It is interesting that the experimental results appear to suggest that
$D_s \sim c \xi$, which is consistent with (\ref{1drot54}) if
$\chi_u \sim T \xi/c^2$ (as is the case with our results 
(\ref{xihight}) and (\ref{chihight}) to leading logarithms).

\section{Conclusions}
\label{conc}
The main results of the paper are already summarized in Section~\ref{intro},
and here we will note some unresolved issues and directions for future work.

All experimental realizations of gapped antiferromagnets 
have additional complications which have not been included in the
model systems studied here. Most important among these are
the spin anisotropies away from perfect Heisenberg symmetry
and the inter-chain couplings which make the system
only quasi-one-dimensional.

Consider first the consequence of anisotropy. The three-fold
degenerate quasiparticle spectrum will now be lifted, and three
resulting particles will have have different energy gaps and masses.
Further, these parameters will depend in a complicated way upon
the external field. Nevertheless, we expect that the simple structure
of the ${\cal S}$-matrix in (\ref{smatrix}) will be retained,
as it only depends upon simple dimensional properties of slowly
moving particles with a quadratic dispersion. Correlations of the
particle density can probably be computed along the semiclassical
lines of this paper: one has to deal with a classical gas of
particles of different masses and average densities. The latter
problem is considerably more complex than the equal mass case,
and there is probably no alternative to numerical simulations.
Correlations of the spin operators appear more problematical--
these will invariable change the labels of the particles when
they act, and therefore lead to differences in the labels
in the forward and backward trajectories. Combined with the
complication that the masses of the different particles are
different, and so their trajectories will have different
velocities, we are faced with what appears to be
a very complex problem with quantum and classical effects
intertwined.

Inter-chain couplings will eventually require us
to consider dynamics in two or three dimensions. If temperatures
are low enough that the
inter-chain motion is coherent, then we have to consider the ${\cal S}$-matrix
for scattering in higher dimensions. In this case the low-momentum behavior
is quite different: in fact the $T$-matrix vanishes at low momenta
for $d \geq 2$. We would then expect all scattering to be
dominated by elastic scattering of impurities which would
control the behavior of the spin diffusion constant
and the quasi-particle broadening. On the other hand, systems
with only incoherent hopping between chains will probably
be dominated by the inelastic scattering along the one-dimensional
chains, and display behavior qualitatively similar to that
discussed in this paper.

\acknowledgements
We are grateful to M.~Takigawa for initially stimulating our
interest in the issue of spin transport in gapped antiferromagnets by informing
us about the convincing experimental evidence for diffusion in Ref~\cite{taki},
and for guiding us through the experimental literature.
We thank G.~Aeppli, I.~Affleck, C.~Broholm, G. Chaboussant, V.~Korepin,
S.~Majumdar, B.~McCoy,
B.~Narozhny, B.~Normand, D.~Reich, T.~Senthil and X.~Zotos
for helpful discussions.
This research was supported by the National Science Foundation 
Grant No. DMR-96-23181.

\appendix

\section{Local conservation laws and spin diffusion}
\label{integrable}
The computation of the spin diffusivity in Section~\ref{hdep}
was carried out using the exact solution a simple classical model of point particles
in one dimension. This model is exactly solvable~\cite{jepson} and 
possesses an infinite number of local conservation laws, as we will show
explicitly below. The existence
of spin diffusion then appears to run counter to the conventional wisdom
that the time evolution of a integrable system is not `chaotic' enough
to be compatible with diffusion. In particular, one might expect that
any non-zero spin current produced in the system will not ultimately
decay to zero because the numerous conservation laws prevent it.
In this appendix we will show that this expectation does not hold
for the particular model being studied, and that an
important `particle-hole'-like symmetry 
allows complete decay of any spin current at $H=0$. 
In a finite magnetic field ($H\neq0$), the particle-hole symmetry is absent,
and then the spin current does not decay completely:
this is consistent with the presence of a purely ballistic component, $F_1$, in 
(\ref{czz1}) which contributes only for $H \neq0$ ($A_1 = 0$ at $H=0$), and
the arguments of Zotos {\em et al}~\cite{zotos}.
A closely related particle-hole
symmetry also played an important role in the appearance of a finite
conductivity in our recent quantum transport analysis in two dimensions~\cite{ds}.

The classical model of Sections~\ref{finitet} and~\ref{hdep} consisted of particles
$k = 1\ldots N$ with spins $m_k$ chosen randomly (at $H=0$) from $1,0,-1$. 
At time $t=0$ the particles had uncorrelated random
positions $x_k (0)$, and subsequently they
occupied `trajectories' $X_k (t) \equiv x_k (0) + v_k t$ where $v_k$ are uncorrelated
random velocities chosen from a Boltzmann distribution. The position $x_k (t)$ of particle $k$
was however a rather complicated function of time, and was chosen
from the set of trajectories, $\{X_k (t)\}$, such that for all $t$, $x_k (t) < x_l (t)$ for 
every $k < l$. 

It is useful at this point to note two discrete symmetries of the above
classical statistical problem at $H=0$. The first is the time-reversal symmetry, ${\bf T}$,
under which both spins and velocities change sign:
\begin{equation}
{\bf T}:~~~v_k \rightarrow -v_k~~~,~~~m_k \rightarrow -m_k.
\end{equation}
The second is the `particle-hole' symmetry ${\bf P}$, under which
only the spins reverse direction:
\begin{equation}
{\bf P}:~~~v_k \rightarrow v_k~~~,~~~m_k \rightarrow -m_k.
\label{integ0}
\end{equation}
These symmetries will be crucial in our discussion below.

Let us now explicitly identify the local conserved quantities of this classical dynamics.
All of the velocities $v_k$ are clearly constants of the motion. However, we would
like to work with locally conserved quantities 
which can be written as the spatial integrals over
local observables, and which are invariant under permutation of the particle
labels; so we define
\begin{eqnarray}
V_n &=& \int dx \left[ \sum_{k=1}^{N}
\left( \frac{dx_k (t)}{dt} \right)^n \delta(x - x_k (t))
\right] \nonumber \\
&=& \sum_{k=1}^{N} v_k^n 
\label{integ1}
\end{eqnarray}
with $n=1\ldots N$ (notice $dx_k (t)/ dt \neq v_k = d X_k (t)/ dt$,
but the result holds
after summation over $k$ because the set $\{x_k (t)\}$ differs
from the set $\{ X_k (t) \}$ only by a renumbering). 
All the $V_n$ are constants of the motion. 
Similarly, with
spins $m_k$ we can define
\begin{equation}
M_p = \sum_{k=1}^{N} m_k^p
\label{integ2}
\end{equation}
with $p=1,2$, as additional locally conserved quantities ($M_1$ is the
spatial integral of $\varrho_z (x,t)$ in (\ref{varrho}), and a similar
result holds of $M_2$). 
We can now easily work out the signature of the $V_n$ and $M_p$ under
the discrete symmetries noted earlier, and tabulate the results:
\begin{equation}
\begin{tabular}{c|c|c}
 & ~~${\bf P}~~$ & ~~${\bf T}$~~ \\
\hline
$V_n$, $n$ odd & 1 & -1 \\
\hline
$V_n$, $n$ even~ & 1 & 1 \\
\hline
$M_1$ &  -1 & -1 \\
\hline
$M_2$ & 1 & 1 \\
\hline
\end{tabular}
\label{integ2a}
\end{equation}

The central quantity in spin transport is the total spin current $J(t)$,
which is not a constant of the motion.
It is also given by an integral over a local quantity as
\begin{eqnarray}
J(t) &=& \int dx \left[
\sum_{k=1}^{N} m_k \frac{dx_k (t)}{dt} \delta(x - x_k (t)) \right] \nonumber \\
&=& \sum_{j,k=1}^{N} m_j v_k A_{jk} (t),
\label{integ3}
\end{eqnarray}
where $A_{jk}$ is defined to be equal to 1 if particle $j$ is on trajectory $k$
at time $t$ and 0 otherwise; we will analytically study the function $A_{jk}$ in
Appendix~\ref{jep}, but here we will be satisfied by a numerical simulation.
Again, as in (\ref{integ2a})
it is useful to note the signature of $J$ under the discrete symmetries:
\begin{equation}
\begin{tabular}{c|c|c}
 & ~~${\bf P}~~$ & ~~${\bf T}$~~ \\
\hline
~~J~~& -1 & 1 \\
\hline
\end{tabular}
\label{integ3a}
\end{equation}
As will become clear shortly,
one of the central points of this appendix
is that the signatures in (\ref{integ3a}) differ
from all of those of the conserved quantities in (\ref{integ2a}).
The current $J(t)$ is the sum of $N$ random numbers of each sign, and so is expected
to be of order $\sqrt{N}$ for a typical initial condition chosen
from the ensemble defined above. We show the {\em deterministic}
time evolution of $J(t)$
for one such initial condition for a system of 400 particles in Fig~\ref{fig23}:
notice that it is rather noisy-looking and repeatedly changes its sign.
Also, among the constants of the motion above, we expect $V_n$ with $n$ odd
and $M_1$ to be of order $\sqrt{N}$ (provided $n$ is not too large), 
and $V_n$ with $n$ even and $M_2$ to be of order $N$ for
a typical initial condition; notice that it is only the conserved quantities of order
$\sqrt{N}$ that can distinguish left movers from right movers, or spin up from down.

Let us now create a macroscopic spin current (of order $N$) in this system.
We do this by hitting the system with a magnetic field gradient impulse
at a time $t=t_0$, and subsequently setting the field to zero.
As a result of the impulse, the velocities of the particles with spin up
are assumed to increase by $v_0$, while those of spin down are assumed
to decrease by $v_0$. Formally, this can be written as
\begin{equation}
v_k \rightarrow v_k + m_l v_0 ~~\mbox{where $l$ is unique solution of $A_{lk} (t_0) = 1$}.
\label{integ4}
\end{equation}
Immediately after the impulse, $J(t)$ will have a macroscopic value 
\begin{equation}
J(t_0^+) = \frac{2}{3} N v_0 + {\cal O}(\sqrt{N})
\label{integ5}
\end{equation}
The subsequent deterministic time evolution of $J(t)$ is also shown in Fig~\ref{fig23}:
it decays in a few collision times
to a value of order $\sqrt{N}$ and then appears
to chaotically oscillate in time! The basic point is now easy to see. Because
$m_k$ is as likely to be $+1$ or $-1$, the impulse on any given particle
is equally likely to be $+v_0$ or $-v_0$. Hence the $V_n$, with $n$ odd,
remain of order $\sqrt{N}$ even after the impulse. This is simply a
manifestation of the fact that the signatures of $J$ under
${\bf P}$ and ${\bf T}$ are different from those of the $V_n$.
A non-zero $J$ is therefore not correlated with an induced
value of a conserved quantity which could prevent the decay
of $J$ to non-macroscopic values.

\section{Numerical computation of the Fourier transform of the correlation
function $C$}
\label{A}
In this appendix we outline the numerical method employed in calculating
$S(q,\omega)$ starting from the numerically determined semiclassical $C(x,t)$
and the procedure used to directly determine the scaling function $\Phi(z)$ 
(see Eqn~(\ref{dirscalfn}))

As the numerical determination of ${\tilde R}({\tilde x},{\tilde t})$ is the most time consuming
part of the entire procedure, we calculated ${\tilde R}$ only at a 
predetermined grid of points in the ${\tilde x}-{\tilde t}$ plane. We chose
${\tilde t}$ values from $0$ to $7.0$ at intervals of $0.2$. For each such
value of ${\tilde t}$, we chose about $20$ points so as to sample ${\tilde R}$
as well as possible in the region in which ${\tilde R} > 5~ \times ~10^{-3}$;
this choice was made to reflect the fact that our absolute error in ${\tilde R}$
was estimated to be about $5~ \times 10^{-4}$. This then defined our grid.
At each ${\tilde t}$, we fit ${\tilde R}$ as a function of ${\tilde x}$ to
the form
\begin{displaymath}
\log ({\tilde R} )= -\frac{a_1a_2+a_3{\tilde x}+a_4{\tilde x}^2+f{\tilde x}^3}{a_2+a_5{\tilde 
x}+{\tilde x}^2}~,
\end{displaymath}
where $f=4/3$ and $a_1=-\log({\tilde R}(0,{\tilde t}))$.
The rationale behind our choice of the value of $f$ is as follows: When
${\tilde x} \gg {\tilde t}$, the complicated correlations between the spin labels of a
given classical trajectory at different times do not matter and $R$ is well
approximated by our `mean-field' theory (see Appendix~\ref{mf}). The mean-field
theory in this limit gives $\log({\tilde R}) \sim -4{\tilde x}/3$ and this
is what determines our choice of $f$. The error
in the fit was estimated to be roughly the same as the error in the original
computation of ${\tilde R}$; thus we did not lose anything by doing the fit.
Having tabulated the fitting parameters for each value of ${\tilde t}$ on
the grid, we evaluated the spatial Fourier transform numerically. The resulting
function of ${\tilde t}$ is expected to be smooth 
as long as $\delta {\tilde \omega}=\left(\omega -\varepsilon(k)\right)L_t$ is not too large. More 
precisely, we do not expect any oscillations on the scale of our grid spacing
in ${\tilde t}$ as long as $0.2\delta{\tilde \omega} \ll 2\pi$. As we
are interested only in $\delta{\tilde \omega} \sim 1$, we can safely interpolate
the resulting function in ${\tilde t}$. In practice we use a cubic-spline
to do the interpolation. Lastly, we do the ${\tilde t}$ integral numerically
to obtain $S(q,\omega)$. The accuracy of both numerical integrations is
quite high and so we expect that the dominant error in our calculation comes
from the interpolation; this is conservatively estimated to be a few percent at the most
for the largest values of $\delta{\tilde \omega}$.

Let us now briefly indicate the procedure used in obtaining the Fourier
transform of ${\tilde R}(0,{\tilde t})$ needed for the calculation of the scaling function
$\Phi(z)$. The available data for ${\tilde R}(0,{\tilde t})$ is fit {\em extremely} well
by the following form:
\begin{displaymath}
\log ({\tilde R} )= -\frac{\alpha {\tilde t}+a{\tilde t}^2+b{\tilde t}^3}{1+c{\tilde t}+d{\tilde 
t}^2}~,
\end{displaymath}
where the choice $\alpha = 4/3\sqrt{\pi}$ is again motivated by `mean-field'
considerations. It is now a simple matter to do the Fourier integral to
a very high accuracy using this fit and we estimate the errors involved to
be less than $0.5\%$ at the most.

\section{Calculation of tagged particle correlations in the classical model}
\label{jep}
In this appendix, we shall attempt to give a self-contained account of
the method devised by Jepsen~\cite{jepson} for the calculation
of the tagged particle correlations in the classical model introduced in
Ref~\cite{jepson}. We will try to adhere to the notation and conventions
of \cite{jepson} as far as possible.

The model is defined as follows: We begin with $N$ particles of
mass $m$
distributed uniformly along a one-dimensional segment of length $L$ with
periodic boundary conditions (we will
eventually take the thermodynamic limit $L \rightarrow \infty $ with
$N/L$ fixed to be equal to the density $\rho$). At time $t=0$ each particle is assigned a
velocity from the classical thermal ensemble defined by the usual
Maxwell-Boltzmann distribution function $g(v)=(m/2\pi T)^{1/2}e^{-mv^2/2T}$. The 
subsequent evolution of the system is purely deterministic; the particles
travel without any change in their velocities until they collide with
another particle. Every collision is elastic and the particles merely exchange
their velocities as a result of the collision.

To begin our analysis, let us label the particles from left to right with an index $i$ running 
from $0$ to $N-1$. Thus the particles are initially at positions $x_i(0)$ such that
$x_i(0)<x_j(0)$ for $i<j$. Actually, it is convenient to identify $i+N$ with $i$ because of
the periodic boundary conditions employed which identify the ends $x=0$ and
$x=L$ of the interval. Note that this labelling of the particles is left
invariant by the dynamics. We also label {\em trajectories} (which follow the
straight line defined by  $X_i(t)=x_i(0) + v_it$ on the
space-time diagram representing the evolution of the system) with an index
$i$, again with the convention that $x_i(0) < x_j(0)$ for $i<j$ (here $v_i$ is
the initial velocity of the $i^{th}$ particle). Let $x_i(t)$ denote
the {\em position} of the $i^{th}$ particle at time $t$. We wish to calculate
the correlator $B(x,t)=\langle \delta(x-x_k(t))\delta(x_k(0))\rangle$ where summation
over the repeated index $k$ is implied and the angular brackets refer to
averaging over the ensemble of initial conditions specified earlier. 

Let us now consider the quantity $A_{jk}(t)$, introduced in Appendix~\ref{integrable},
which is defined to be equal to $1$ if particle $j$ is on
trajectory $k$ at time $t$ and $0$ otherwise. Another useful quantity is
the number $n_k$ of ({\em signed}) crossings suffered by the $k^{th}$ trajectory upto time $t$. 
Every time this trajectory is hit from the left, $n_k$ decreases
by $1$ and every time it is hit from the right $n_k$ increases by $1$. Clearly,
$A_{jk}(t) = 1$ for $j=k+n_k(t)$ and zero otherwise. We may probe the
dynamics a bit more by defining another quantity $r_n(h,k,t)$ which equals
$1$ if trajectory $h$ has crossed trajectory $k$ precisely $n$ times upto
time $t$ and zero otherwise. Here too, we are talking of {\em signed} crossings;
if trajectory $h$ crosses from the left this is a negative crossing and if
it crosses from the right it is a positive crossing. Clearly $r_n$ has the
interpretation of a probability when averaged over any ensemble of initial
conditions. Let us also define the corresponding `generating function' as
\begin{equation}
s(u;h,k,t)=\sum_{n=-\infty}^{\infty}r_n(h,k,t)e^{inu}~.
\label{defgenfn}
\end{equation}

The reason for introducing $r_n$ and $s(u)$ is that $A_{jk}(t)$, which
is clearly a central quantity of interest, may be very conveniently expressed
in terms of $s(u)$ as
\begin{equation}
A_{jk}(t)=\frac{1}{N}\sum_{l=0}^{N-1}e^{-\frac{2\pi i}{N}(j-k)l}\prod_{m=0}^{N-
1}s\left(\frac{2\pi l}{N};m,k,t\right)~;
\label{Agivenbys}
\end{equation}
here we are using the convention that $s(u;k,k,t) \equiv 1$.
This is quite easy to check from the defintions of $s(u)$ and $A_{jk}$.
Moreover, it is possible to write down a fairly explicit expression for
$s(u;h,k,t)$. This takes a slightly different form depending on whether $h$ is
greater or less than $k$. If $h > k$, we have
\begin{equation}
s(u;h,k,t)=S\left [u,w_{kh}\right ]~,
\label{hgreaterk}
\end{equation}
while if $h < k$, we have
\begin{equation}
s(u;h,k,t)=e^{-iu}S\left [u,w_{kh}\right ]~.
\label{hlessk}
\end{equation}
$S\left[u,w_{kh}\right]$ used above is defined as:
\begin{equation}
S\left[u,w_{kh}\right] = e^{inu}~,
\label{definebigs}
\end{equation}
where $w_{kh} \equiv x_k(0)-x_h(0) +(v_k-v_h)t$, and $n$ is the integer that
satisfies $(n-1)L < w_{kh} < nL$. Using this definition, we can write the
following compact expression for $A_{jk}(t)$ in terms of $S$:
\begin{equation}
A_{jk}(t) = \frac{1}{N}\sum_{l=0}^{N-1}e^{-\frac{2\pi i lj}{N}}\prod_{h=0}^{N-1}
S\left[\frac{2\pi l}{N},w_{kh}\right]~.
\label{finalA}
\end{equation}

With all this machinery in place, it is a relatively straightforward matter
to calculate the correlation function we need. We begin by explicitly writing
out the ensemble averages involved:
\begin{equation}
B(x,t)=\frac{N!}{L^{N}} \int \left \{dx\right \}\int \left [dv\right ]\left (\prod_{l=0}^{N-1} 
g(v_l)\right)\delta(x_0(t)-x)~,
\label{explicitav}
\end{equation}
where we have used the definitions
$\int\left \{ dx\right \}  \equiv  \int_{0}^{L}dx_{N-1}\int_{0}^{x_{N-1}}dx_{N-2} \ldots 
\int_{0}^{x_2}dx_1~$ and $\int \left [ dv \right ] \equiv  \int_{-\infty}^{\infty}dv_{N-1}\int_{-
\infty}^{\infty}dv_{N-2}\ldots \int_{-\infty}^{\infty}dv_{0}~$ 
with $x_k(0) \equiv x_k$,
and it is understood that $x_0(0)$ is set equal to $0$ when evaluating the
right hand side of (\ref{explicitav}). Now we can
transform from particle positions to trajectories by writing
\begin{displaymath}
\delta(x_0(t)-x) = \sum_kA_{0k}\delta(X_k(t)-x)~.
\end{displaymath}
Using this and writing $A_{0k}$ in terms of $S\left[u,w_{kh}\right]$ allows us
to express our correlation function as
\begin{eqnarray}
B(x,t) & = & \frac{N}{L^{N}} \int \left [dx\right ]\int \left [dv\right ]\left (\prod_{l=0}^{N-1} 
g(v_l)\right)\frac{1}{N}\sum_{k,l=0}^{N-1}\prod_{h=0}^{N-1}S\left[\frac{2\pi 
l}{N},w_{kh}\right]\delta(X_k(t)-x)~.
\label{Btrajrep}
\end{eqnarray}
Here we have also used the fact that the integrand in this representation
is explicitly symmetric in the spatial integration variables to change the
spatial integration to $\int \left [dx\right ] \equiv\int_{0}^{L}dx_{N-1}\int_{0}^{L}dx_{N-
2}\ldots \int_{0}^{L}dx_{1}$.

It is now convenient to define $R\left[u,x_k+v_kt-x_h\right] \equiv 
\int_{-\infty}^{\infty}dv_hg(v_h)S\left[u,w_{kh}\right]$. Using this
we can rewrite our expression for the correlation function as
\begin{eqnarray}
B(x,t) & = & \frac{\rho}{N}\sum_{u}\biggl [ ~\int dv_0g(v_0)\delta(X_0(t)-x)\left(
\frac{1}{L}\int_{0}^{L}dx_hR(u,X_0(t)-x_h)\right)^{N-1} \nonumber \\
&& +~\frac{N-1}{L}\int_{0}^{L}dx_k\int_{-\infty}^{\infty}dv_kg(v_k)R(u,X_k(t))\delta(X_k(t)-x) 
\nonumber \\
&& ~~~~~~~~~~~~~~~~~~~~~~~~~~\times ~\left(\frac{1}{L}\int_{0}^{L}dx_hR(u,X_k(t)-x_h)\right)^{N-
2}~\biggr ]~,
\label{onemore}
\end{eqnarray}
where $k \neq h$, $k$,$h \neq 0$, $u \equiv 2\pi l/N$ and $\sum_{u} \equiv
\sum_{l=0}^{N-1}$.
To proceed further we need to work out $R\left[u,X_k(t)\right]$ and
$(1/L)\int_{0}^{L}dx_hR\left[u,X_k(t)-x_h\right]$. This is quite
straightforward to do in the limit of large $L$ and we only give the
final results below:
\begin{eqnarray}
R\left[u,X_k(t)\right] &=& \frac{1}{2}E_c(y)+\left(1-\frac{1}{2}E_c(y)\right)e^{iu}~, \nonumber 
\\
\frac{1}{L}\int_{0}^{L}dx_hR\left[u,X_k(t)-x_h\right] &=& 1+\frac{1}{L}\left(1-e^{-
iu}\right)T\left[u,X_k(t)\right]~,
\label{calcRintR}
\end{eqnarray}
where
\begin{eqnarray}
T\left[u,X_k(t)\right] &=& \sqrt{\frac{2T}{m}}t\biggr[ye^{iu}+\frac{e^{iu}-
1}{2}\left(\frac{1}{\sqrt{\pi}}e^{-y^2}-yE_c(y)\right)\biggr]~, \nonumber \\
E_c(y) &=& \frac{2}{\sqrt{\pi}}\int_{y}^{\infty}dze^{-z^2}~, \nonumber \\
y &=& \sqrt{\frac{m}{2T}}\frac{X_k(t)}{t}~.
\label{defTEcy}
\end{eqnarray}
Now, in the the thermodynamic limit specified earlier we can write
\begin{displaymath}
\left(1+\frac{1}{L}(1-e^{-iu})T\left[u,X_k(t)\right]\right)^{N-\nu} =
\exp\left(\rho(1-e^{-iu})T\left[u,X_k(t)\right]\right)~,
\end{displaymath}
valid for any finite number $\nu$.
Using this and (\ref{calcRintR}) in the expression (\ref{onemore}) for the
correlation function and doing the remaining integrals over positions and
velocities gives us
\begin{eqnarray}
B(x,t) &=& \rho \int_{0}^{2\pi}\frac{du}{2\pi}\biggl [\rho 
\left(2f_1(w)f_2(w)+e^{iu}f_2^2(w)+e^{-iu}f_1^2(w)\right)\exp\left(\rho(1-e^{-
iu})T\left[u,x\right]\right) \nonumber \\
 & & ~~~~~~~~~~~~~~~~~~~~~~~~~~+\frac{1}{t}\sqrt{\frac{m}{2\pi T}}~e^{-w^2}\exp\left(\rho(1-e^{-
iu})T\left[u,x\right]\right)\biggr]~;
\label{finally}
\end{eqnarray}
here we have replaced the sum over $u$ by the corresponding integral in the
thermodynamic limit and used the definitions $f_1(w) \equiv E_c(w)/2$, $f_2(w)
\equiv 1-f_1(w)$ and $w \equiv (m/2T)^{1/2}x/t$. To do the $u$ integral, we
note that $T$ may be expressed as $Ge^{iu} -A$ where $A$ and $G$ are functions
purely of $x$ and $t$. This allows us to use the standard Bessel function identity,
\begin{eqnarray}
\frac{1}{2\pi}\int_{0}^{2\pi}due^{-inu}\exp\left(\rho(1-e^{-iu})(Ge^{iu}-A)\right) 
& =& \left(\frac{G}{A}\right)^{\frac{n}{2}}e^{-\rho(A+G)}I_n(2\rho\sqrt{AG})~,
\label{tisdone}
\end{eqnarray}
to finally arrive at the results quoted in (\ref{czz3}) of Section~\ref{hdep}
upon using the appropriate values for $\rho$ and $m$.
 
\section{Approximate analytical calculation of the relaxation function}
\label{mf}

In this appendix, we briefly outline our approximate `mean-field' theory
for the relaxation function $R(x,t)$.

We begin by noting that the classical model defined in Section~\ref{finitet}
has been solved exactly in Ref~\cite{ssyoung} for the special case in which
there is only one possible value for the spin label $m$. All of the
difficulties we encounter in attempting to generalize this solution
to the case of interest here stem from the fact that there are complicated
correlations between the $m_k(t)$ (defined in Section~\ref{finitet})
at different times.

\begin{figure}
\epsfxsize=4.5in
\centerline{\epsffile{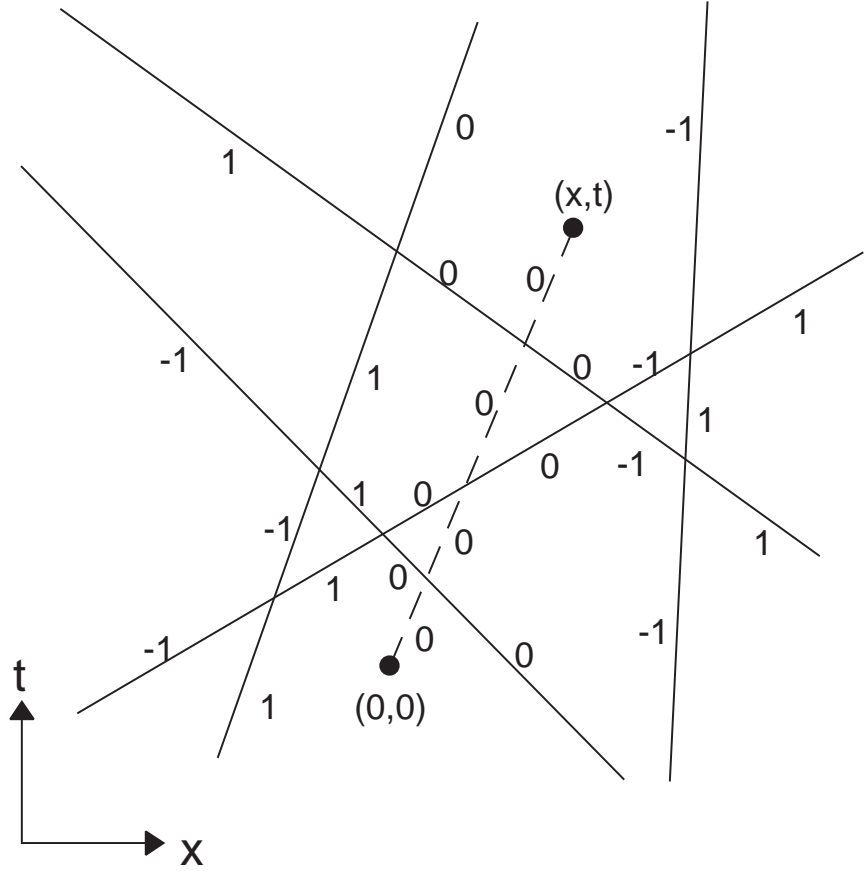}}
\vspace{0.2in}
\caption{A typical set of particle trajectories contributing to $C(x,t)$.
Each full line represents paths traversed by particles moving both forward and backward in time.
The dashed line is traversed only going forward in time.
Shown on the trajectories are the values of the particle spins $m_k$ which
are independent of $t$ in the low $T$ limit. Notice that all the trajectories
intersecting the dashed line have a spin equal to that of the dashed line:
only such configurations contribute to the $C(x,t)$. }
\label{fig9}
\end{figure}

\begin{figure}
\epsfxsize=4.5in
\centerline{\epsffile{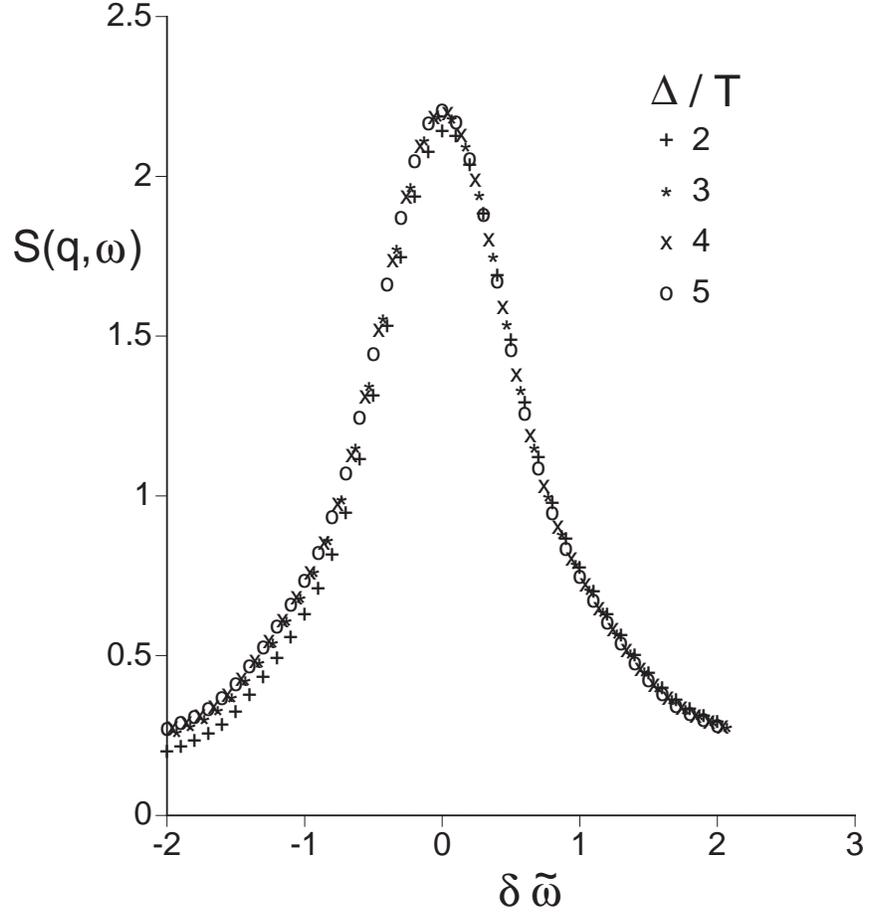}}
\vspace{0.2in}
\caption{$S(q,\omega )$ rescaled by a factor of ${\cal  A}cL_t/(\pi^2 \Delta)$
plotted against $\delta {\tilde \omega}=L_t(\omega -\Delta)$ with $q=\pi/a$ for $\Delta/T=2$, 
$3$, $4$, and $5$. Note the scaling collapse of the curves corresponding to
the three lowest temperatures.}
\label{fig10}
\end{figure}

\begin{figure}
\epsfxsize=4.5in
\centerline{\epsffile{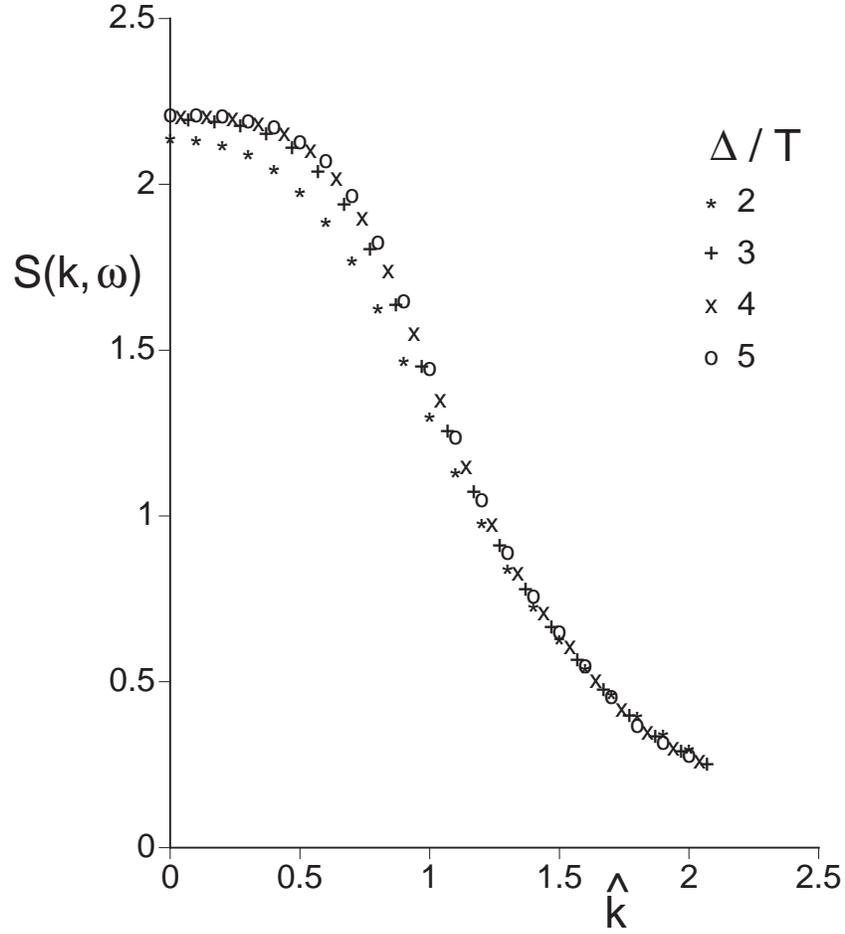}}
\vspace{0.2in}
\caption{$S(\pi/a+k,\omega)$ rescaled by a factor of ${\cal A}cL_t/(\pi^2 \Delta)$
plotted against the rescaled variable ${ {\hat k}}=kc\left(L_t/\Delta \right)^{1/2}$ with 
$\omega=\Delta$ for $\Delta/T=2$, $3$, $4$, and $5$.
Again, note the scaling collapse of the curves corresponding to
the three lowest temperatures.}
\label{fig11}
\end{figure}

\begin{figure}
\epsfxsize=4.5in
\centerline{\epsffile{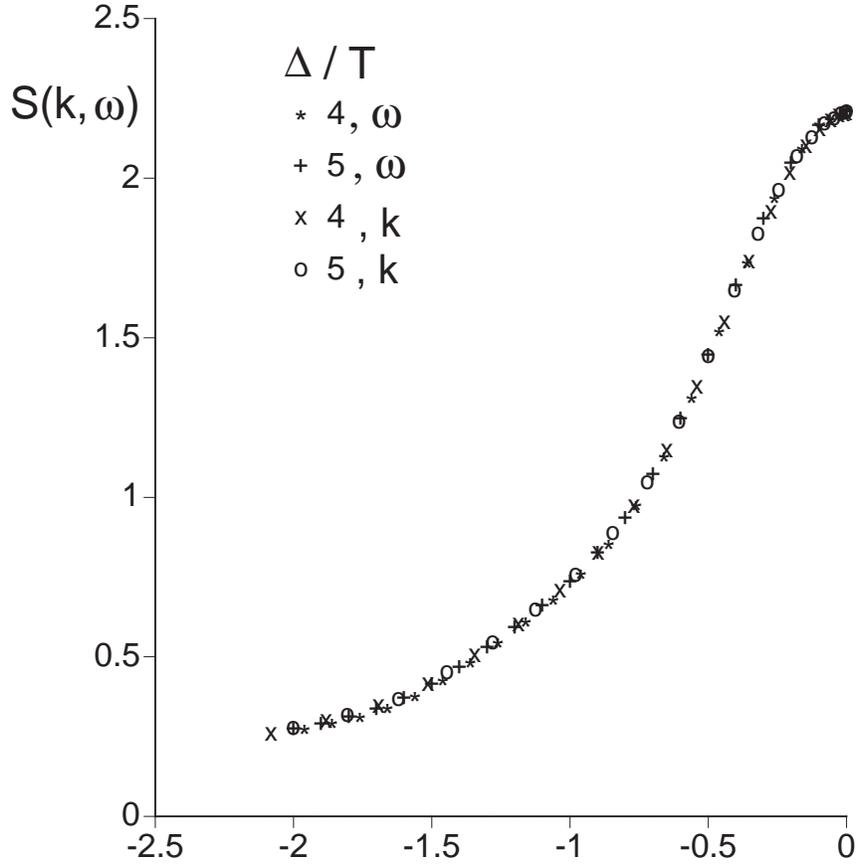}}
\vspace{0.2in}
\caption{The scaling curve of Fig~{\protect \ref{fig11}} (defined by the data
for $\Delta/T = 4$ and $\Delta/T = 5$) plotted against the independent variable
$(-{\hat k}^2/2)$ compared to the scaling curve of Fig~{\protect \ref{fig10}}
(again defined by the data for $\Delta/T=4$ and $\Delta/T =5$) for the corresponding negative 
values of $\delta {\tilde \omega}$. The two
coincide within our numerical errors.}
\label{fig12}
\end{figure}

\begin{figure}
\epsfxsize=4.5in
\centerline{\epsffile{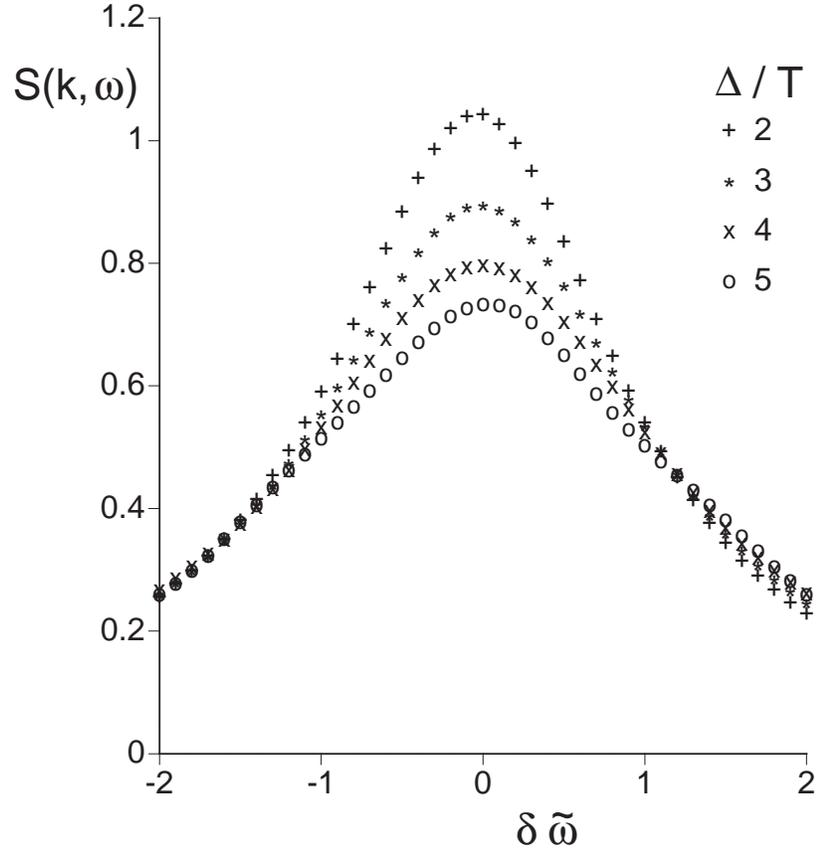}}
\vspace{0.2in}
\caption{$S(k+\pi/a,\omega )$ rescaled by a factor of ${\cal  A}cL_t/(\pi^2 \Delta)$
plotted against ${\delta{\tilde  \omega}}=L_t(\omega -2^{1/2}\Delta)$ with $k=\Delta/c$ for 
$\Delta/T=2$, $3$, $4$, and $5$.}
\label{fig13}
\end{figure}

\begin{figure}
\epsfxsize=4.5in
\centerline{\epsffile{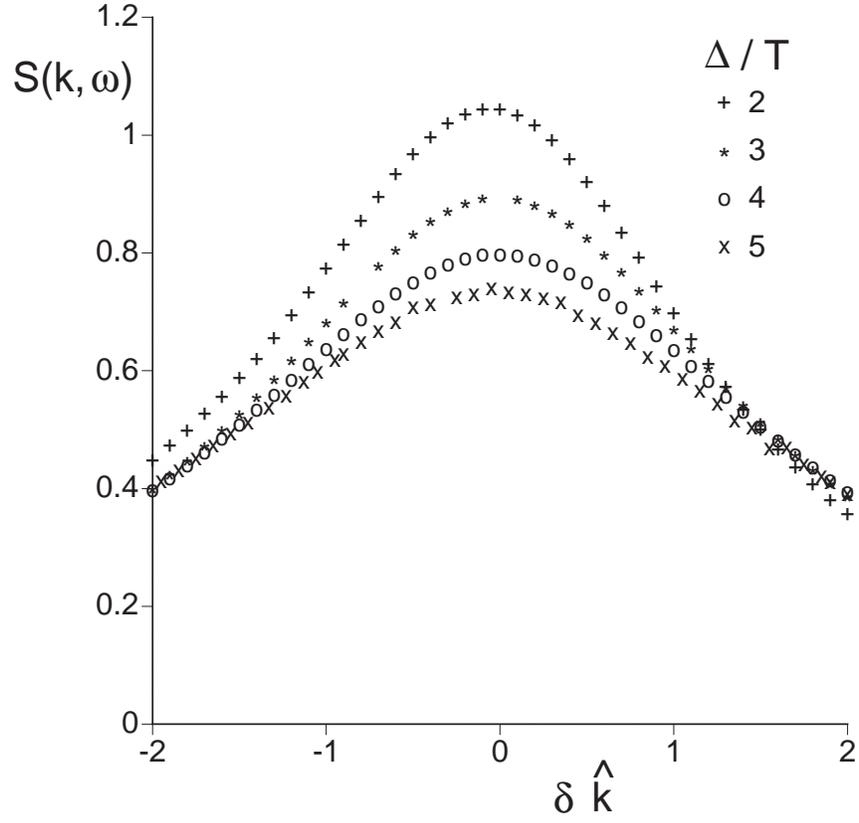}}
\vspace{0.2in}
\caption{$S(k+\pi/a,\omega)$ rescaled by a factor of ${\cal A}cL_t/(\pi^2 \Delta)$
plotted against the rescaled variable ${\delta{\hat k}}=cL_t(k-\Delta/c)$ with 
$\omega=2^{1/2}\Delta$ for $\Delta/T=2$, $3$, $4$, and $5$.}
\label{fig14}
\end{figure}

\begin{figure}
\epsfxsize=4.5in
\centerline{\epsffile{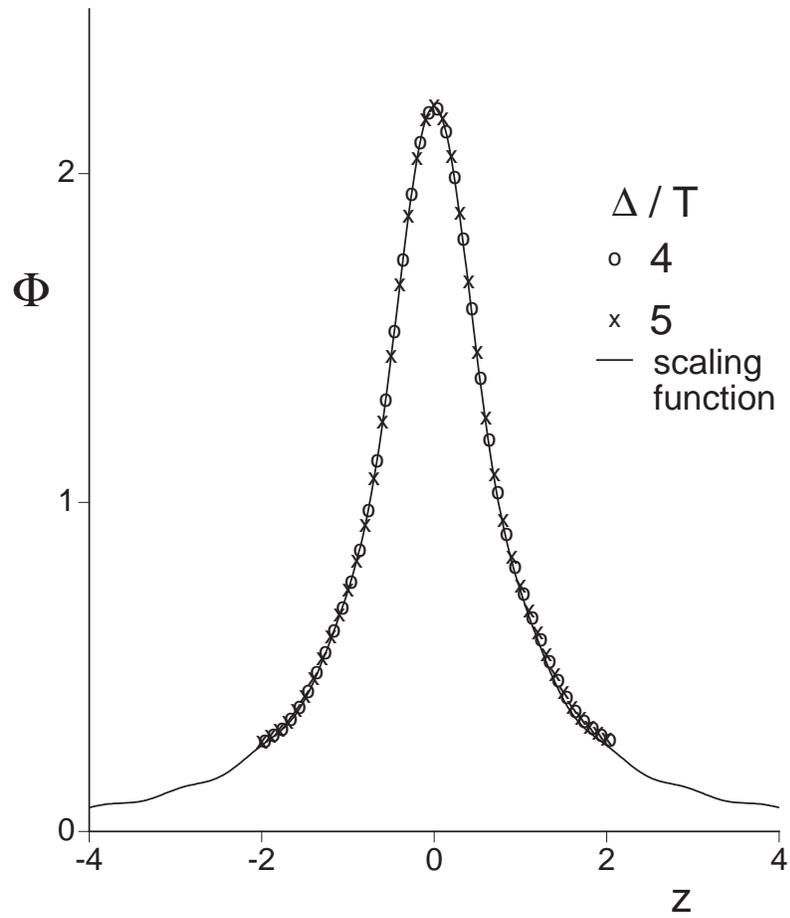}}
\vspace{0.2in}
\caption{The scaling function $\Phi(z)$ determined directly from Eqn~{\protect \ref{dirscalfn}} 
compared with the scaling curve defined by the results already
shown in Fig~{\protect \ref{fig10}} and Fig~{\protect \ref{fig12}}}
\label{fig15}
\end{figure}

\begin{figure}
\epsfxsize=4.5in
\centerline{\epsffile{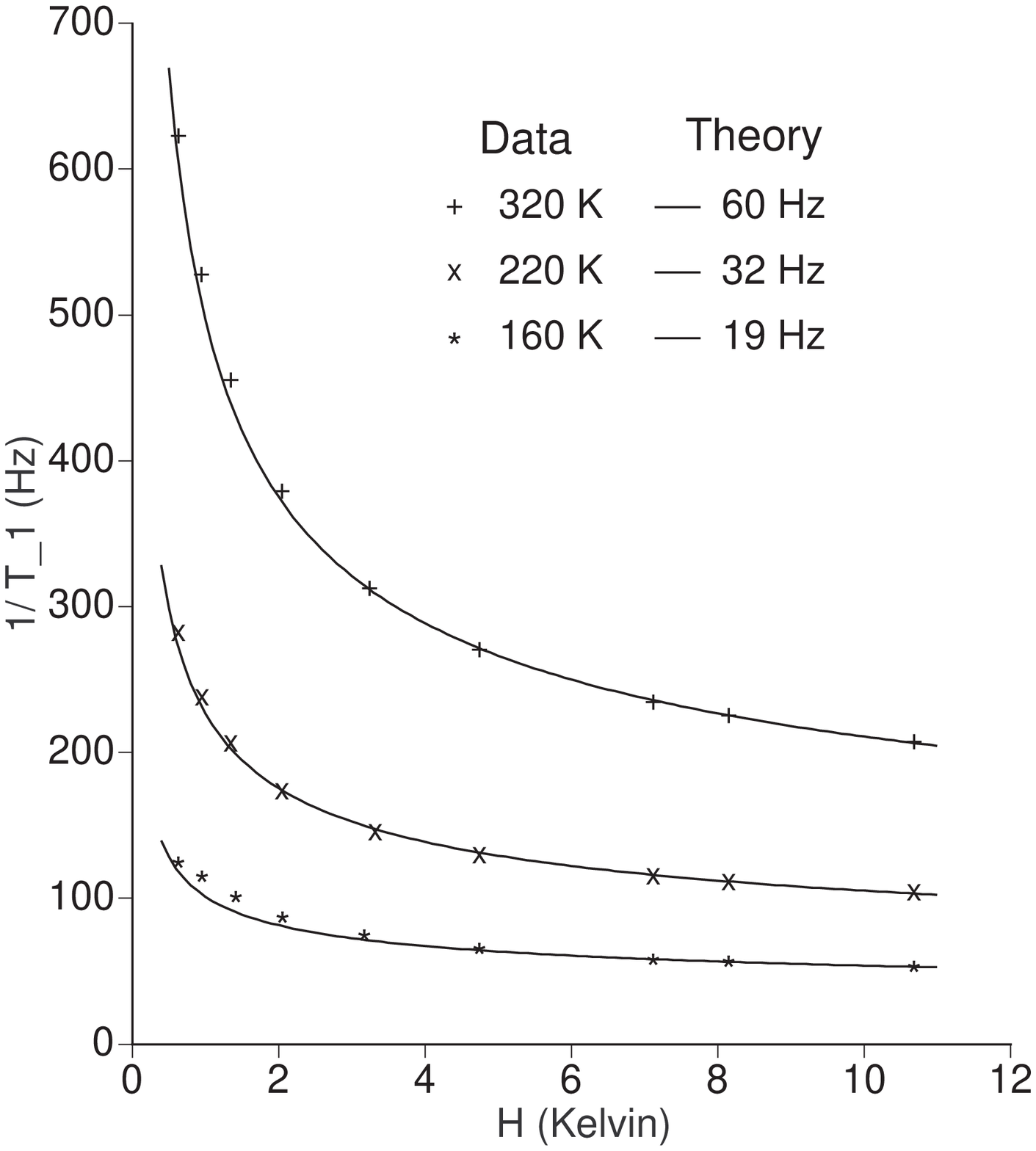}}
\vspace{0.2in}
\caption{Field dependence of $1/T_1$ for $T > 120$ K. The experimental data of
{\protect \cite{taki}}
is compared with the theoretical predictions offset by a field-independent background
rate $R_b$ which is the only free parameter of the fit; the fit value of $R_b$ is
shown under the theory column.}
\label{fig16}
\end{figure}

\begin{figure}
\epsfxsize=4.5in
\centerline{\epsffile{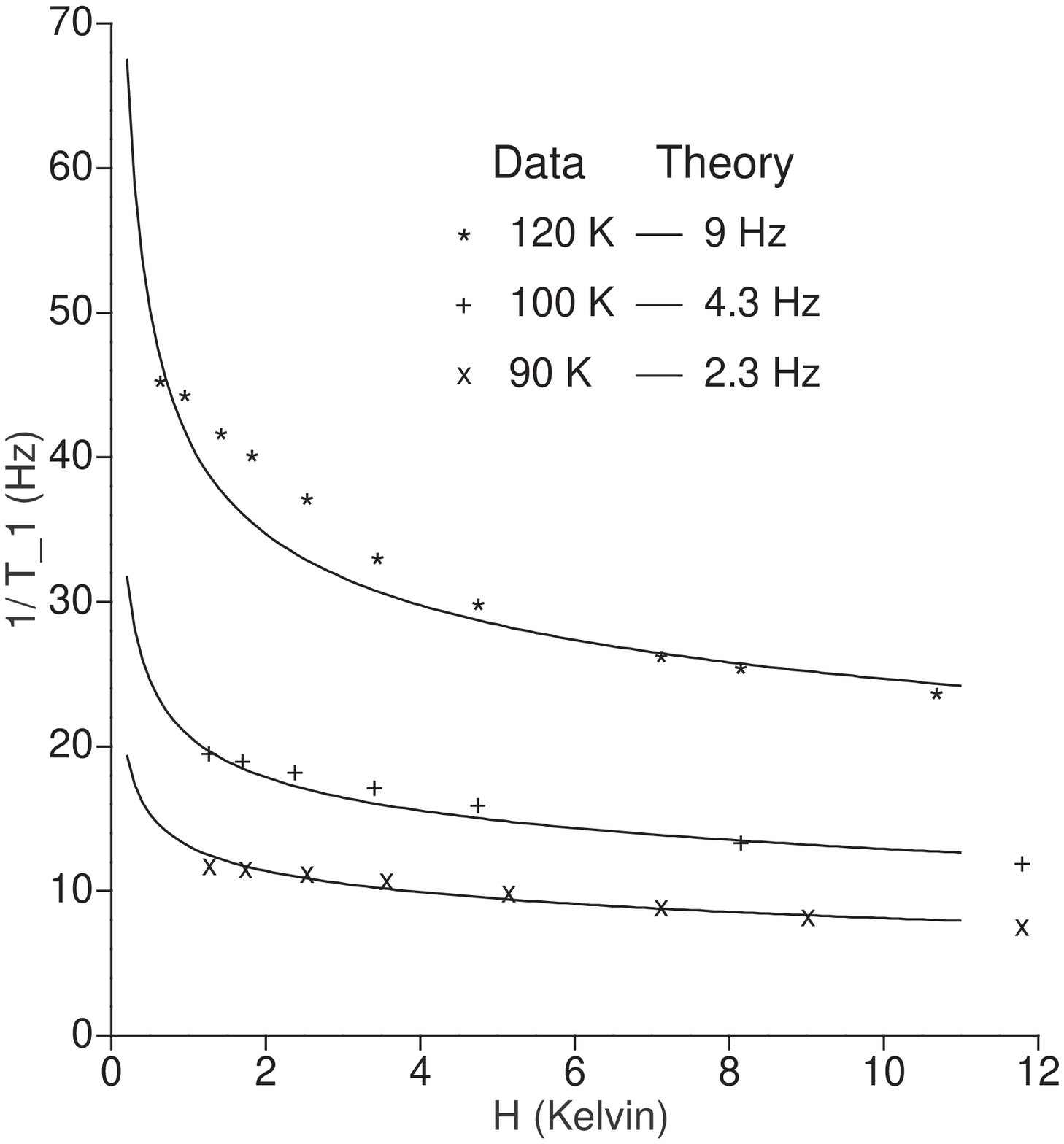}}
\vspace{0.2in}
\caption{Field dependence of $1/T_1$ for a few temperatures $T <120$ K. The experimental data of 
{\protect \cite{taki}} is
compared with the theoretical predictions offset by a field-independent background
rate $R_b$ which is the only free parameter of the fit; the fit value of $R_b$ is
shown under the theory column.}
\label{fig17}
\end{figure}

\begin{figure}
\epsfxsize=4.5in
\centerline{\epsffile{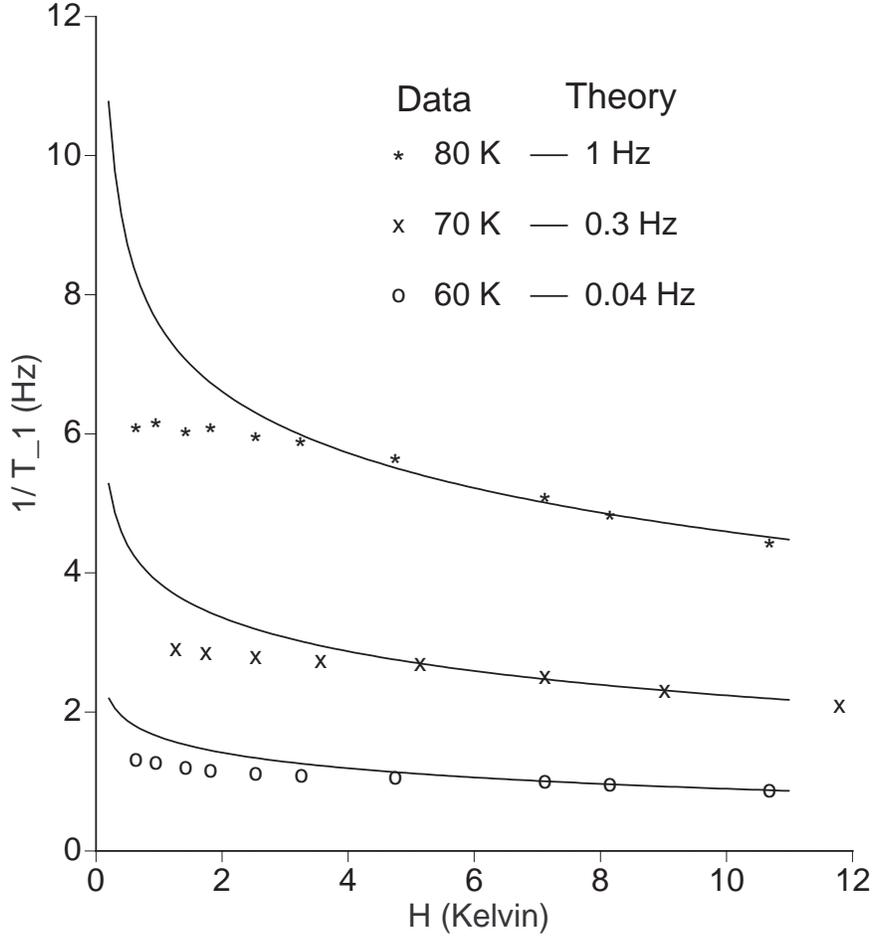}}
\vspace{0.2in}
\caption{Field dependence of $1/T_1$ for the lowest temperatures for which
data is available. The experimental data of {\protect \cite{taki}} is
compared with the theoretical predictions offset by a field-independent background
rate $R_b$ which is the only free parameter of the fit; the fit value of $R_b$ is
shown under the theory column.}
\label{fig18}
\end{figure}

\begin{figure}
\epsfxsize=4.5in
\centerline{\epsffile{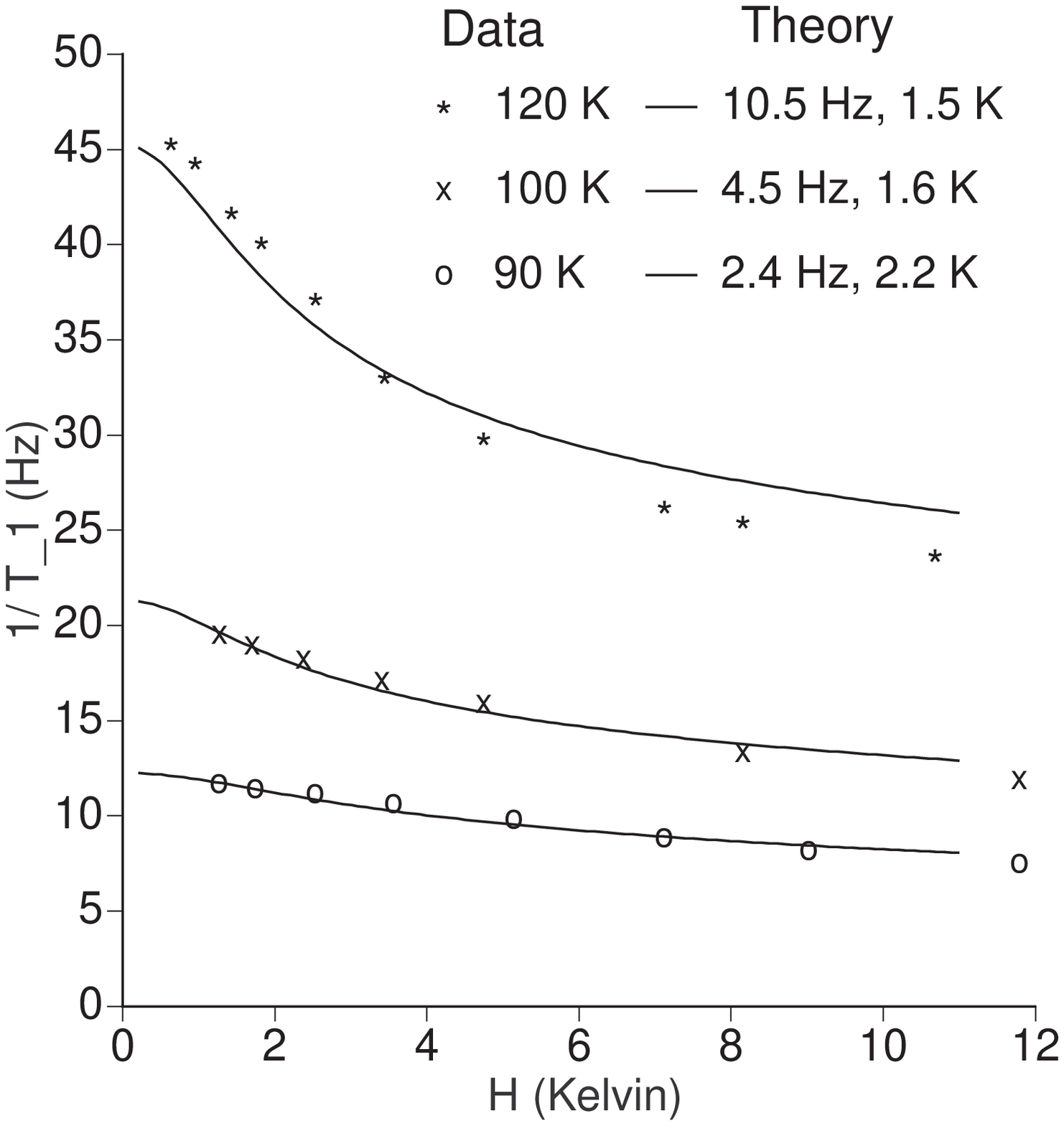}}
\vspace{0.2in}
\caption{Field dependence of $1/T_1$ fit to the phenomenological form described
in the text. The experimental data of {\protect \cite{taki}} at $T=120$, $100$, and $90$ K is 
compared to
our phenomenological form
that incorporates a spin dissipation rate $\gamma $ in addition to a
field independent background rate $R_b$. The values of $R_b, \gamma$ are listed
under the theory column.}
\label{fig19}
\end{figure}

\begin{figure}
\epsfxsize=4.5in
\centerline{\epsffile{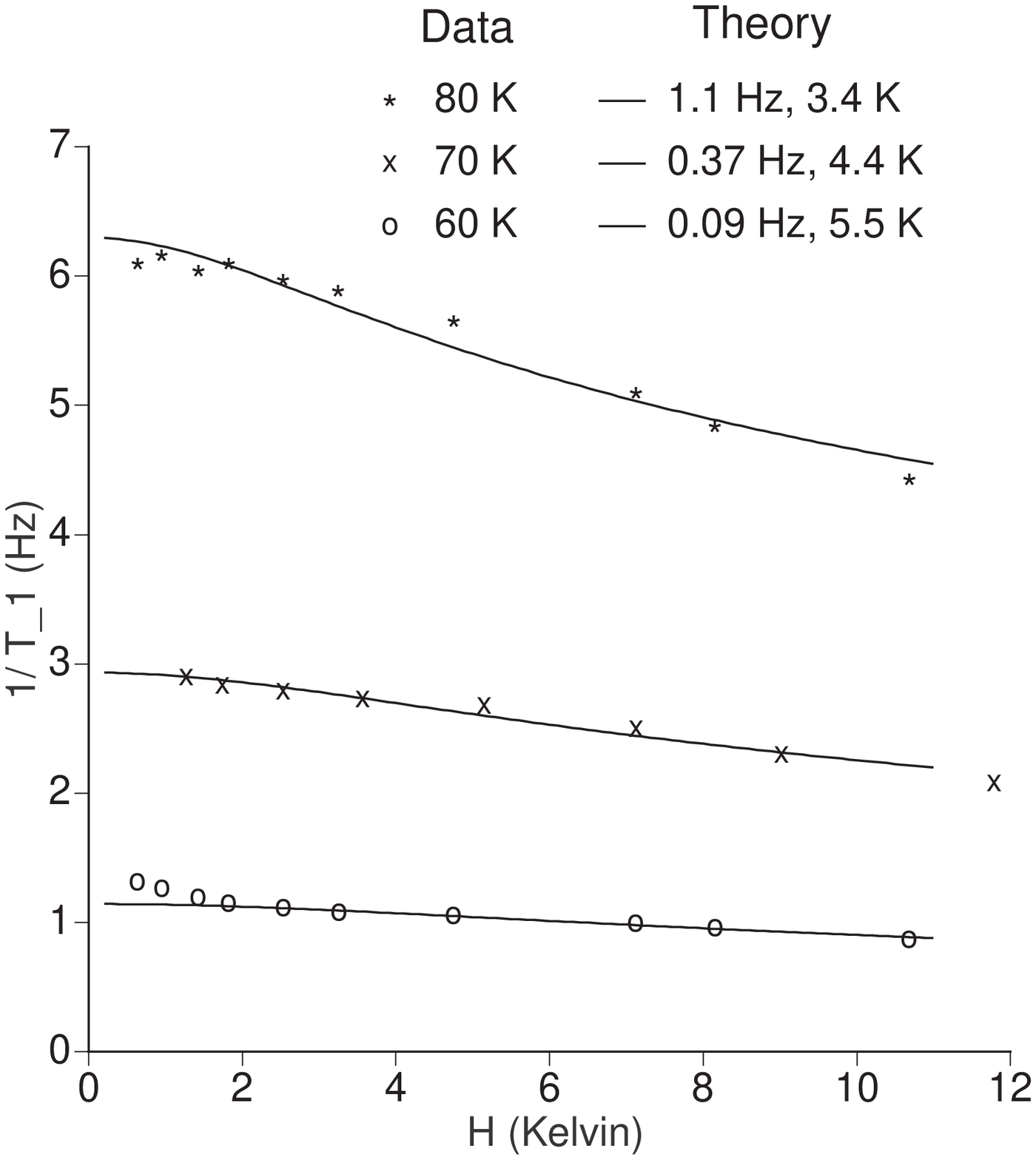}}
\vspace{0.2in}
\caption{Field dependence of $1/T_1$ fit to the phenomenological form described
in the text. The experimental data of {\protect \cite{taki}} at $T=80$, $70$, and $60$ K is 
compared to
our phenomenological form
that incorporates a spin dissipation rate $\gamma$ in addition to a
field independent background rate $R_b$. The values of $R_b, \gamma$ are listed
under the theory column.}
\label{fig20}
\end{figure}

\begin{figure}
\epsfxsize=4.5in
\centerline{\epsffile{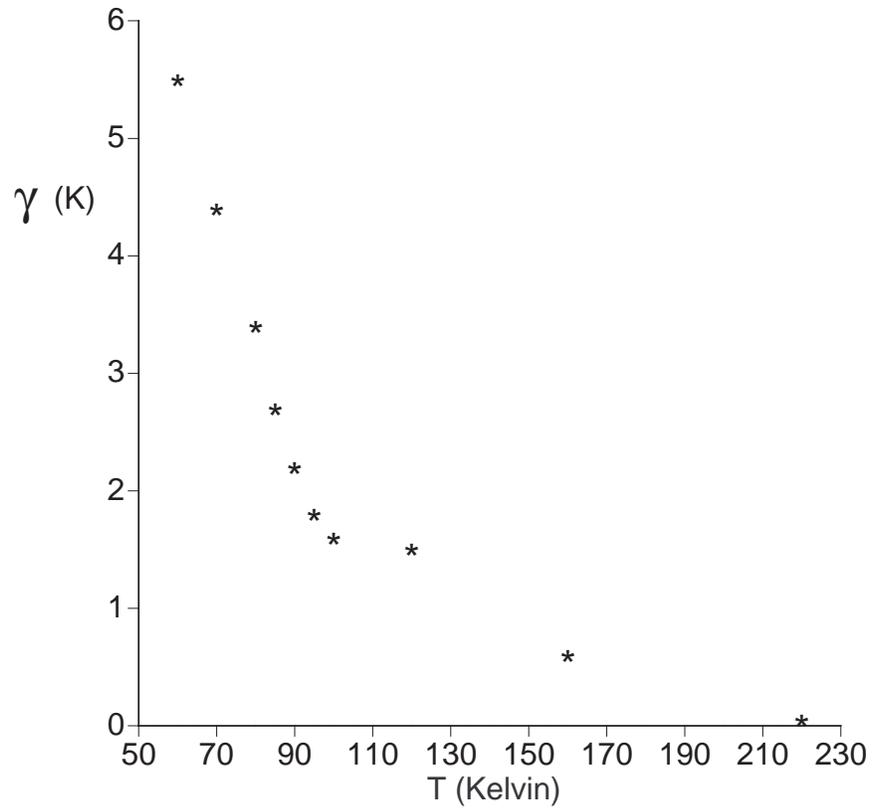}}
\vspace{0.2in}
\caption{Temperature dependence of the spin-dissipation rate $\gamma$ determined by fitting our 
phenomenological
form for $1/T_1$ to the experimental data of \protect{\cite{taki}}.}
\label{fig21}
\end{figure}

\begin{figure}
\epsfxsize=4.5in
\centerline{\epsffile{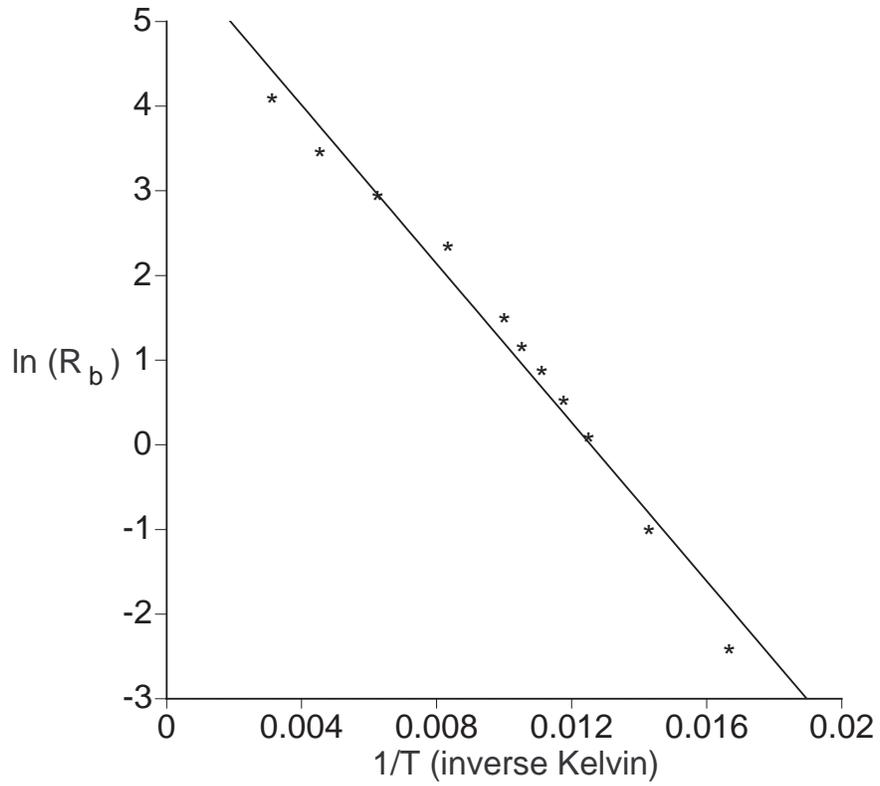}}
\vspace{0.2in}
\caption{Temperature dependence of the background rate $R_b$ determined by fitting our 
phenomenological
form for $1/T_1$ to the experimental data of \protect{\cite{taki}}. We plot
$\ln(R_b)$ against $1/T$ to check for activated behaviour and indeed find an approximate linear 
relation, the best fit for the slope being $468$ K.}
\label{fig22}
\end{figure}

\begin{figure}
\epsfxsize=4.5in
\centerline{\epsffile{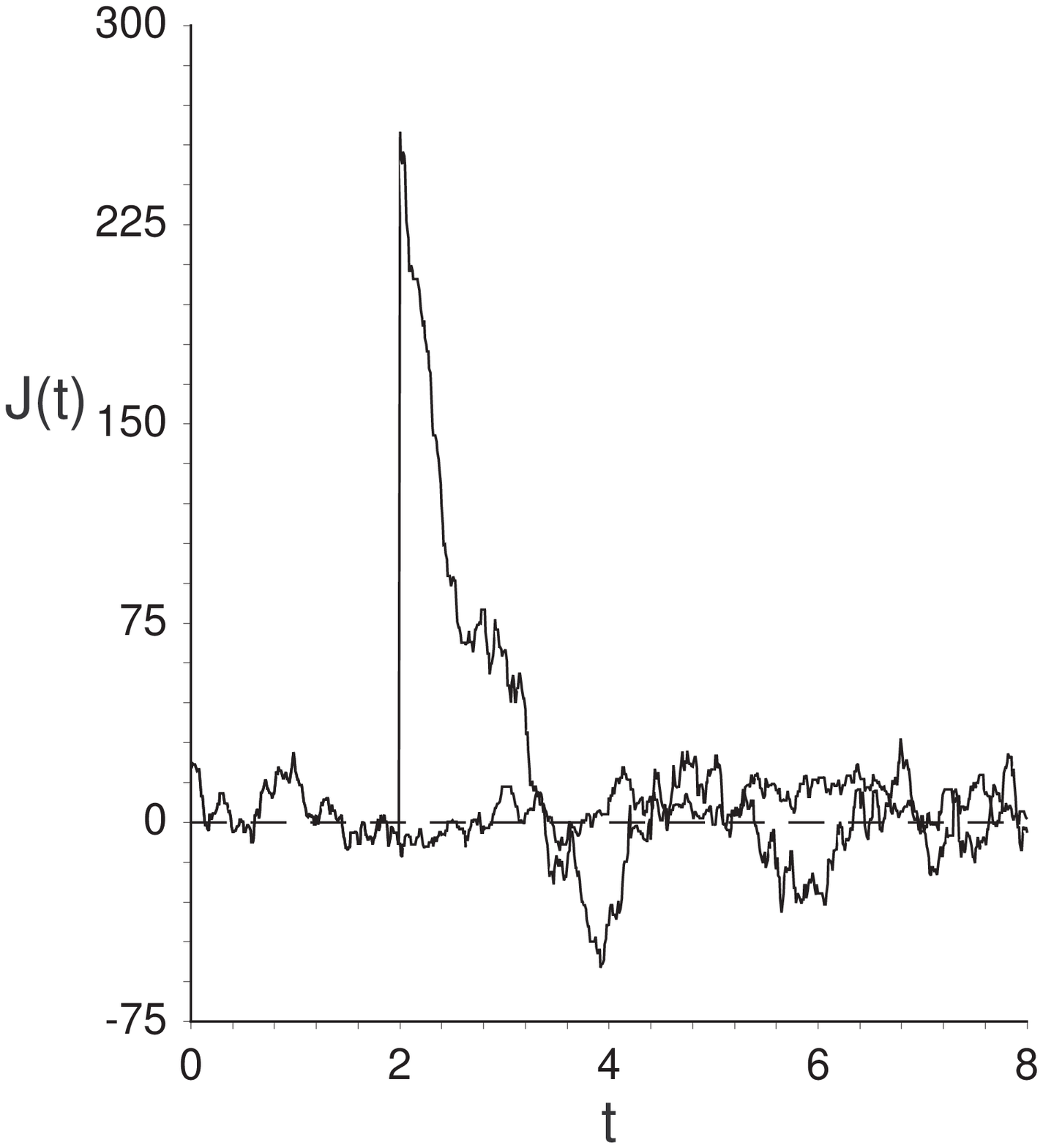}}
\vspace{0.2in}
\caption{Deterministic time evolution of the spin current $J(t)$ (defined in 
(\protect\ref{integ3})) for two systems of 400 particles
on a circle with the same initial condiions; the value of $J(t)$ changes in
discrete steps at each collision between a pair of particles.  For one of systems, 
there is an impulse in velocities
given by (\protect\ref{integ4}) at a time $t_0 = 2$. This produces a
macroscopically significant $J(t)$, which however decays away in a few collision
times. The only remnant of the impulse is a `heating' of the system,
reflected in the larger amplitude of the order $\protect\sqrt{N}$ fluctuations in $J(t)$
for the impacted system.
}
\label{fig23}
\end{figure}

Our `mean-field' approximation consists of simply ignoring these correlation
effects (hence our choice of terminology to describe our approximation).
Having made this uncontrolled approximation, it is now a fairly straightforward
matter to obtain a closed form expression for $R(x,t)$ in analogy with
the corresponding discussion in \cite{ssyoung}. The actual calculation
proceeds as follows: Let $q$ be the probability that any given
solid line in Fig~\ref{fig9} intersects the dotted line. If we ignore the
correlations between the $m_k(t)$ at different times, then the probability that
this line carries a spin label equal to the spin label of the dotted line is
$1/3$. So given that the line intersects the dotted line, this intersection contributes
a factor of $-1$ to $R(x,t)$ with probability $1/3$ and a factor of $0$
with probability $2/3$ (if the line does not intersect at all, we of course
get a factor of $1$). Within our mean field theory, $R$ is just a product of such factors, one 
from each solid line. This gives $R(x,t)=(1-q -q/3)^N$, where
$N$ is the total number of thermally excited particles in the system. Now, using
$q = \langle |x-vt|\rangle /L$~~\cite{ssyoung} (where the angular brackets
denote averaging over the Maxwell-Boltzmann distribution function for $v$ and
$L$ is the length of the system)
and taking the thermodynamic limit, we obtain $R(x,t)=\exp (-4\rho\langle |x-vt|\rangle/3)$. We 
can now do our
usual rescalings and write down the main result of our mean field theory:
\begin{equation}
{\tilde R}({\tilde x},{\tilde t}) = \exp (-4\langle |{\tilde x}-{\tilde v}{\tilde t}| \rangle 
/3)~,
\label{mfRtilde}
\end{equation}
where the angular brackets now denote averaging over the distribution
${\tilde {\cal P}}({\tilde v}) = \frac {1}{\sqrt \pi } e^{-{\tilde v}^2}$
and ${\tilde x}$ and ${\tilde t}$ are defined as in Section~\ref{finitet}.
In particular, note that this implies ${\tilde R}(0,{\tilde t})=e^{-4|{\tilde t}|/3\sqrt{\pi}}$; 
this
turns out to be reasonably accurate for some purposes (see the discussion on
the approximate form of the scaling function $\Phi(z)$ in Section~\ref{finitet}).

\newpage

\begin{references}
\bibitem{note1}
The simplest non-local 
conserved charge in Ref.~\protect\cite{luscher2}, $Q^{ab}$,
does have the same quantum numbers as the spin current.
The discussion in Appendix~\protect\ref{integrable}
would then suggest that $Q^{ab}$ could prevent complete decay
of the spin current. However the non-local nature of $Q^{ab}$ 
suggests to us that this is unlikely. In a system with $N$
thermally excited particles, $Q^{ab}$ consists of two components:
a local piece with order $N$ terms which is proportional to the
spin current at low momentum, and a second non-local piece with order
$N^2$ terms.
In a typical thermal state, the naive expectation is that
the value of the local piece will be of order $\sqrt{N}$,
while that of the non-local piece will be of order $N$.
If we now set up a spin current with an impulse, as
in Appendix~\protect\ref{integrable}, the local piece
will jump to a term
of order $N$. However this does not change the magnitude
of the total value of $Q^{ab}$, and it is reasonable to expect
that, within a few collision times, there will be
a redistribution
within the two pieces of $Q^{ab}$ so that their orders of 
magnitude are
similar to those in a typical thermal state.
\end{references}
\end{document}